%% file: main.tex
\def\draft{1}
\documentclass[11pt]{article}
\input{header}
\usepackage[normalem]{ulem}
\newcommand{\CSP}{\ensuremath{\mathsf{CSP}}\xspace}

\newcommand{\MinCSP}{\ensuremath{\mathsf{Min}\text{-}\mathsf{CSP}}\xspace}

\newcommand{\MaxCSP}{\ensuremath{\mathsf{Max}\text{-}\mathsf{CSP}}\xspace}

\newcommand{\ExactCSP}{\ensuremath{\mathsf{Exact}\text{-}\mathsf{CSP}}\xspace}

\newcommand{\Core}{\ensuremath{\operatorname{core}}\xspace}
\newcommand{\Aut}{\ensuremath{\operatorname{Aut}}\xspace}

\newcommand{\Sat}{\ensuremath{\mathrm{Sat}}\xspace}
\newcommand{\Index}{\ensuremath{\mathsf{INDEX}}\xspace}

\newcommand{\XOR}{\ensuremath{\mathsf{XOR}}\xspace}
\newcommand{\cdeg}{\deg_{\mathbb Z}}
\newcommand{\sketchlowerboundstatement}{For a fixed \amat{nonempty, non-constant-satisfiable} Boolean \amat{relation} $R$ with $d=\cdeg(R)$, there is a constant $F_R$ such that, for every sufficiently large $n$, there is a distribution on instances on $n$ variables with frequencies at most $F_R$ for which every public-coin linear sketch deciding satisfiability with average error at most $1/3$ has \amat{size} $\amat{\Omega_k}(n^d)$. Every constraint in these instances has distinct variables within its scope.}
\newcommand{\unregularizedsparsificationstatement}{%
Let $B\in\Z^{m\times D}$ satisfy the sparsity, norm, and row-access assumptions of \cref{thm:ub-recovery}, and let a strict-turnstile stream maintain integer frequencies $0\leq f_i\leq\mu$. Set $G=B^\top\diag(f)^2B$. Suppose a known $\kappa\geq1$ satisfies $\lambda_{\max}(G)/\lambda_{\min}^+(G)\leq\kappa$ whenever $G\ne0$, where $\lambda_{\min}^+(G)$ is the smallest positive eigenvalue. Given $0<\varepsilon_0<1$ and $0<\delta<1/2$, a randomized one-pass algorithm returns active indices $S$ and positive weights forming $W=\sum_{i\in S}w_ib_ib_i^\top$ such that, with probability at least $1-\delta$,
\[
 (1-\varepsilon_0)G\preceq W\preceq(1+\varepsilon_0)G.
\]
For $Q_\kappa=2+mD\mu\kappa/(\varepsilon_0\delta)$, the output contains $O_{s_0,b_0}(D\varepsilon_0^{-2}\polylog Q_\kappa)$ indices, and the algorithm uses $O_{s_0,b_0}(D\varepsilon_0^{-4}\polylog Q_\kappa)$ bits, including seeds, output, and working space. The time per update and the time after the stream are polynomial in $m,D,\mu,\varepsilon_0^{-1},\delta^{-1},\kappa$ and the time needed to obtain a row. If $G=0$, the successful output is empty after zero rows are ignored.
}
\newcommand{\snote}[1]{\ifnum\draft=1\textcolor{red}{[\textbf{Santhoshini:} #1]}\fi}
\title{A New Characterization of Streaming Decidability of Boolean CSPs Under Deletions%
}
\author{%
\begin{tabular}{@{}c@{\hspace{1.3em}}c@{\hspace{1.3em}}c@{}}
Aditya Bandekar & Amatya Sharma & Santhoshini Velusamy\thanks{Supported in part by NSERC Discovery grant RGPIN-2026-04689 and ECR supplement DGECR-2026-00421.} \\[0.4em]
{\footnotesize University of Waterloo} &
{\footnotesize University of Michigan, Ann Arbor} &
{\footnotesize University of Waterloo} \\[0.2em]
{\scriptsize \href{mailto:aditya.bandekar@uwaterloo.ca}{\nolinkurl{aditya.bandekar@uwaterloo.ca}}} &
{\scriptsize \href{mailto:amatya@umich.edu}{\nolinkurl{amatya@umich.edu}}} &
{\scriptsize \href{mailto:santhoshini.velusamy@uwaterloo.ca}{\nolinkurl{santhoshini.velusamy@uwaterloo.ca}}}
\end{tabular}%
}
\date{}
\hypersetup{
  pdftitle={A New Characterization of Streaming Decidability of Boolean CSPs Under Deletions},
  pdfauthor={Aditya Bandekar, Amatya Sharma, Santhoshini Velusamy}
}

\begin{document}
\hypersetup{pageanchor=false}
\maketitlewithparagraphnotes

\begin{abstract}
We characterize the single-pass streaming complexity of deciding satisfiability for Boolean Constraint Satisfaction Problems (CSPs) in the strict-turnstile model, for every fixed constraint language $\Gamma$ that is not trivially satisfiable. Here $\Gamma$ is a finite set of $k$-ary relations over $\{0,1\}$, and an instance consists of constraints $R(x_{i_1},\ldots,x_{i_k})$, with $R\in\Gamma$, on $n$ Boolean variables. In the strict-turnstile model, constraints may be inserted and deleted while their multiplicities remain nonnegative, and the goal is to decide whether all constraints of positive multiplicity at the end are simultaneously satisfiable.

Sharma and Velusamy [ESA 2026] characterized the insertion-only space complexity by \emph{non-redundancy}, up to a logarithmic factor. We show that the strict-turnstile space complexity is instead characterized, up to polylogarithmic factors, by the \emph{integer capturing degree}. For a relation $R\subseteq\{0,1\}^k$, its integer capturing degree $d_R$ is the minimum degree of a polynomial with integer coefficients that vanishes exactly on $R$ within the Boolean hypercube. For a constraint language $\Gamma$, let $d_\Gamma=\max_{R\in\Gamma}d_R$. On polynomial-length streams, we give an $O_k(n^{d_\Gamma}\operatorname{polylog} n)$-space randomized algorithm and prove an $\Omega_k(n^{d_\Gamma})$ lower bound against all algorithms. Our algorithm goes beyond deciding satisfiability: it constructs a sparse instance that preserves the entire set of satisfying assignments of the original instance.

To our knowledge, this gives the first separation between insertion-only and turnstile complexity in the streaming CSP literature. Our characterization also differs from the Fourier-degree characterization of \emph{exact optimization}—computing the maximum number of constraints that can be satisfied simultaneously—by Kol, Paramonov, Saxena, and Yu [ITCS 2023]. In particular, for $k\geq 3$, $k$-XOR with literals exhibits three distinct space complexities: $\widetilde{\Theta}(n)$ for insertion-only satisfiability, $\widetilde{\Theta}(n^{\lceil k/2\rceil})$ for strict-turnstile satisfiability, and $\widetilde{\Theta}(n^k)$ for exact optimization in either model.

Our upper bound combines capturing polynomials with spectral sparsification, using integrality to turn approximate spectral guarantees into exact preservation of satisfying assignments. This approach extends to CSPs over arbitrary fixed finite domains. For Boolean relations, we establish an exact duality between the capturing degree and the minimum degree of a polynomial that is nonzero at exactly one unsatisfying tuple. 
We use the coefficients of this polynomial to combine linear sketches in a reduction from INDEX, obtaining a nearly tight sketching lower bound.
The framework of Jiang, Liu, and Yu [FOCS 2026] then transfers this bound to arbitrary strict-turnstile streaming algorithms on polynomial-length streams with only a constant-factor loss in space.

\end{abstract}

\newpage

\tableofcontents

\newpage
\hypersetup{pageanchor=true}

\input{intro}

\input{technicaloverview}

\paragraph*{Acknowledgements.}
We gratefully acknowledge Aaron (Louie) Putterman and Joshua Brakensiek for helpful discussions in the initial stages of this project. In particular, Louie generated a proof of an $\Omega(n^2)$-space lower bound for $3$-XOR using GPT-6 Astra and shared it with us, which served as a key inspiration for our own hardness proof and subsequent generalizations. He also pointed us to relevant work in the streaming sparsification literature, deepening our understanding of this area. SV also thanks Samson Zhou for raising the question of deciding satisfiability in the turnstile setting following her talk at WOLA 2026.
\sant{We also thank Bart M. P. Jansen for bringing the notion of nondeterministic degree of a CSP to our attention and for pointing out its connection to our work.}

\paragraph*{AI disclosure.}
GPT-6 Astra was used in the initial stages of this project to develop a lower-bound argument for $3$-XOR. We subsequently generalized this argument to the Boolean CSPs studied in this paper without LLM assistance. Our contributions in this phase included identifying capturing degree as the governing complexity parameter and establishing its duality with selection degree.

For the upper bound, we were aware of a simple $O(n^{\lceil k/2\rceil})$-space algorithm for $k$-XOR through a reduction to the connectivity problem in \cite{ahn2012analyzing}. We suspected that existing ideas in the streaming sparsification literature might suffice to generalize this algorithm to all CSPs. GPT-6 Astra then produced a proof of \cref{cor:overview-matrix}, essentially adapting techniques from \cite{CKL22, KLMMS17, ahn2012analyzing}, thus completing the proof of the upper bound.

Codex assisted with writing and literature searches under our direction. All proofs are written and verified by us, and we take full responsibility for the results and their presentation.
\input{prelims}
\input{lowerbound}

\input{upperbound}
\input{larger_domains}

\printbibliography
\newpage

\appendix

\input{upperbound-proof}

\end{document}

%% file: header.tex
\usepackage[margin=1in]{geometry}

\usepackage[utf8]{inputenc}

\usepackage{amsthm,amsmath,amssymb,mathtools}
\usepackage{bm}
\usepackage{graphicx}
\usepackage{color,xcolor}
\usepackage{paralist,enumitem}
\usepackage[normalem]{ulem}
\usepackage{tabularx}
\usepackage{tikz}
\usepackage{caption,comment}
\usepackage{algorithmicx}
\usepackage{algorithm}
\usepackage{algpseudocode}
\algrenewcommand{\algorithmicrequire}{\textbf{Input:}}
\algrenewcommand{\algorithmicensure}{\textbf{Output:}}
\usepackage{lipsum}
\usepackage{booktabs,needspace}
\usepackage{subcaption}
\usepackage{xspace}
\usepackage{lineno}

\usepackage[para]{manyfoot}
\DeclareNewFootnote[para]{Title}
\nolinenumbers

\usepackage{microtype}
\usepackage{silence}
\newcommand{\defref}[2]{%
  {\hypersetup{linkcolor=green!50!black}%
  \hyperref[#2]{\emph{#1}}}%
}

\newcounter{algsubstate}
\renewcommand{\thealgsubstate}{\alph{algsubstate}}

\algnewcommand\algorithmicinput{\textbf{Input:}}
\algnewcommand\Input{\item[\algorithmicinput]}
\algnewcommand\algorithmicoutput{\textbf{Output:}}
\algnewcommand\Output{\item[\algorithmicoutput]}
\algnewcommand\algorithmicgoal{\textbf{Goal:}}
\algnewcommand\Goal{\item[\algorithmicgoal]}

\usepackage[colorinlistoftodos,textsize=tiny,textwidth=2cm,color=green!50!gray]{todonotes}
\newif\iftodoss
\todossfalse %

\newcommand{\DeclareAuthorCommands}[3]{\iftodoss\expandafter\newcommand\csname #1\endcsname[1]{\textcolor{#3}{##1}}\expandafter\newcommand\csname #1i\endcsname[1]{\textcolor{#3}{[#2: ##1]}}\expandafter\newcommand\csname #1r\endcsname[1]{\todo[color=#3!20]{#2: ##1}}\else\expandafter\newcommand\csname #1\endcsname[1]{##1}\expandafter\newcommand\csname #1i\endcsname[1]{}\expandafter\newcommand\csname #1r\endcsname[1]{}\fi}
\DeclareAuthorCommands{amat}{AS}{blue!90!black}
\DeclareAuthorCommands{sant}{SV}{red}
\DeclareAuthorCommands{loui}{AP}{blue}
\DeclareAuthorCommands{jos}{JB}{green!50!black}
\DeclareAuthorCommands{adi}{AB}{red!50!black}

\usepackage[style=alphabetic,backend=biber,minalphanames=3,maxalphanames=7,maxbibnames=99,maxcitenames=10]{biblatex}

\usepackage[colorlinks=true,allcolors=blue,linktocpage]{hyperref}
\usepackage[capitalize,noabbrev,nameinlink]{cleveref}

\makeatletter
\newcommand{\titlefootnotelabel}[1]{%
  \ifnum#1=2\relax
    \TextOrMath{\textasteriskcentered}{*}%
  \else
    \TextOrMath{\textdagger}{\dagger}%
  \fi
}
\newcommand{\maketitlewithparagraphnotes}{%
  \begingroup
  \let\@fnsymbol\titlefootnotelabel
  \renewcommand{\thefootnoteTitle}{\titlefootnotelabel{\value{footnoteTitle}}}
  \long\def\footnotetext[##1]##2{%
    \begingroup
    \def\@makefnmark{\hbox{\@textsuperscript{\normalfont\@thefnmark}}}%
    \footnotetextTitle[##1]{##2}%
    \endgroup
  }
  \maketitle
  \endgroup
}
\makeatother
\makeatletter
\providecommand{\theHALG@line}{}
\renewcommand{\theHALG@line}{\thealgorithm.\arabic{ALG@line}}
\newcommand{\alglabel}[1]{%
  \begingroup
  \def\@currentcounter{ALG@line}%
  \protected@edef\@currentlabel{\theALG@line}%
  \protected@edef\cref@currentlabel{%
    [line][\arabic{ALG@line}][\arabic{algorithm}]\theALG@line}%
  \label[line]{#1}%
  \endgroup
}
\newcommand{\detailedsectionheadings}{%
  \c@secnumdepth=4\relax
  \renewcommand\paragraph{\@startsection{paragraph}{4}{\z@}%
    {-2.5ex \@plus -1ex \@minus -.2ex}{1ex \@plus .2ex}%
    {\normalfont\normalsize\bfseries}}%
}
\makeatother
\crefname{line}{line}{lines}
\Crefname{line}{Line}{Lines}
\numberwithin{equation}{section}
\newtheorem{theorem}[equation]{Theorem}
\newtheorem{lemma}[equation]{Lemma}
\newtheorem{corollary}[equation]{Corollary}
\newtheorem{claim}[equation]{Claim}
\newtheorem{observation}[equation]{Observation}
\theoremstyle{definition}
\newtheorem{definition}[equation]{Definition}
\theoremstyle{plain}
\newtheorem{example}[equation]{Example}

\NewDocumentEnvironment{restate}{m o}{%
  \par\medskip\noindent\textbf{\Cref{#1} (restated).}\IfValueT{#2}{\ #2}\ \itshape
}{\par\medskip}

\crefname{lemma}{Lemma}{Lemmas}
\crefname{definition}{Definition}{Definitions}
\crefname{example}{Example}{Examples}
\crefname{claim}{Claim}{Claims}
\crefname{observation}{Observation}{Observations}
\crefname{fact}{Fact}{Facts}
\crefname{corollary}{Corollary}{Corollaries}
\crefname{theorem}{Theorem}{Theorems}
\crefname{table}{table}{tables}
\Crefname{theorem}{Theorem}{Theorems}
\Crefname{claim}{Claim}{Claims}
\Crefname{observation}{Observation}{Observations}
\Crefname{fact}{Fact}{Facts}
\Crefname{corollary}{Corollary}{Corollaries}
\Crefname{table}{Table}{Tables}

\newcommand{\R}{\mathbb{R}}
\newcommand{\Z}{\mathbb{Z}}

\newcommand{\supp}{\operatorname{supp}}

\newcommand{\polylog}{\operatorname{polylog}}
\newcommand{\diag}{\operatorname{diag}}
\newcommand{\tr}{\operatorname{tr}}
\newcommand{\rank}{\operatorname{rank}}

\newcommand{\one}{\mathbf 1}

\DeclareMathOperator{\im}{im}

\allowdisplaybreaks

%% file: intro.tex
\section{Introduction}
\label{sec:intro}

A Constraint Satisfaction Problem ($\CSP(\Gamma)$) is defined by a \emph{constraint language} $\Gamma$ that is a fixed finite collection of $k$-ary relations over a finite domain $D$.\footnote{Relations of different arities can be given a common arity by adding arguments that do not affect satisfaction.} An instance of $\CSP(\Gamma)$ consists of $n$ variables $V=\{x_1,\ldots,x_n\}$ taking values in $D$ and a multiset of constraints $C=R(x_{i_1},\ldots,x_{i_k})$, where $R\in\Gamma$. An assignment $\sigma:V\to D$ satisfies $C$ if $(\sigma(x_{i_1}),\ldots,\sigma(x_{i_k}))\in R$, and satisfies the instance if it satisfies every constraint.

This is the standard general formulation of finite-domain CSPs. It includes both the well-studied \emph{single-relation model} and the \emph{Boolean model with literals}. For a single relation $R$, we write $\CSP(R)$ for the special case $\Gamma=\{R\}$. In the model with \emph{literals}, constraints may be applied to variables or their negations. For each pattern of negated arguments, complementing the corresponding coordinates of a relation gives another Boolean relation. Including these relations in the language lets us express every constraint with literals as a constraint applied directly to variables.

Graph coloring, satisfiability of Boolean formulas, and systems of equations over finite fields are examples of CSPs. Their study includes classifications of computational complexity~\cite{Schaefer78,jeavons1997closure,feder1998computational,bulatov2005classifying,Bulatov17,zhuk2020proof}, approximation algorithms and hardness of approximation~\cite{khanna2001approximability,Hastad01,alon2003random,raghavendra2008optimal,austrin2009approximation,DBLP:conf/stoc/KhotTW14}, and learning constraints from examples and queries~\cite{BessiereKLO17,bessiere2020chain,carbonnel2022redundancy,BessierePartial23,BessiereCH23}. Other work studies exact sparsification and kernelization, which reduce instance size while preserving satisfiability~\cite{dell2014satisfiability,jansen2019optimal,chen2020best,lagerkvist2020sparsification,brakensiek2025richness}, and sparsifying instances while approximately preserving the number of constraints satisfied by every assignment~\cite{kogan2015sketching,filtser2017sparsification,KPS24,khanna2025efficient,brakensiek2025redundancy,KhannaPS26}. We focus on streaming algorithms, which process constraints sequentially while retaining limited memory, a setting studied for both approximation and exact computation~\cite{KKS15,GVV17,KK19,CGV20,CGS+22-linear-space,KPSY23,SSSV23-random-ordering,SSSV23-dicut,CGSV24,SSSV25,FMW25,FMW26,ABFS26,ABF26,FMW26-tight,STV26,sharma2026characterizingstreamingdecidabilitycsps,BPSV26}.%

Two standard models for streaming algorithms are \emph{insertion-only} and \emph{strict-turnstile}. In an insertion-only stream, constraints are added one at a time. A strict-turnstile stream also permits deleting copies of constraints, provided that the number of copies of each constraint remains nonnegative after every update. A \emph{one-pass} algorithm processes the updates sequentially and answers at the end using only its retained memory. Space is measured in bits as a function of the number of variables $n$, with \amat{$D$ and $\Gamma$ fixed. Computation time is unrestricted.} The comparisons below concern streams of length polynomial in $n$.

There are four natural objectives for a CSP instance. $\MaxCSP(\Gamma)$ asks us to approximate the maximum number of constraints that can be satisfied simultaneously, while $\ExactCSP(\Gamma)$ asks us to compute this maximum exactly. $\CSP(\Gamma)$ asks only whether all constraints can be satisfied simultaneously. Finally, $\MinCSP(\Gamma)$ asks us to approximate the minimum number of unsatisfied constraints. We compare the space needed for these objectives in insertion-only and strict-turnstile streams. The insertion-only results provide a starting point for understanding the effect of deletions on these four objectives. %

\paragraph{Insertion-only streams.}
Max-CSP has received considerable attention in the insertion-only setting. For every fixed $\varepsilon>0$, sampling constraints gives a $(1-\varepsilon)$-approximation in $O(n\polylog n)$ space. Much of the work therefore concerns the approximation factors achievable in sublinear space. Results for Max-Cut~\cite{KKS15,KK19} and other Boolean CSPs~\cite{GVV17,CGV20} led to a classification for sketching algorithms, whose summaries can be combined across instances~\cite{CGSV24} and linear-space lower bounds for broad classes of CSPs~\cite{CGS+22-linear-space}. Recent work characterizes the approximation threshold for single-pass sublinear-space algorithms in terms of the basic linear programming relaxation~\cite{STV25,FMW26,ABF26,FMW26-tight,STV26}. Related results consider multiple passes~\cite{FMW26-tight,FMW26} and random arrival order~\cite{SSSV23-random-ordering, KKS15}.

For Exact-CSP, the parameter that governs space complexity turns out to be a polynomial degree. The \emph{indicator} of a Boolean relation is the function that equals one on accepted tuples and zero on all other tuples. It has a unique multilinear polynomial representation, where multilinear means that each variable has exponent at most one; the degree of this polynomial is also called its \emph{Fourier degree}.\footnote{\sant{This classical degree measure was studied in parallel computation by \textcite{DKR90} and in decision-tree complexity by \textcite{NS94}. The indicator polynomial has also been studied in other works, for example \cite{JW24}.}} Kol, Paramonov, Saxena, and Yu~\cite{KPSY23} give bounds of \amat{$\Omega_k(n^r)$ and $O_k(n^r\log n)$} for Fourier degree $r$. Their algorithm sums the indicator polynomials of the constraints and stores the resulting \amat{$O_k(n^r)$ coefficients, each using $O_k(\log n)$ bits} for polynomial-length streams. Evaluating this sum on an assignment gives its number of satisfied constraints, so maximizing over all assignments gives the exact optimum. Inserting a constraint adds its polynomial to the sum, and deleting it subtracts that polynomial; the same summary therefore also supports deletions. However, their lower bound applies only to languages that require constraints to be applicable to literals and constants, \amat{which as we discussed already are included in our constraint-language formulation above.}
Brakensiek, Putterman, Sharma, and Velusamy~\cite{BPSV26} extend this characterization to single-relation constraint families %
by reducing the relation to its \emph{core}.\footnote{\amat{The core of $\Gamma$ is obtained by restricting every relation to a smallest common subdomain onto which $D$ can be mapped while preserving every relation and fixing the values in that subdomain. This restriction preserves the optimum.}}

For satisfiability, the characterization is instead given by \emph{non-redundancy}. An instance is nonredundant if each constraint has an assignment violating that constraint and satisfying all the others. Thus removing any constraint changes the set of solutions. The non-redundancy of \amat{a constraint language} is the largest number of constraints in such an instance on $n$ variables. Sharma and Velusamy~\cite{sharma2026characterizingstreamingdecidabilitycsps} characterize insertion-only space through this parameter for Boolean CSPs. Brakensiek, Putterman, Sharma, and Velusamy~\cite{BPSV26} extend this to \amat{every fixed finite constraint language by taking its common core.}
An algebraic characterization of the asymptotic growth of non-redundancy remains a major open problem. Recent work \cite{bessiere2020chain,carbonnel2022redundancy,brakensiek2025redundancy,brakensiek2025richness,BrakensiekGP26,SV26b,brakensiek2026shrinking, brakensiek2026upperboundscspnonredundancy} \amat{has} substantially advanced this question, but it remains open even for \amat{some} Boolean relations of arity four.
\amat{\cite{BPSV26} also characterize the space complexity in the insertion-only setting for $\MinCSP(\Gamma)$, which approximates the minimum number of unsatisfied constraints. For every fixed $\varepsilon>0$, a $(1+\varepsilon)$-approximation has the same characterization by non-redundancy of the core, up to polylogarithmic factors.}

\paragraph{Turnstile streams.}
Several approximation guarantees survive the introduction of deletions. The $O(n\polylog n)$-space $(1-\varepsilon)$-approximation, for fixed $\varepsilon>0$, extends by sampling constraints according to their current multiplicities using standard turnstile sampling algorithms~\cite{JST11}. Independent samples from the final instance estimate all assignments' satisfied counts simultaneously, just as in the insertion-only argument. At smaller space bounds, the approximation algorithms of Chou, Golovnev, Sudan, and Velusamy~\cite{CGSV24} also support strict-turnstile updates: their linear sketches add a constraint's contribution on insertion and subtract it on deletion. Vu~\cite{Vu-2208.09160} gives dynamic Max-SAT approximation algorithms for \amat{\emph{binary streams}}, where every constraint frequency remains in $\{0,1\}$ after every update.
\amat{For Boolean $\ExactCSP(\Gamma)$, deletions preserve the same bounds $\Omega_k(n^r)$ and $O_k(n^r\log n)$, where $r$ is the maximum Fourier degree of the relations in the core.} Recall that \amat{after restricting to the core,} the upper bound stores the sum of the constraints' indicator polynomials. Deleting a constraint subtracts its polynomial from this sum, so the same space bound holds under strict-turnstile updates~\cite{KPSY23,BPSV26}.
The lower bounds hold even against insertion-only streams, so the characterization is the same in the strict-turnstile setting.

For satisfiability, prior strict-turnstile algorithms address particular problems rather than giving a general characterization. Ahn, Guha, and McGregor~\cite{ahn2012analyzing} decide graph bipartiteness in $O(n\polylog n)$ space under edge insertions and deletions. Each edge requires its endpoints to receive different values in $\{0,1\}$. Thus testing bipartiteness is exactly deciding satisfiability for the Boolean relation $R=\{01,10\}$, or equivalently for equations $x_u\oplus x_v=1$, where $\oplus$ denotes addition modulo two. %

This leads us to the central question addressed in this paper:
\begin{quote}
\emph{Can we characterize the single-pass strict-turnstile space complexity of $\CSP(\amat{\Gamma})$ for every Boolean constraint language $\Gamma$, and is non-redundancy still the right characterization?}
\end{quote}

\paragraph{Applications.}
The problem also has practical motivation in systems and verification, where changing requirements must remain jointly satisfiable. Incremental SAT solvers for circuit verification and test generation reuse a circuit description while adding and removing constraints for different tests~\cite{WKS01}. Network verification systems such as VeriFlow~\cite{VeriFlow13} and APKeep~\cite{APKeep20} check network properties such as reachability as forwarding rules are installed, modified, or withdrawn. 
The database management software Cobra~\cite{Cobra20} continuously checks whether observed database transactions are consistent with a serial execution, while discarding information. 
Stream-reasoning systems maintain scheduling requirements as jobs arrive and old requirements expire~\cite{GGKOSS13}. These systems have different representations and performance goals, but share the need to reason about requirements that change over time. Motivated by these applications and the complexity comparison above, we study the memory needed to decide satisfiability when constraints can be inserted and deleted.

\subsection{Our results}
Our main result characterizes the space complexity of deciding satisfiability of \amat{every fixed nontrivial Boolean constraint language} in the strict-turnstile setting, up to polylogarithmic factors.

The parameter governing our characterization is the \emph{capturing degree} of the constraint language. An integer polynomial \emph{captures} a $k$-ary Boolean relation $R$ if it vanishes exactly on $R$ within $\{0,1\}^{k}$. The minimum degree of such a polynomial is its \emph{integer capturing degree}, denoted $\cdeg(R)$. For a language $\Gamma$, define $\cdeg(\Gamma)=\max_{R\in\Gamma}\cdeg(R)$. \sant{By \cref{obs:coefficient-fields}, allowing rational or real coefficients
instead of integer coefficients does not change the minimum degree
of a polynomial that vanishes exactly on the satisfying tuples.
Consequently, our integer capturing degree coincides with the
non-deterministic degree introduced by \textcite{deWolf03},
with the convention that the polynomial vanishes on satisfying
tuples; see \cref{sec:related_work} for further discussion.}

Our result identifies the integer capturing degree as the governing parameter under strict-turnstile deletions, by giving a solution-preserving streaming algorithm and a matching space lower bound up to polylogarithmic factors.

\begin{theorem}[Main result; \cref{thm:upper,thm:stream-lower,cor:lb-languages,cor:ub-languages,cor:lb-exact-sparsification}]
\label[theorem]{thm:intro-classification}

For every fixed nontrivial Boolean language $\Gamma$,\footnote{Here every relation is nonempty and proper, and neither constant assignment satisfies every constraint permitted by $\Gamma$. Full relations can be ignored, and constraints using the empty relation are handled by a counter. If a common constant satisfies the remaining language, its decision problem is trivial.} let $d=\cdeg(\Gamma)$. Deciding $\CSP(\Gamma)$ on $n$ variables admits a randomized one-pass strict-turnstile algorithm using $O_k(n^d\polylog n)$ bits and requires $\Omega_k(n^d)$ bits.

Moreover, the algorithm also solves the \textit{exact sparsification} problem, i.e., constructs a subset of at most $O_k(n^d\polylog n)$ active constraints that preserves the entire set of satisfying assignments of the original instance. For this problem, we can extend the lower bound above to constant-satisfiable Boolean languages.

\end{theorem}

The randomized algorithm has error probability $1/3$ and we assume polynomially bounded stream lengths. The subscripts in $O_k$ and $\Omega_k$ allow constants depending on the fixed maximum arity $k$. \amat{For a domain of size $q$, the notation $O_{k,q}$ also allows dependence on $q$.}

To our knowledge, our result gives the first polynomial separation between insertion-only and strict-turnstile space for CSP satisfiability. It also separates strict-turnstile satisfiability from exact optimization. 

\amat{Both comparisons are strict for an infinite family of constraint languages. For instance, consider the $k$-XOR predicate with literals: every constraint is of the form \adi{$\ell_{i_1}\oplus \cdots\oplus\ell_{i_k}=1$}, where each $\ell_{i_j}$ is either the variable $x_{i_j}$ or its negation $\neg x_{i_j}$. For every $k\geq2$, insertion-only satisfiability has bounds $\Omega_k(n)$ and $O_k(n\log n)$}~\cite{sharma2026characterizingstreamingdecidabilitycsps}. Our strict-turnstile bounds are $\Omega_k(n^{\lceil k/2\rceil})$ and $O_k(n^{\lceil k/2\rceil}\polylog n)$ (\sant{see \cref{cor:xor-degree}}), whereas exact optimization has bounds $\Omega_k(n^k)$ and $O_k(n^k\log n)$ in either update model~\cite{KPSY23}. Both polynomial separations hold for every $k\geq3$.

Our lower bound also applies to $\MinCSP(\Gamma)$. \amat{Recall that this problem asks to approximate the minimum number of unsatisfied constraints, and that its insertion-only space complexity is characterized by non-redundancy, up to polylogarithmic factors \cite{BPSV26}.}
Indeed, any finite multiplicative approximation for $\MinCSP(\Gamma)$ must return zero on a satisfiable instance and a positive value otherwise, and therefore decide satisfiability.
\amat{Consequently, every randomized one-pass strict-turnstile algorithm with this guarantee requires $\Omega_k(n^d)$ bits. Our upper bound preserves satisfying assignments but does not provide an approximation guarantee for the minimum number of violations. Whether Min-CSP has the same strict-turnstile space characterization as satisfiability remains open.}

\subsection{Related work}\label{sec:related_work}
Low-degree polynomial representations of \CSP constraints have previously been used for sparsification and kernelization. \textcite{jansen2019optimal} and \textcite{chen2020best} introduced the notion of a polynomial \emph{capturing} an unsatisfying tuple $u\notin R$. Their definition allows the polynomial to be over either $\mathbb{Q}$ or $\mathbb{Z}/q\mathbb{Z}$ for some integer $q>1$, and therefore in particular includes finite fields $\mathbb{F}_p$, unlike our definition over the integers.\footnote{For example, the capturing degree of $k$-\XOR under their definition is $1$, while its integer capturing degree is $\lceil k/2\rceil$.} They showed that if every unsatisfying tuple of a fixed Boolean relation admits such a capturing polynomial of degree at most $d$, then every $n$-variable instance admits an equivalent sub-instance with $O(n^d)$ constraints that preserves all satisfying assignments.

The integer capturing degree is closely connected to the
non-deterministic degree of Boolean functions, introduced by
\textcite{deWolf03}. A non-deterministic polynomial for a Boolean
function $f$ is a real polynomial $p$ satisfying
$p(x)\neq 0$ if and only if $f(x)=1$, and $\mathsf{ndeg}(f)$ is the
minimum degree of such a polynomial. In our terminology, the
polynomial vanishes on satisfying tuples, so the corresponding
Boolean function is the complement of the relation's characteristic
function. \cref{obs:coefficient-fields} shows that restricting the coefficients to integers does not change this minimum degree. Thus, our integer
capturing degree is precisely non-deterministic degree, with this
choice of convention.

Non-deterministic degree has been studied through its connections
to several other complexity measures. De Wolf~\cite{deWolf03}
showed that it exactly characterizes non-deterministic quantum
query complexity. It is also closely related to rational degree,
the minimum of $\max\{\deg p,\deg q\}$ over rational representations
$f=p/q$ whose denominator is nonzero on the Boolean cube:
\[
    \mathsf{rdeg}(f)
    =\max\{\mathsf{ndeg}(f),\mathsf{ndeg}(\neg f)\}.
\]
\textcite{MW15} connected rational approximation
degree to quantum query complexity with postselection.
\textcite{IJK+23} studied rational degree for symmetric
functions, unate functions, and read-once formulas, as well as
composition properties and separations from other complexity
measures. More recently, \textcite{KKWY26} proved that the Fourier degree of every Boolean
function is at most $\widetilde O(\mathsf{rdeg}(f)^3)$, resolving
a longstanding question about the relationship between rational
degree and Fourier degree.

\textcite{RS26} study extremal properties of
non-deterministic degree and related questions concerning effective
Nullstellensatz on the Boolean cube. They show that almost all
Boolean functions on $k$ variables have non-deterministic degree
at most $k/2+O(\sqrt{k\log k})$, confirming a conjecture of
\textcite{IJK+23}. 

Our work gives a streaming
characterization in terms of this established degree notion:
for every fixed non-trivial Boolean constraint language, its
non-deterministic degree determines the space complexity of
strict-turnstile satisfiability up to polylogarithmic factors.

\subsection{Organization}
In \Cref{sec:tech-overview}, we explain the key technical ideas using $4$-XOR as a running example. \Cref{sec:prelims} collects the preliminaries needed for our proofs. We prove the lower bound in \Cref{sec:hardness} and the upper bound in \Cref{sec:upper}, with the full streaming row-sparsification algorithm and analysis in Appendix~\ref{sec:ub-algorithm}. In \Cref{sec:larger-domains}, we extend the upper bound to larger finite domains and discuss barriers to extending our lower-bound techniques to other settings.

%% file: technicaloverview.tex
\section{Technical Overview}
\label{sec:tech-overview}

We explain our techniques using the $4$-XOR predicate without literals. Satisfiability for $2$-XOR without literals is equivalent to graph bipartiteness, for which Ahn, Guha, and McGregor~\cite{ahn2012analyzing} give a near-linear space dynamic algorithm. In contrast, $3$-XOR without literals is constant-satisfiable, since assigning one to every variable satisfies every constraint. Thus, we choose $4$-XOR\footnote{Henceforth, we will assume that literals are not allowed without explicitly mentioning it.} as a natural example to explain our results. We first explain the quadratic upper bound in \cref{sec:overview-upper}. We then develop the matching lower bound in \cref{sec:overview-lower}, starting from the known lower bound for exact optimization.

\paragraph{From insertions to deletions.}
Each $4$-XOR constraint, and more generally, each $k$-XOR constraint, is a linear equation over $\mathbb F_2$. Under insertions, we can retain a basis, namely a subset of at most $n+1$ original equations whose linear combinations include every constraint received so far. Each retained equation specifies only $k$ variables. For fixed $k$, the basis therefore can be stored using $O(n\log n)$ bits of memory and preserves all satisfying assignments, giving a near-linear space insertion-only algorithm for satisfiability.

Sharma and Velusamy~\cite{sharma2026characterizingstreamingdecidabilitycsps} extend this approach to general CSPs through \emph{non-redundancy}, a parameter that plays a role similar to the size of a basis. However, deletions make maintaining this summary difficult. An equation that we did not store because it was implied by the retained basis may become necessary after one of those basis equations is deleted. For example, consider the following four equations over $\mathbb F_2$.
\[
\begin{aligned}
 E_1:\quad x_1+x_2+x_3+x_4&=1,
 & E_2:\quad x_1+x_2+x_5+x_6&=1,\\
 E_3:\quad x_3+x_4+x_7+x_8&=1,
 & E_4:\quad x_5+x_6+x_7+x_8&=1.
\end{aligned}
\]
Adding the first three gives the fourth, so a basis can omit $E_4$. If $E_1$ is later deleted, however, $E_4$ is still present and is no longer implied by $E_2,E_3$. Indeed, the assignment $x_1=x_3=1$, with all other variables zero, satisfies $E_2,E_3$ but leaves $E_4$ unsatisfied. The algorithm must retain enough information to recover an equation that was previously omitted.

A naive solution is to store the multiplicity of every possible constraint, using $O(n^4\log n)$ bits. Indeed, this also supports the stronger objective of \emph{exact optimization}, which asks to compute the minimum number of constraints left unsatisfied by any assignment.\footnote{For an instance with $m$ constraints, this minimum is $m$ minus the maximum number of simultaneously satisfied constraints. Thus it is equivalent to the maximization objective studied in \cite{KPSY23,BPSV26}.} At the end of the stream, we can enumerate all the assignments and count the constraints each leaves unsatisfied.

More generally, this upper bound follows from the polynomial representation used by Kol, Paramonov, Saxena, and Yu~\cite{KPSY23}, and Brakensiek, Putterman, Sharma, and Velusamy~\cite{BPSV26}. Every Boolean relation $R$ has a unique multilinear polynomial $g_R$ that evaluates to zero on satisfying tuples and one on unsatisfying tuples. We call $g_R$ its \emph{indicator polynomial}. Its coefficients are integers, and its degree is called the \emph{Fourier degree}. For $4$-XOR, $g_R(z)=(1+\prod_{a=1}^4(1-2z_a))/2$, which has degree four.
For each constraint $C$, substituting its variables into $g_R$ gives a polynomial $g_C(x)$ that is one when $C$ is unsatisfied and zero otherwise. Thus $\sum_C f_Cg_C(x)$ counts the unsatisfied constraints, including their multiplicities, where $f_C$ is the number of copies of $C$. If the indicator polynomial has degree $\ell$, this polynomial has only $\amat{O_k}(n^\ell)$ monomials. We can store their integer coefficients, adding or subtracting the coefficients of $g_C$ when a copy of $C$ is inserted or deleted. Each coefficient takes $\amat{O_k}(\log n)$ bits on polynomial-length streams, and evaluating the polynomial on every assignment gives the exact optimum.

Understanding satisfiability in the strict-turnstile model, which allows deletions, therefore raises two questions. First, can we decide satisfiability using less space than exact optimization requires? Second, can we match the nearly linear space achievable in the insertion-only model, or do deletions necessarily increase the space requirement? We address these questions in turn.

\subsection{Upper Bound}
\label{sec:overview-upper}

\paragraph{Capturing Polynomial.}
Let us start with the first question. The degree-four indicator polynomial already gives an $O(n^4\log n)$-space algorithm for $4$-XOR. However, it retains more information than satisfiability needs. In particular, a constraint's polynomial need not evaluate to exactly one when the constraint is unsatisfied. We only need it to be zero on satisfying tuples and a non-zero integer on unsatisfying tuples. We call such an integer polynomial a \emph{capturing polynomial}. Observe that squaring its value gives a non-negative integer that is zero precisely when the constraint is satisfied. Consequently, the sum of these squares is zero exactly when every constraint is satisfied. This is the property we will exploit to come up with our algorithm.%

In general, the minimum degree of a capturing polynomial is at most the Fourier degree, since the indicator polynomial is also, by definition, a capturing polynomial. For $4$-XOR, this inequality is strict. \amat{Let $r=z_1+z_2+z_3+z_4$ be the number of ones in the tuple. The polynomial $(r-1)(r-3)$ vanishes exactly on satisfying tuples. Since we care only about Boolean inputs, we can replace each $z_a^2$ by $z_a$, obtaining the multilinear capturing polynomial $p(z)=3-3\sum_a z_a+2\sum_{a<b}z_az_b$ of degree two.}

\paragraph{Matrix Representation.} \amat{For each constraint $C$, let $p_C$ be the polynomial obtained by substituting the variables in $C$ in $p$. Let
$\phi(x)=(1,(x_i)_{i\in[n]},(x_ix_j)_{1\leq i<j\leq n})^\top$
be the column vector of all multilinear monomials of degree at most two, with dimension $D=1+n+\binom n2=O(n^2)$.} For each constraint $C$, let $b_C$ be the coefficient vector of $p_C$ in this basis, so that $p_C(x)=b_C^\top\phi(x)$. For a constraint on four distinct variables, $b_C$ has at most eleven non-zero entries, each bounded in magnitude by a constant. More generally, for a fixed relation $R$, both the number and the magnitudes of the non-zero coefficients are $O_k(1)$, independent of $n$. We now encode the constraints as rows of a matrix, with each row scaled by the corresponding constraint's frequency. Let $f_C$ denote the frequency of constraint $C$, and call $C$ \emph{active} if $f_C>0$. Enumerating all possible constraint types as $C_1,\ldots,C_m$, define
\[
A_f=
\amat{\bigl[f_{C_1}b_{C_1}\;\cdots\;f_{C_m}b_{C_m}\bigr]^\top}.
\]%
Then, for any assignment $\alpha$,
\[
A_f\phi(\alpha)=
\bigl(f_{C_1}p_{C_1}(\alpha),\ldots,f_{C_m}p_{C_m}(\alpha)\bigr)^\top.
\]

To see why this representation is useful, define $G=A_f^\top A_f$. The \emph{energy} of an assignment $\alpha$ is the quadratic form
\begin{equation}
\label{eq:overview-upper-energy}
 E(\alpha)=\phi(\alpha)^\top G\phi(\alpha)
          =\sum_C f_C^2p_C(\alpha)^2.\nonumber
\end{equation}
This energy is zero for a satisfying assignment, and at least one otherwise. Indeed, every active constraint that is unsatisfied contributes the square of a non-zero integer.

We use sparsification to retain a small weighted subset of rows while approximately preserving the energy. Specifically, suppose we can recover a set $S$ of active constraints and positive weights $w_C$ such that $W=\sum_{C\in S}w_Cb_Cb_C^\top$ satisfies
\begin{equation}
\label{eq:overview-sparsification}
(1-\varepsilon)(G+\lambda I)
\preceq W+\lambda I
\preceq(1+\varepsilon)(G+\lambda I),
\end{equation}
where $0<\varepsilon\leq1/8$ and $0<\lambda<1/(100D)$\footnote{\sant{A direct spectral approximation to $G$ would also preserve its kernel, but the space bound of our recovery procedure would depend on its \emph{condition number}, i.e., the ratio of its largest eigenvalue to its smallest positive eigenvalue. The integer energy gap allows us to stop at $\lambda=\Theta(1/D)$, avoiding the need to resolve potentially much smaller positive eigenvalues while preserving all satisfying assignments exactly.}}. Here $I$ is the identity matrix, and $U\preceq V$ means that $y^\top Uy\leq y^\top Vy$ for every real vector $y$.

We first show that $S$ has exactly the same satisfying assignments as the original instance. Every satisfying assignment of the original instance satisfies $S$, since $S$ contains only active constraints. Conversely, suppose $\alpha$ satisfies every constraint in $S$. Then $p_C(\alpha)=0$ for every $C\in S$, so $\phi(\alpha)^\top W\phi(\alpha)=0$. Moreover, each coordinate of $\phi(\alpha)$ is a Boolean monomial and hence lies in $\{0,1\}$, giving \amat{$\|\phi(\alpha)\|_2^2\leq D$}. Applying the left inequality \amat{in \eqref{eq:overview-sparsification}} to $\phi(\alpha)$ therefore yields
\[
(1-\varepsilon)E(\alpha)
\leq \phi(\alpha)^\top(W+\lambda I)\phi(\alpha)
=\lambda\amat{\|\phi(\alpha)\|_2^2}
\leq\lambda D.
\]
Thus $E(\alpha)\leq\lambda D/(1-\varepsilon)<1/4$. Since $E(\alpha)$ is either zero or at least one, it must be zero, and $\alpha$ satisfies every original constraint. Recovering such a set $S$ in $O(D\operatorname{polylog}n)$ space would therefore yield the desired $O(n^2\operatorname{polylog}n)$-space algorithm.

\paragraph{Matrix Sparsification.}
We obtain the required recovery guarantee by adapting the graph sparsification techniques of \textcite{KLMMS17} and \textcite{CKL22}. Let $E$ be the edge incidence matrix, with a row $e_u-e_v$ for each possible edge $\{u,v\}$, where $e_u$ is the unit vector corresponding to vertex $u$. For a vector $f$ of nonnegative edge weights, with $f_e=0$ for absent edges, the graph Laplacian is $L_f=E^\top\operatorname{diag}(f)E$. Spectral sparsification seeks a weighted subset of edges whose Laplacian $\widetilde L$ satisfies
\[
 (1-\varepsilon)y^\top L_fy
 \leq y^\top\widetilde Ly
 \leq(1+\varepsilon)y^\top L_fy
\]
for every real vector $y$. Our goal is analogous: we seek a weighted subset of constraint rows that approximately preserves the quadratic form of $G=A_f^\top A_f$, in which each row's contribution is weighted by the square of its constraint's frequency.

Both works recover sparsifiers from linear sketches. The algorithm of~\cite{KLMMS17} handles unweighted graphs, where $f_e\in\{0,1\}$ indicates whether edge $e$ is present. It maintains sketches under edge insertions and deletions and recovers a sparsifier of $L_f$ using space nearly linear in the number of vertices. For nonnegative edge weights, the procedure in~\cite[Lemma 6.5]{CKL22} sketches $f$ and recovers a sparsifier of
\[
 L_{f^2}=E^\top\operatorname{diag}(f)^2E,
\]
the Laplacian of the graph with squared edge weights. This formulation matches the squared frequencies in our matrix $G$. The linearity of the sketches allows updates to $f$ to be processed directly during the stream. We adapt these arguments to our integer coefficient rows, using their constant number of non-zero entries and bounded coefficient magnitudes, to obtain the following guarantee.

\begin{theorem}[Informal; \cref{cor:ub-active,thm:ub-recovery}]
\label{cor:overview-matrix}
Fix a constant $0<\varepsilon\leq1/8$ and a power of two $\lambda=\Theta(1/D)$ satisfying $\lambda<1/(100D)$. There is a randomized one-pass algorithm that processes any polynomial-length strict-turnstile stream of constraints using $O(D\operatorname{polylog}n)$ bits of space. With probability at least $2/3$, it recovers a set $S$ of $O(D\operatorname{polylog}n)$ constraints active at the end of the stream and positive weights $(w_C)_{C\in S}$ such that the matrix $W=\sum_{C\in S}w_Cb_Cb_C^\top$ satisfies \cref{eq:overview-sparsification}.
\end{theorem}

\sant{We give a technical overview of the proof of this theorem in \cref{sec:ub-overview} and the full algorithm and analysis in \cref{sec:ub-algorithm}.}

\paragraph{General Boolean Relations.}
The same argument extends to every fixed Boolean relation $R$. Choose a capturing polynomial for $R$ with integer coefficients and degree $d$. Substituting the variables of each constraint yields polynomials of degree at most $d$, all expressible in a common monomial basis of size $O_k(n^d)$. Since these polynomials arise from a single fixed polynomial, each coefficient row has only constantly many non-zero integer entries, all of bounded magnitude. Thus the rows retain the sparsity, boundedness, and integrality properties used above. Moreover, every unsatisfied active constraint contributes at least one to the energy. The same sparsification argument therefore yields an algorithm using $O_k(n^d\operatorname{polylog}n)$ bits of space. Choosing a capturing polynomial of minimum degree gives our upper bound in terms of the \emph{capturing degree}, $d=\cdeg(R)$. We prove this bound formally in \cref{thm:upper}.

\paragraph{Extensions to Larger Domains and Constraint Languages.}
The same matrix representation yields both extensions. For larger domains, let $R$ be a fixed nonempty proper relation of arity $k$ over $\{0,\ldots,q-1\}$, and choose a capturing polynomial with integer coefficients and degree $d=\cdeg(R)$. The resulting constraint polynomials share a monomial basis of size $D=O_{k,q}(n^d)$. On this domain, the monomial evaluations are bounded by constants depending only on $k,q$, so $\|\phi(\alpha)\|_2^2=O_{k,q}(D)$. Choosing $\lambda=\Theta_{k,q}(1/D)$ with a sufficiently small implicit constant keeps the regularization term below the integer energy gap. The same sparsification argument therefore yields an algorithm using $O_{k,q}(n^{\cdeg(R)}\operatorname{polylog}n)$ bits of space.\footnote{In fact, we can first replace $R$ by its core, $\Core(R)$: the restriction of $R$ to a smallest subdomain admitting a relation-preserving map from the original domain that fixes the subdomain pointwise. Applying this map coordinatewise to assignments preserves satisfiability, yielding the stronger space bound $O_{k,q}(n^{\cdeg(\Core(R))}\operatorname{polylog}n)$. We prove this in \cref{sec:ub-finite-domain}.}

For constraint languages, let $\Gamma$ be a fixed finite set of nonempty proper Boolean relations of arity $k$. Choose a minimum-degree integer capturing polynomial for each $R\in\Gamma$, and let $d=\max_{R\in\Gamma}\cdeg(R)$. All constraint polynomials then share a monomial basis of size $O_k(n^d)$. Since $\Gamma$ is fixed, each coefficient row still has only constantly many non-zero integer entries, all of bounded magnitude. Combining all constraint types in a single matrix and applying the same sparsification argument therefore yields an algorithm using $O_k(n^d\operatorname{polylog}n)$ bits of space. We state and prove this bound in \cref{cor:ub-languages}.

\subsection{Lower Bound}
\label{sec:overview-lower}
We have answered the first question: satisfiability can be decided in $O(n^2\operatorname{polylog}n)$ space, improving on the $O(n^4\log n)$ bound for exact optimization. We now address the second question by showing that this quadratic dependence on $n$ is necessary. This establishes a separation between strict-turnstile satisfiability and its insertion-only counterpart, which admits a near-linear space algorithm.

\paragraph{Exact optimization and Fourier degree.}
We begin with the exact-optimization reduction of~\cite{BPSV26}, using $4$-XOR as a running example\footnote{\sant{\textcite{KPSY23}
also obtain the same lower bound but their reduction is different and crucially uses literals which we want to avoid.}}. Recall that the value of an assignment is the number of constraints it leaves unsatisfied, and exact optimization asks for the minimum value over all assignments. The indicator polynomial
\[
g_R(z)=\frac{1+\prod_{a=1}^4(1-2z_a)}{2}
\]
is zero on satisfying tuples and one on unsatisfying tuples. It has degree four, with coefficient $8$ on $z_1z_2z_3z_4$. Summing the substituted copies of $g_R$ over all constraints gives a polynomial whose evaluation at each assignment equals that assignment's value. The reduction uses evaluations of this polynomial to isolate the contribution of a chosen constraint.

The reduction starts from the one-way communication problem $\Index$. Alice receives bits $X=(X_t)_{t\in[n]^4}$, where \amat{$[n]=\{1,\ldots,n\}$}, and Bob receives an index $j=(j_1,j_2,j_3,j_4)\in[n]^4$. Alice sends a single message, after which Bob must recover $X_j$. Even with shared randomness and constant error probability bounded away from $1/2$, this requires $\Omega(n^4)$ bits of communication~\cite{Index1,Index2}.

The construction uses four groups of \amat{variables $x_{a,u}$, for $u\in[n]$, one for each coordinate $a\in[4]$}. Here $n$ is the size of each group, so the instance has $4n=O(n)$ variables in total. For each $t=(t_1,t_2,t_3,t_4)\in[n]^4$, define
\[
C_t=R(x_{1,t_1},x_{2,t_2},x_{3,t_3},x_{4,t_4}).
\]
Alice inserts $C_t$ if and only if $X_t=1$, then sends the streaming algorithm's memory state to Bob. The resulting instance has value polynomial
\[
F_X(x)=\sum_{t\in[n]^4}X_t \cdot g_R(x_{1,t_1},x_{2,t_2},x_{3,t_3},x_{4,t_4}).
\]

Bob wants to determine whether Alice inserted $C_j$. Let $z_a$ denote the value of $x_{a,j_a}$, and obtain $H_j(z)$ from $F_X(x)$ by setting all other variables to zero. Every degree-four monomial except the one corresponding to $C_j$ contains a variable set to zero and therefore vanishes. Hence
\[
 H_j(z)=8X_jz_1z_2z_3z_4+Q_j(z),
 \qquad \deg Q_j\leq3.
\]
Since $H_j$ is multilinear, its values on $\{0,1\}^4$ determine all its coefficients, including $8X_j$. It therefore suffices for Bob to recover these sixteen values.

Bob obtains these evaluations using the pinning lemma of~\cite[Lemma 4.6]{BPSV26}. For any chosen assignment $\alpha$, he can append a gadget ensuring that every optimal assignment agrees with either $\alpha$ or its complement on the original variables and satisfies all the added constraints. Complementing all four inputs preserves their parity, so $\alpha$ and its complement violate exactly the same $4$-XOR constraints. Since the added constraints contribute no violations, the optimum of the extended instance is exactly $F_X(\alpha)$. The gadget introduces only $O(n)$ additional variables.

For each $z\in\{0,1\}^4$, Bob starts from a copy of Alice's memory state and appends the gadget for the assignment that sets $x_{a,j_a}=z_a$ for each $a\in[4]$ and all other original variables to zero. These sixteen continuations yield the values $H_j(z)$ on $\{0,1\}^4$. Inclusion--exclusion then gives
\[
 \sum_{z\in\{0,1\}^4}(-1)^{4-|z|}H_j(z)=8X_j,
\]
where $|z|$ is the number of ones in $z$. Every lower-degree monomial omits a coordinate, so its contributions cancel between pairs of assignments differing only in that coordinate. Bob therefore recovers $X_j$. The communication lower bound for $\Index$ then implies an $\Omega(n^4)$ space lower bound.

\paragraph{The obstruction for satisfiability.}
The exact-optimization reduction recovers $X_j$ by taking an alternating sum of the values $H_j(z)$. A satisfiability algorithm tells Bob only whether each of these values is zero, which need not reveal the queried bit. For example, suppose Alice inserted a constraint involving only variables outside $C_j$. This constraint receives $0000$ under all sixteen assignments, making every pinned instance unsatisfiable regardless of $X_j$. In exact optimization, its contribution cancels in the alternating sum. For satisfiability, however, Bob receives the same answer in all sixteen cases and learns nothing about $X_j$.

One might try to avoid this problem by choosing an assignment that violates $C_j$ but satisfies every other possible constraint. Even for $4$-XOR, such an assignment need not exist. Choose one additional variable in each of the first two groups. Keeping the last two queried variables fixed, the two choices in each of the first two groups give four constraints, including $C_j$. Every variable appears an even number of times across these four odd-parity equations, so any assignment must violate an even number of them. Thus an assignment that violates $C_j$ necessarily violates at least one of the other three constraints as well.

Our approach is to choose an assignment violating $C_j$ and remove every other constraint that it violates. Bob can identify which potential constraints his assignment violates, but he does not know which of them Alice inserted. Deleting all of them could therefore create negative frequencies. He needs a way to cancel only the contributions actually present.

We implement this cancellation using linear sketches. The theorem of \textcite[Corollary 7.2.2]{JLY26} transfers lower bounds for linear sketches to the strict-turnstile model with only a constant-factor loss in space\amat{.}\footnote{\textcite{LNW14} established a streaming-to-sketching equivalence for turnstile streams, and \textcite{AHLW16} extended it to strict-turnstile streams. These results assumed correctness on arbitrarily long streams, with reductions requiring doubly exponential stream lengths. \textcite{JLY26} obtain a transfer for algorithms required to handle all streams only up to a sufficiently large polynomial length.}
The key property of linear sketches is their additivity. Let $f$ be the frequency vector of an instance, with $f_C$ counting the copies of $C$, and write $L(f)$ for its sketch. For integer vectors $f,h$,
\[
L(f+h)=L(f)+L(h).
\]
Thus adding sketches adds the underlying frequency vectors, and subtracting sketches subtracts them. Alice will send sketches of selected subsets of her constraints, chosen without knowing Bob's query. Bob can then combine these sketches according to his query to cancel the other violated constraints while retaining $C_j$ exactly when Alice inserted it.

Suppose the sketch uses at most $B$ bits and its decoder decides satisfiability with constant error. Alice and Bob share the randomness defining $L$, so all sketches use the same map. We include the auxiliary variables needed for pinning in the variable set from the outset.

\paragraph{Combining sketches to remove unsatisfied constraints.}
A naive approach is for Alice to send $X_tL(e_{C_t})$ for every $t\in[n]^4$, where $e_C$ is the frequency vector representing one copy of $C$. Bob selects the sketch indexed by $j$. This is the sketch of either the empty instance, if $X_j=0$, or the single constraint $C_j$, if $X_j=1$.

Let $\alpha_j$ assign one to the four queried variables and zero to every other original variable. By the pinning lemma, Bob can add a gadget that allows precisely $\alpha_j$ and its complement on the original variables, each with a satisfying extension to the auxiliary variables. These two assignments satisfy exactly the same $4$-XOR constraints, so it suffices to analyze $\alpha_j$. Bob adds the sketch of the pinning gadget to his selected sketch. If $X_j=0$, only the pinning constraints remain, and the instance is satisfiable. If $X_j=1$, the query constraint receives either $1111$ or $0000$ under every assignment allowed by the gadget. Both tuples violate $C_j$, so the instance is unsatisfiable. Bob can therefore recover $X_j$.

The drawback is that Alice sends $n^4$ sketches. The communication lower bound for $\Index$ then gives $n^4B=\Omega(n^4)$, yielding only the trivial bound $B=\Omega(1)$.

To reduce communication, we relax the goal of isolating $C_j$. Bob needs to isolate it only among the constraints violated by $\alpha_j$; constraints satisfied by $\alpha_j$ can safely remain. Recall that $\alpha_j$ sets the four queried variables to one and all other original variables to zero. A constraint matching the query in exactly $r$ coordinates therefore receives $r$ ones and is violated precisely when $r\in\{0,2,4\}$. The case $r=4$ identifies $C_j$, which Bob wants to retain exactly when Alice inserted it. His task is therefore to eliminate the constraints with $r=0$ or $r=2$.

The case $r=0$ is easy to exclude: Bob selects only sketches of constraints incident to at least one queried variable. The remaining challenge is to cancel constraints with exactly two matches. Such a constraint belongs to two subsets obtained by fixing a single queried coordinate and one subset obtained by fixing both matching coordinates. This suggests combining sketches of these two kinds of subsets.

Accordingly, Alice sends
\begin{equation}
\label{eq:overview-subset-sketches}
 M_{a,u}=\sum_{t:t_a=u}X_tL(e_{C_t}),
 \qquad
 M_{ab,(u,v)}=\sum_{\substack{t:t_a=u\\t_b=v}}X_tL(e_{C_t}),
\end{equation}
where the first family ranges over $a\in[4]$ and $u\in[n]$, and the second over $1\leq a<b\leq4$ and $u,v\in[n]$. These are sketches of Alice's constraints incident to one specified variable or to two specified variables from different groups, respectively. There are $4n+6n^2=O(n^2)$ sketches in total, all computed without knowing Bob's query.

Bob selects the sketches indexed by his queried variables and forms
\begin{equation}
\label{eq:overview-bob}
 M_{\mathrm{Bob}}
 =2\sum_{1\leq a<b\leq4}M_{ab,(j_a,j_b)}
  -\sum_{a=1}^4M_{a,j_a}.
\end{equation}
Let $r=r(t,j)$ be the number of coordinates in which $t$ agrees with $j$. The constraint $C_t$ appears in $r$ selected single-index sketches and $\binom r2$ selected pair sketches. By linearity, its frequency in the vector represented by $M_{\mathrm{Bob}}$ is
\[
 X_t\left(2\binom r2-r\right)=X_t\,r(r-2).
\]
Under $\alpha_j$, the constraint receives exactly $r$ ones and is violated precisely when $r\in\{0,2,4\}$. The multiplier $r(r-2)$ vanishes when $r=0$ or $r=2$. The case $r=4$ occurs only for $C_j$, whose frequency becomes $8X_j$. Every other constraint with non-zero frequency has $r=1$ or $r=3$ and is therefore satisfied by $\alpha_j$.

The resulting vector may still have negative entries, so Bob must repair it before decoding. Negative frequencies occur only when $r=1$, where the frequency is $-X_t$. Bob adds one copy of every constraint with $r=1$, changing its frequency to $1-X_t\geq0$. He can compute the sketch of this correction from his query alone. Since all constraints added in this step are satisfied by $\alpha_j$, the correction makes the instance valid without introducing any further violated constraints.

Bob then adds the sketch of the pinning gadget for $\alpha_j$, \sant{as in \cite[Lemma 4.4]{BPSV26}}. Each pinning constraint contains an auxiliary variable and is therefore distinct from every $C_t$, so the corrected frequencies remain unchanged. The gadget uses only \sant{$O_k(1)$} auxiliary variables,
keeping the total number of variables at $O(n)$. If $X_j=0$, all remaining original constraints are satisfied by $\alpha_j$, which extends to a satisfying assignment of the gadget. If $X_j=1$, the query constraint has positive frequency and is violated by both $\alpha_j$ and its complement, the only assignments permitted by the gadget on the original variables. The final instance is therefore unsatisfiable exactly when $X_j=1$, allowing Bob to recover the queried bit.

Alice sends $O(n^2)$ sketches of at most $B$ bits each. The communication lower bound for $\Index$ therefore implies $n^2B=\Omega(n^4)$, and hence $B=\Omega(n^2)$. Applying the transfer theorem of \textcite{JLY26} yields the $\Omega(n^2)$ space lower bound in the strict-turnstile model.

\paragraph{Selection polynomials and the general lower bound.}
To generalize this construction, we express the cancellation rule as a polynomial in the tuple received by a constraint. Let $z\in\{0,1\}^4$ be the tuple received by $C_t$ under $\alpha_j$. In the case of $4$-\XOR, since $\alpha_j$ sets the queried variables to one and all other variables to zero, $z_a=1$ exactly when $t_a=j_a$. Thus $\sum_a z_a$ counts the selected single-index sketches containing $C_t$, while $\sum_{a<b}z_az_b$ counts the selected pair sketches containing it. Bob's combination therefore assigns $C_t$ frequency $X_tq(z)$, where
\[
 q(z)=2\sum_{a<b}z_az_b-\sum_a z_a.
\]
As the calculation above shows, $q$ vanishes on the unsatisfying tuples with zero or two ones, while $q(1111)=8$. It therefore singles out $1111$ among the unsatisfying tuples.

This is the property needed for the general reduction. For an unsatisfying tuple $b\notin R$, we call a polynomial $q_b$ with integer coefficients a \emph{selection polynomial} if it is non-zero at $b$ and zero at every other tuple outside $R$. Its values on satisfying tuples are unrestricted, since constraints satisfied by Bob's chosen assignment can remain.

Fix a degree-$s$ selection polynomial $q_b$, choosing its sign so that $q_b(b)>0$. In the general reduction, we use $k$ groups of variables and constraints
\[
 C_t=R(x_{1,t_1},\ldots,x_{k,t_k}),
 \qquad t\in[n]^k,
\]
with Alice inserting $C_t$ exactly when $X_t=1$. For his query $j$, Bob assigns $b_a$ to $x_{a,j_a}$ and $1-b_a$ to every other variable in group $a$. Thus only $C_j$ receives the tuple $b$.

We now rewrite $q_b$ as a polynomial in formal variables $z_1,\ldots,z_k$ by replacing its \amat{$a$-th} variable with $z_a$ when $b_a=1$ and with $1-z_a$ when $b_a=0$. For a fixed constraint $C_t$, the value assigned to its \amat{$a$-th} variable by the pinning assignment is
\[
\begin{cases}
b_a, & t_a=j_a,\\
1-b_a, & t_a\neq j_a,
\end{cases}
\;=\;
\begin{cases}
\mathbf{1}_{\{t_a=j_a\}}, & b_a=1,\\
1-\mathbf{1}_{\{t_a=j_a\}}, & b_a=0.
\end{cases}
\]
The expressions on the right are precisely our substitutions evaluated at $z_a=\mathbf{1}_{\{t_a=j_a\}}$. Thus evaluating the rewritten polynomial at these indicators gives the value of $q_b$ on the tuple received by $C_t$ under Bob's assignment. These substitutions do not increase degree. After multilinearization, each term therefore has the form $c_T\prod_{a\in T}z_a$ with $|T|\leq s$.

For each non-zero coefficient $c_T$ and every tuple $u\in[n]^T$, Alice sends the sketch of her constraints with $t_a=u_a$ for all $a\in T$. There are $O_k(n^s)$ such sketches. Bob selects the sketch with $u_a=j_a$ for every $a\in T$, multiplies it by $c_T$, and sums over the terms. The constant term corresponds to $T=\varnothing$ and uses the sketch of Alice's entire instance.

For a fixed constraint $C_t$, the subset selected for $T$ contains $C_t$ precisely when $t_a=j_a$ for every $a\in T$. Its frequency in the combined sketch is therefore
\[
 X_t\sum_T c_T\prod_{a\in T}\mathbf{1}_{\{t_a=j_a\}}.
\]
As we argued above, this equals $X_t$ times the value of $q_b$ on the tuple received by $C_t$ under Bob's assignment.

Only $C_j$ receives $b$, so its frequency is $X_jq_b(b)$, which is positive exactly when $X_j=1$. Every other violated constraint has frequency zero. Negative frequencies can occur only on constraints satisfied by Bob's assignment. Bob corrects them without knowing Alice's bits: if $q_b$ evaluates to $-\beta<0$ on the tuple received by $C_t$, he adds $\beta$ copies of that constraint, changing its frequency from $-\beta X_t$ to $\beta(1-X_t)\geq0$. This correction depends only on the query and introduces no constraints violated by the chosen assignment.

Bob then adds the pinning gadget, making the final instance unsatisfiable exactly when $X_j=1$. If each sketch uses $B$ bits, Alice's message has size $O_k(n^sB)$. Since her input contains $n^k$ bits, the communication lower bound for $\Index$ implies $B=\Omega_k(n^{k-s})$.

The key to matching our upper bound is an exact relation between the capturing degree and the minimum selection degree that we prove in \cref{sec:prelims}.

\begin{lemma}[Informal; \cref{lem:capturing-selection}]
\label[lemma]{lem:overview-degrees}
Let $R\subseteq\{0,1\}^k$ be nonempty and proper. Let $d=\cdeg(R)$, and let $s$ be the minimum degree of a selection polynomial over all choices of the unsatisfying tuple $b\notin R$. Then $d+s=k$.
\end{lemma}

Applying the reduction with $s=k-d$ gives an $\Omega_k(n^d)$ sketching lower bound for every nonempty Boolean relation $R$ that is not constant-satisfiable. The streaming-to-sketching transfer then yields an $\Omega_k(n^d)$ strict-turnstile lower bound, matching our $O_k(n^d\operatorname{polylog}n)$ upper bound up to polylogarithmic factors.

To see why $d+s\geq k$, take a degree-$d$ capturing polynomial and a degree-$s$ selection polynomial for some tuple $b\notin R$. Their product is nonzero at $b$ and zero at every other Boolean tuple. Its unique multilinear representation is therefore a nonzero multiple of
\[
 \prod_{i:b_i=1}z_i\prod_{i:b_i=0}(1-z_i),
\]
which has degree $k$. Since the product has degree at most $d+s$ and multilinearization cannot increase degree, we obtain $d+s\geq k$.

For the reverse inequality, consider all polynomials of degree less than $d$ that vanish on $R$. We claim that these polynomials have a common zero $b\notin R$. Otherwise, for each tuple outside $R$, we could choose such a polynomial that is nonzero at that tuple. A suitable rational linear combination would then vanish on $R$ and be nonzero at every tuple outside $R$. Clearing denominators would give a capturing polynomial of degree less than $d$, a contradiction.

To construct a selection polynomial for $b$, we start with a function $f$ that equals one at $b$ and zero at every other tuple outside $R$. Its values on $R$ remain free to choose. A linear-algebra argument using the common-zero property of $b$ shows that we can choose rational values so that every Fourier term of degree less than $d$ vanishes.

We then multiply $f$ by the full parity character $\chi_{[k]}$. This replaces each Fourier character $\chi_T(z)=\prod_{i\in T}(1-2z_i)$ by the character on the complementary coordinates:
\[
 \chi_{[k]}(z)\chi_T(z)=\chi_{[k]\setminus T}(z)
 \qquad\text{on Boolean inputs}.
\]
Since every character appearing in $f$ has $|T|\geq d$, every resulting term has degree $k-|T|\leq k-d$. Moreover, the parity character is always $1$ or $-1$, so multiplication preserves exactly where $f$ is zero. The resulting polynomial therefore has degree at most $k-d$, is nonzero at $b$, and vanishes at every other tuple outside $R$. Clearing denominators gives an integer selection polynomial, proving $s\leq k-d$.

\sant{For a non-constant-satisfiable Boolean constraint language $\Gamma$, we choose $R\in\Gamma$ of maximum capturing degree and encode Alice's input using only constraints applied on $R$. Although $R$ need not itself be a core, $\Gamma$ is, so Bob can use constraints from $\Gamma$ to pin his assignment up to a common automorphism of $\Gamma$. Such automorphisms preserve satisfaction of every constraint, so the same reduction yields the lower bound with exponent $\deg_{\mathbb Z}(\Gamma)$.}

We give the full lower-bound proof, including its extension to Boolean constraint languages, in \cref{sec:hardness}.

\paragraph{Bottleneck for larger domains.}
The degree identity above can fail over larger domains, even for relations that are already cores. Consider
$
 R=\{(x,y,z)\in\{0,1,2\}^3:x+y+z\equiv1\pmod3\}.
$
Its capturing degree is two, while its minimum selection degree is four, attained at the tuple $000$. The two degrees therefore sum to six, rather than the arity three. Our upper bound still gives an $O(n^2\operatorname{polylog}n)$-space algorithm, but the Boolean degree identity no longer provides a matching lower bound. This example shows that extending our lower-bound argument to larger domains requires a different approach. We discuss this barrier in \cref{sec:ternary-counterexample}.

\paragraph{Binary streams.}
In a \emph{binary stream}, every constraint has frequency zero or one after each update. Thus insertions add constraints that are currently absent, and deletions remove constraints that are currently present. Binary streams form a restricted class of strict-turnstile streams. Putterman~\cite{PuttermanPC} pointed us to an $O_k(n^2\operatorname{polylog}n)$-space algorithm for deciding the satisfiability of $k$-\XOR with literals on binary streams, for every fixed $k$~\cite[Theorem 5.1.3]{kuchlous2025}.
For every fixed $k\geq5$, this upper bound is polynomially smaller than our strict-turnstile lower bound of $\Omega_k(n^{\lceil k/2\rceil})$, obtained from \cref{cor:lb-languages}. Thus binary streams can admit substantially smaller space bounds than general strict-turnstile streams. In a related result, Kallaugher and Price~\cite{KallaugherPrice2020} exhibit a problem solvable in $O(\log n)$ space on binary streams for which every linear sketch requires $\Omega(n^{1/3}/\log^{1/3}n)$ bits. Determining the optimal space bound for $k$-\XOR satisfiability with literals on binary streams and obtaining a corresponding classification for general Boolean relations remain interesting open problems.

%% file: prelims.tex
\section{Preliminaries}
\label{sec:prelims}

\amat{Let $R\subseteq\{0,1\}^k$ be a fixed Boolean relation.} The arity $k$, the number of coordinates of $R$, is fixed, \amat{and constants in $O_k$ and $\Omega_k$ may depend only on $k$.} For a positive integer $m$, we define $[m]=\{1,\ldots,m\}$. The notation $\polylog(t)$ denotes a fixed power of $\log(t+2)$.

\subsection{Relations, instances, and streams}
\label{sec:instances}
A Boolean relation is \emph{proper} if it is not the full Boolean cube.
A tuple is \emph{satisfying} if it belongs to $R$, and \emph{unsatisfying} otherwise. We define $\bar R=\{0,1\}^k\setminus R$. The \emph{Hamming weight} of a Boolean tuple $a$ is $\sum_{i=1}^k a_i$. A constraint has the form $R(x_{i_1},\ldots,x_{i_k})$ and is satisfied by an assignment $\alpha$ when $(\alpha(x_{i_1}),\ldots,\alpha(x_{i_k}))\in R$. Its \emph{scope} is the ordered tuple $(x_{i_1},\ldots,x_{i_k})$. Constraints apply $R$ directly to variables, without negations or prescribed constants. The model permits repeated variables within a scope, although every constraint in our reduction has distinct variables.

An instance is a multiset of constraints. On $n$ variables, we define $\mathcal U_n$ to be the universe of constraint types, indexed by ordered scopes in $[n]^k$. We represent an instance by its frequency vector $f\in\Z_{\geq0}^{\mathcal U_n}$. A constraint $C$ is \emph{active} if $f_C>0$. We define $\supp(f)=\{C:f_C>0\}$ as the set of active constraints. An assignment satisfies the instance if it satisfies every active constraint. We define $\Sat_R(f)$ to be the set of such assignments and $\CSP(R)$ to be the problem of deciding whether this set is nonempty. For a set $S$ of constraint types, we also write $S$ for the frequency vector with value one on $S$ and zero elsewhere. This convention for sets does not restrict the multiplicities in a general multiset instance. %

\paragraph{Constraint languages.}
A \emph{constraint language} $\Gamma$ is a fixed finite collection of relations over a common finite domain $D$.  We assume without loss of generality that all its relations have a common arity $k$.
\footnote{\amat{We may assume a common arity by padding smaller-arity relations with coordinates that impose no conditions. This preserves satisfiability and capturing degree.}} A constraint in a $\Gamma$-instance has the form $R(x_{i_1},\ldots,x_{i_{k}})$ for some $R\in\Gamma$, and $\CSP(\Gamma)$ asks whether all its constraints can be satisfied simultaneously. The universe $\mathcal U_n^\Gamma$ specifies both a relation and an ordered scope, so $|\mathcal U_n^\Gamma|=\sum_{R\in\Gamma}n^{k}\leq|\Gamma|n^k$. We write $\Sat_\Gamma(f)$ for the satisfying assignments of an instance $f$. The frequency-vector, stream, and sketch definitions apply with this universe as well. A language is \emph{Boolean} when $D=\{0,1\}$, and we use $O_k$ and $\Omega_k$ in this case. Over a domain of size $q$, we use $O_{k,q}$ and $\Omega_{k,q}$. For fixed $k,q$, there are only finitely many relations of arity at most $k$ and constraint languages, including their restrictions to core domains. We fix the polynomial and gadget choices for each of them, so the constants in our bounds can be chosen to depend only on $k,q$.

\paragraph{Streams.}
A unit update inserts or deletes one copy of a constraint. A stream starts from the zero frequency vector and is \emph{strict turnstile} if every prefix frequency vector is nonnegative. Both strict-turnstile and signed streams permit updates of $-1$. A signed stream may delete a constraint more times than it has been inserted, making its accumulated frequency negative; the strict-turnstile model forbids this.
A one-pass algorithm processes the updates once and decides satisfiability at the end. We allow unrestricted computation time and count all retained memory, including decoding workspace.

\paragraph{Non-constant-satisfiable \CSP.}
We call $R$, or $\CSP(R)$, \emph{constant-satisfiable} if there is a value $a\in\{0,1\}$ such that assigning $a$ to every variable satisfies every instance. This holds exactly when $0^k\in R$ or $1^k\in R$. We call $R$ \emph{non-constant-satisfiable} if neither tuple belongs to $R$. If $R$ is constant-satisfiable, an algorithm can ignore all updates and always report satisfiable, using constant space. If $R=\varnothing$, every active constraint is unsatisfiable, so it suffices to maintain the counter $\sum_{C\in\mathcal U_n}f_C$ and report satisfiable exactly when it is zero. When every prefix multiplicity is at most $\mu$, this counter uses $\amat{O_k}(\log(n\mu+2))$ bits. Thus the only nontrivial cases are the nonempty, non-constant-satisfiable relations.

A Boolean language $\Gamma$ is \emph{constant-satisfiable} if there is one value $a\in\{0,1\}$ such that $a^{k}\in R$ for every $R\in\Gamma$, and \emph{non-constant-satisfiable} otherwise. In the former case, the constant assignment with value $a$ satisfies every $\Gamma$-instance, so decision uses constant space. The value must be common to all relations; the individual relations may be constant-satisfiable without the language being so. Empty relations can be handled separately by a counter for the total number of their active copies, using $\amat{O_k}(\log(n\mu+2))$ bits when prefix multiplicities are at most $\mu$. If this counter is positive the instance is unsatisfiable; otherwise only the nonempty relations matter. Full relations impose no restrictions. We therefore state the language extensions for collections of nonempty proper relations.

\amat{Unless otherwise specified, $R\subseteq\{0,1\}^k$ is assumed to be a fixed, \emph{proper} (i.e., $R\neq \{0,1\}^k$), nonempty, non-constant-satisfiable Boolean relation, as assumed in the introduction.}

\subsection{Linear sketches}
\label{sec:sketches}

For the lower bound, we consider the following finite-image form of a \emph{linear sketch}, following \cite[Definition 3.1.3]{JLY26}. \amat{A linear sketch is specified by a public random seed $\omega$ and a sketching map}
\(
L_\omega:\Z^{\mathcal U_n}\longrightarrow\Gamma_\omega
\)
into a finite abelian group. \amat{The map is a homomorphism, and $L_\omega(f)$ is the sketch of $f$.} Additivity means $L_\omega(f+h)=L_\omega(f)+L_\omega(h)$ for all integer vectors $f,h$, including signed vectors. The decoder receives $\omega$ and $L_\omega(f)$ and decides whether the instance $f$ is satisfiable. We define the \amat{sketch size} to be
\(
B=\sup_\omega\log_2|\im L_\omega|.
\)
Thus one \amat{sketch} can be encoded in at most $B+1$ bits; the public seed is not charged to this bound. For a constraint $C$, we define $e_C$ to be the frequency vector corresponding to one copy of $C$.

\subsection{Cores and pinning}
\label{sec:cores}
\amat{Our proofs work directly with non-constant-satisfiable Boolean relations and do not require first replacing a relation by its core. We nevertheless recall cores and their consequences, particularly the pinning lemma that we use below, and retain the general reduction to a core for completeness. In this subsection, $R$ may be any nonempty Boolean relation.}

\begin{definition}[Automorphisms and cores]
An \emph{automorphism} of $R$ is a bijection $g:\{0,1\}\to\{0,1\}$ such that $a\in R$ if and only if $g(a)\in R$, where $g$ is applied coordinatewise. We define $\Aut(R)$ to be the set of these maps. The relation $R$ is a \emph{core} if every map $e:\{0,1\}\to\{0,1\}$ with $e(R)\subseteq R$ is an automorphism.
\end{definition}
\begin{observation}[\amat{Boolean cores and constant assignments}]
\label[observation]{lem:boolean-core}
\amat{For a nonempty Boolean relation $R$, the relation $R$ is a core if and only if $\CSP(R)$ is non-constant-satisfiable.}
\end{observation}
\begin{proof}
The only nonbijective maps from $\{0,1\}$ to itself are the two constant maps. Since $R$ is nonempty, applying the constant map with value $a$ to $R$ gives the singleton set $\{a^k\}$. This map therefore preserves $R$ exactly when $a^k\in R$. If either constant tuple belongs to $R$, a nonbijective map preserves $R$, so $R$ is not a core. Conversely, if neither tuple belongs to $R$, every map preserving $R$ must be a permutation. Such a permutation maps $R$ onto itself, since $R$ is finite, and hence is an automorphism. This proves the equivalence.
\end{proof}

\amat{We also retain the notation for the core of an arbitrary nonempty Boolean relation.} If either constant tuple belongs to $R$, choose such a value $a$ and define $\Core(R)=R\cap\{a\}^k$, considered as a relation on the one-element domain $\{a\}$. Otherwise, define $\Core(R)=R$. Thus the core removes any domain value unnecessary for satisfiability. We use $R^\star=\Core(R)$ for this relation. \amat{ However, as implied by \cref{lem:boolean-core}, for the nontrivial cases we consider throughout this paper, i.e., if $R$ is nonempty and non-constant-satisfiable, then $\Core(R)=R$.}

The reduction of \cite{BPSV26} lets us replace $R$ by its core. Intuitively, applying the map onto the core to any satisfying assignment gives a satisfying assignment over the core domain; an assignment over that domain is already an assignment for $R$.

\begin{lemma}[Reduction to the core, {\cite[Lemma 4.2]{BPSV26}}]
\label[lemma]{lem:core-reduction}
Replacing each $R$-constraint by the $\Core(R)$-constraint on the same scope preserves satisfiability in both directions. This replacement preserves variables, multiplicities, and the update sequence. Consequently, the two decision problems have the same strict-turnstile streaming space complexity, up to constant extra space.
\end{lemma}

For a core relation, we can add constraints so that every satisfying assignment agrees with a chosen assignment on the original variables, up to applying one common automorphism. This common change of values preserves whether any constraint on the original variables is satisfied. The following variant of \cite[Lemma 4.4]{BPSV26} gives this guarantee with distinct variables within every scope.

\begin{lemma}[Core pinning, {\cite[Lemma 4.4]{BPSV26}}]
\label[lemma]{lem:pinning}
If $R$ is a core, then for every variable set $V$ and every assignment $\alpha:V\to\{0,1\}$, there are a set $Y$ of $4k$ variables disjoint from $V$ and an $R$-instance $P_\alpha$ on $V\cup Y$ with $\amat{O_k}(|V|+1)$ constraints such that
\[
\{\gamma|_V:\gamma\in\Sat_R(P_\alpha)\}
=\{g\circ\alpha:g\in\Aut(R)\}.
\]
Here $\gamma|_V$ denotes the restriction of $\gamma$ to $V$. Every scope contains distinct variables and at least one variable from $Y$. The variables in $Y$ may be shared by several constraints. No ordered scope is repeated. If $R$ is symmetric, meaning that membership is invariant under coordinate permutations, unordered scopes can also be deduplicated.
\end{lemma}

\paragraph{Cores of constraint languages.}
A map $e:\amat{D}\to\amat{D}$ is an \emph{endomorphism} of $\Gamma$ if it preserves every relation in $\Gamma$. An \emph{automorphism} is a bijection preserving membership in every relation in both directions; these maps form $\Aut(\Gamma)$. The language is a \emph{core} if every endomorphism is an automorphism. For a Boolean language of nonempty relations, a constant map preserves the language exactly when its value satisfies every relation. Since constant maps are the only nonbijective Boolean maps, $\Gamma$ is a core if and only if it is non-constant-satisfiable, by the same argument as in \cref{lem:boolean-core}. If a common satisfying value $a$ exists, define $\Core(\Gamma)$ by restricting every relation to the domain $\{a\}$; otherwise, define $\Core(\Gamma)=\Gamma$. Thus the core is taken for the entire language, with one common domain restriction. For an $r$-ary relation $R$, a map of the domain preserves $R\times\amat{D}^{k-r}$ if and only if it preserves $R$. Thus padding leaves the language's endomorphisms, automorphisms, and common constant satisfying assignments unchanged.

Brakensiek, Putterman, Sharma, and Velusamy~\cite[Appendix A]{BPSV26} extend the core reduction and pinning argument to constraint languages. In particular, when $\Gamma$ is a core, any assignment $\alpha$ on a variable set $V$ can be pinned, up to one common automorphism in $\Aut(\Gamma)$, using constraints from $\Gamma$. The pinning instance has $\amat{O_{k,q}}(|V|+1)$ constraints and $\amat{O_{k,q}}(1)$ additional variables, and the assignments on $V$ extending to its solutions are exactly $\{g\circ\alpha:g\in\Aut(\Gamma)\}$.

\subsection{Capturing and selection polynomials}
\label{sec:polynomials}
We now define the notion of a \emph{capturing polynomial} which is zero precisely on the satisfying tuples. The definitions and lemmas in this subsection apply to every nonempty proper Boolean relation $R$, whether or not it is constant-satisfiable.

\begin{definition}[Capturing polynomial]
\label[definition]{def:capturing}
An integer polynomial $p\in\Z[z_1,\ldots,z_k]$ captures $R$ if, for every Boolean tuple $a$,
\[
p(a)=0\quad\Longleftrightarrow\quad a\in R.
\]
The \emph{integer capturing degree} $d=\cdeg(R)$ is the minimum degree of such a polynomial. For an unsatisfying tuple $b$, we also define
\[
d_R(b)=\min\{\deg p:p\in\Z[z_1,\ldots,z_k],\ p(a)=0\text{ for every }a\in R,\ p(b)\ne0\}.
\]
A polynomial in the latter minimum may be zero at other unsatisfying tuples.
\end{definition}

For a one-element core, we set $\cdeg(\Core(R))=0$.

\begin{observation}[\amat{Integer, rational, and real coefficients}]
\label[observation]{obs:coefficient-fields}
\amat{Let $D\subseteq\Z$ be finite and let $R\subsetneq D^k$ be nonempty. The minimum degree of a polynomial whose zeros on $D^k$ are exactly $R$ is the same for integer, rational, and real coefficients. The same holds for the minimum degree of a polynomial vanishing on $R$ and nonzero at a specified tuple $b\in D^k\setminus R$.}
\end{observation}
\begin{proof}
\amat{Let $p$ be a real polynomial of degree $d$ whose zeros on $D^k$ are exactly $R$. Vanishing on $R$ imposes homogeneous linear equations on the coefficients of a polynomial of degree at most $d$. These equations have integer coefficients, since every element of $D$ is an integer. Gaussian elimination therefore gives a basis $p_1,\ldots,p_m$ of the real solution space consisting of polynomials with rational coefficients. Write $p=\sum_i a_i p_i$ with $a_i\in\R$.}

\amat{Replace the $a_i$ by sufficiently close rational numbers. The resulting polynomial still vanishes on $R$, because every $p_i$ does. It remains nonzero at every tuple outside $R$, because there are finitely many such tuples and evaluation at each is continuous in the coefficients. Clearing denominators then gives an integer polynomial of degree at most $d$ with the same zeros on $D^k$. The reverse inequality follows because every integer polynomial is also a rational and a real polynomial. For the second statement, the same argument only requires preserving the nonzero value at $b$.}
\end{proof}

\amat{For Boolean relations,} without loss of generality, we assume that the polynomials in these definitions are multilinear, meaning that every variable has exponent at most one: replacing every positive power $z_i^r$ by $z_i$ preserves all Boolean evaluations and cannot increase degree.

We also have
\begin{equation}
\label{eq:max-degree}
d=\cdeg(R)=\max_{b\in\bar R}d_R(b).
\end{equation}
Every capturing polynomial separates each unsatisfying tuple from $R$, giving $d\geq\max_b d_R(b)$. For the reverse inequality, enumerate $\bar R=\{b_1,\ldots,b_\ell\}$ and choose an integer polynomial $p_i$ attaining $d_R(b_i)$. For each $b_j$, the expression $\sum_{i=1}^\ell T^{i-1}p_i(b_j)$ is a nonzero polynomial in $T$, since its coefficient of $T^{j-1}$ is $p_j(b_j)\ne0$. Choose an integer $T$ that is not a root of any of these finitely many polynomials. Then $\sum_i T^{i-1}p_i$ captures $R$ and has degree at most $\max_b d_R(b)$.

\begin{definition}[Capturing degree of a constraint language]
\label[definition]{def:language-degree}
For a fixed finite Boolean language $\Gamma$ of nonempty proper relations, its \emph{integer capturing degree} is $\cdeg(\Gamma)=\max_{R\in\Gamma}\cdeg(R)$. Equivalently, it is the least $d$ such that every $R\in\Gamma$ has an integer capturing polynomial of degree at most $d$. We set this degree to zero for the empty language and for a language on a one-element core domain. Empty and full relations do not contribute to the maximum and are handled as described above.
\end{definition}

Padding preserves capturing degree. If $R$ has arity $r$ and $R'=R\times\{0,1\}^{k-r}$, a polynomial capturing $R$ also captures $R'$ when viewed as a polynomial in $k$ variables. Conversely, setting the added coordinates to zero in a polynomial capturing $R'$ gives one capturing $R$, without increasing its degree. Hence $\cdeg(R')=\cdeg(R)$, so padding does not change the language's capturing degree.

We now define a \emph{selection polynomial} which we use to separate an unsatisfying tuple from all the other unsatisfying tuples.
\begin{definition}[Selection polynomial]
\label[definition]{def:selection}
For an unsatisfying tuple $b$, an integer polynomial $q\in\Z[z_1,\ldots,z_k]$ selects $b$ if
\[
q(b)\ne0,\qquad q(a)=0\quad\text{for every unsatisfying }a\ne b.
\]
Its values on satisfying tuples are unrestricted. We define $s_R(b)$ to be the minimum degree of a selection polynomial for $b$.
\end{definition}

\begin{lemma}[Capturing and selection degrees]
\label[lemma]{lem:capturing-selection}
For every unsatisfying tuple $b\notin R$,
\[
    d_R(b)+s_R(b)=k.
\]
Consequently, if $d=\deg_{\mathbb Z}(R)$, then some
unsatisfying tuple $b$ has an integer multilinear
selection polynomial $q$ of degree $k-d$ with $q(b)>0$.
\end{lemma}

\begin{proof}
Fix $b\notin R$ and define $r=d_R(b)$.

We first show that $r+s_R(b)\ge k$.
Let $p$ be a degree-$r$ polynomial that is zero on $R$
and nonzero at $b$, and let $q$ be a minimum-degree
selection polynomial for $b$.
On the Boolean cube, the product $pq$ is nonzero
exactly at $b$, and so its unique multilinear representation is
\[
 p(b)q(b)\prod_{i:b_i=1}z_i\prod_{i:b_i=0}(1-z_i).
\]
Observe that the coefficient of $z_1\cdots z_k$ is nonzero, and hence it has degree $k$.
Since multilinearization cannot increase degree, we get
\[
    k\le\deg(pq)\le r+s_R(b).
\]

For the reverse inequality, we construct a selection
polynomial of degree at most $k-r$.
We work over $\mathbb Q$, since rescaling to integer coefficients
does not change the degree.

\amat{By the minimality of $r$, every polynomial of degree less than $r$ that is zero on $R$ must also be zero at $b$. Let $v(a)$ be the vector of all multilinear monomials of degree less than $r$, evaluated at $a$. We claim that $v(b)$ lies in the rational span of the vectors $v(a)$ for $a\in R$. Otherwise, there would be a rational vector $u$ with $u^\top v(a)=0$ for every $a\in R$ but $u^\top v(b)\ne0$. The polynomial $h(z)=u^\top v(z)$ would then have degree less than $r$, vanish on $R$, and be nonzero at $b$, a contradiction.}\amati{is it too dense though?}

\amat{Consequently, $v(b)=\sum_{a\in R}\lambda_a v(a)$ for rational coefficients $\lambda_a$. Taking the inner product with the coefficient vector of any multilinear polynomial $h$ of degree less than $r$ gives}
\begin{equation}\label{eq:evaluation-dependence}
    h(b)=\sum_{a\in R}\lambda_a h(a)
    \qquad\text{for every }h\text{ with }\deg h<r.
\end{equation}
\amat{The coefficients $\lambda_a$ depend only on these evaluation vectors, so the same coefficients work for every $h$. Multilinearizing $h$ preserves its Boolean evaluations and cannot increase its degree, so the identity also holds when $h$ is not multilinear.}

We use these coefficients to construct a selection polynomial for $b$ of degree at most $k-r$. For each Boolean tuple $a$, define
\[
    c_a=
    \begin{cases}
        1, & a=b,\\
        -\lambda_a, & a\in R,\\
        0, & \text{otherwise}.
    \end{cases}
\]
By \cref{eq:evaluation-dependence},
\begin{equation}\label{eq:low-degree-cancellation}
    \sum_{a\in\{0,1\}^k}c_a h(a)=0
    \qquad\text{for every }h\text{ with }\deg h<r.
\end{equation}

For each $a\in\{0,1\}^k$, define its indicator polynomial
\[
    \delta_a(z)=
    \prod_{i:a_i=1}z_i
    \prod_{i:a_i=0}(1-z_i).
\]
On Boolean inputs, $\delta_a$ equals one at $a$ and zero at every other tuple.
Consider the rational multilinear polynomial
\[
    q(z)=\sum_{a\in\{0,1\}^k}
          (-1)^{|a|}c_a\delta_a(z).
\]
Here $|a|=\sum_{i=1}^k a_i$ is the Hamming weight of $a$.
For every $a\in\{0,1\}^k$, we have $q(a)=(-1)^{|a|}c_a$.
Thus $q(b)\ne0$ and $q$ is zero at every other unsatisfying tuple.
Clearing denominators will give an integer polynomial with these same zeros.

It remains to show that $\deg q\le k-r$.
Fix $S\subseteq[k]$ and define $z_S=\prod_{i\in S}z_i$.
The coefficient $\gamma_S$ of $z_S$ in $q$ is
\[
\begin{aligned}
    \gamma_S
    &=(-1)^{|S|}
      \sum_{\substack{a\in\{0,1\}^k\\
                      a_i=0\ \forall i\notin S}}c_a\\
    &=(-1)^{|S|}
      \sum_{a\in\{0,1\}^k}
      c_a\prod_{i\notin S}(1-a_i).
\end{aligned}
\]

This is because in the expansion of $\delta_a$, the coefficient
of $z_S$ is zero unless every coordinate where
$a_i=1$ belongs to $S$, in which case it equals
$(-1)^{|S|-|a|}$.

Now, note that if $|S|>k-r$, the polynomial
\[
    h_S(z)=\prod_{i\notin S}(1-z_i)
\]
has degree $k-|S|<r$. By
\eqref{eq:low-degree-cancellation}, the coefficient $\gamma_S$ is therefore zero.

Thus the coefficient of every monomial of degree greater than $k-r$
is zero, proving that $\deg q\le k-r$.
Rescaling the coefficients gives an integer selection
polynomial of the same degree. Consequently,
\(
    s_R(b)\le k-r.
\)
Combining the two inequalities yields
\(
    d_R(b)+s_R(b)=k.
\)

Finally, by \cref{eq:max-degree}, there exists an unsatisfying
tuple $b$ with $d_R(b)=d$. The equality just proved
gives a multilinear integer selection polynomial of degree $k-d$.
Changing its sign if necessary ensures $q(b)>0$.
\end{proof}

As an example, we show that the capturing and selection degrees of $k$-XOR are $\lceil k/2 \rceil$ and $\lfloor k/2 \rfloor$, respectively.

\begin{corollary}[Capturing degree of parity]
\label[corollary]{cor:xor-degree}
\amat{For every $k\geq2$ and $\beta\in\{0,1\}$, the relation
\[
    R_\beta=\left\{z\in\{0,1\}^k:
        \sum_{i=1}^k z_i\equiv\beta\pmod2\right\}
\]
has integer capturing degree $\lceil k/2\rceil$. The same bound holds
when any subset of the coordinates is complemented.}\amatr{added k xor degree proof}\santr{changed some wording}
\end{corollary}
\begin{proof}
First consider $R_1$ and let $w=z_1+\cdots+z_k$. The polynomial
\[
    p(z)=\prod_{j=0}^{\lceil k/2\rceil-1}\bigl(w-(2j+1)\bigr)
\]
vanishes exactly on the odd-weight tuples, so
$\cdeg(R_1)\leq\lceil k/2\rceil$. On the other hand,
\[
    q(z)=\prod_{j=1}^{\lfloor k/2\rfloor}(w-2j)
\]
is nonzero at $0^k$ and vanishes at every other even-weight tuple.
It therefore selects $0^k$, giving $s_{R_1}(0^k)\leq\lfloor k/2\rfloor$.
By \cref{lem:capturing-selection},
\[
    \cdeg(R_1)\geq d_{R_1}(0^k)
    =k-s_{R_1}(0^k)\geq\lceil k/2\rceil.
\]
This proves equality. Complementing the coordinates is an invertible affine
substitution on the Boolean cube, so it preserves the minimum capturing
degree. Complementing one coordinate exchanges $R_0$ and $R_1$,
proving the remaining assertions.
\end{proof}

\subsection{Linear-algebra notation}
\label{sec:linear-algebra}
For symmetric matrices, $U\preceq V$ means $x^\top Ux\leq x^\top Vx$ for every real vector $x$. A symmetric matrix $U$ is positive semidefinite if $x^\top Ux\geq0$ for all $x$, and positive definite if the inequality is strict for nonzero $x$. Its operator norm $\|U\|_{\mathrm{op}}$ is its largest absolute eigenvalue. Let $I$ be the identity matrix, and let $\diag(f)$ be the diagonal matrix with entries $f_i$.
For a square matrix $G$, $\tr G$ is the sum of its diagonal entries, and $\rank G$ is the dimension of its column span. For positive semidefinite $G$, its \emph{pseudoinverse} $G^\dagger$ is obtained by inverting each positive eigenvalue and leaving zero eigenvalues at zero.

%% file: lowerbound.tex
\section{Lower bound}
\label{sec:hardness}
We prove that deciding satisfiability under our strict turnstile model requires $\amat{\Omega_k}(n^d)$ bits of space, where $d$ is the integer capturing degree of \amat{$R$} and $n$ is the number of variables. This bound applies to algorithms required to handle all legal streams up to a sufficiently large polynomial length.

\begin{theorem}[Main lower bound]
\label[theorem]{thm:stream-lower}
Let $R$ be a fixed \amat{nonempty, non-constant-satisfiable} Boolean relation and \amat{let $d=\cdeg(R)$}. There is a constant $c_R$ such that every randomized one-pass algorithm deciding $\CSP(R)$ with error at most $1/3$ on every strict-turnstile stream of length at most $n^{c_R}$ on $n$ variables uses
\[
 \amat{\Omega_k}(n^d)
\]
bits of space, for all sufficiently large $n$.
\end{theorem}

To prove this theorem, we combine two statements. The first, proved in \cref{sec:reduction}, is the lower bound for linear sketches.
\begin{theorem}[Sketching lower bound]
\label[theorem]{thm:sketch-lower}
\sketchlowerboundstatement
\end{theorem}

The second statement follows from the strict-turnstile streaming-to-sketching theorem of Jiang, Liu, and Yu~\cite[Corollary 7.2.2]{JLY26}.
\begin{theorem}[Strict-turnstile streaming to sketching, {\cite[Corollary 7.2.2]{JLY26}}]
\label[theorem]{thm:transfer}
Fix $R$ and a constant integer $F\geq1$. There is a constant $c_{R,F}$ such that any randomized $S$-bit one-pass algorithm deciding $\CSP(R)$ with error at most $1/3$ on every strict-turnstile stream of length at most $n^{c_{R,F}}$ yields a public-coin linear sketch with the same error bound on every instance in $\{0,\ldots,F\}^{\mathcal U_n}$ and \amat{size} $\amat{O_{k,F}}(S)$, for all sufficiently large $n$.
Although the final frequency of each constraint is at most \(F\), its intermediate frequency during the stream may be polynomially large in $n$.
\end{theorem}
Every final instance in $\{0,\ldots,F\}^{\mathcal U_n}$ has at most $Fn^k$ constraint copies. We first reduce the streaming error to $1/100$ using a constant number of independent copies. For sufficiently large $c_{R,F}$, we have $n^{c_{R,F}}\geq(n+Fn^k+100)^{C_R}$, where $C_R$ is the constant in their corollary. Their frequency-vector argument then gives a public-coin sketch with image size $2^{O_{k,F}(S)}$, hence size $O_{k,F}(S)$ bits, and error at most $3/100$.

These statements give the streaming lower bound directly.
\begin{proof}[Proof of \cref{thm:stream-lower}]
Choose $c_R$ large enough to apply \cref{thm:transfer} with the constant $F_R$ from \cref{thm:sketch-lower}. An $S$-bit streaming algorithm then gives a linear sketch of \amat{size} $\amat{O_k}(S)$, whereas \cref{thm:sketch-lower} requires $\amat{\Omega_k}(n^d)$ bits. Therefore, $S=\amat{\Omega_k}(n^d)$.

\end{proof}

It remains to prove \cref{thm:sketch-lower}, which we do in the next subsection.

\subsection{\Index{} to sketching reduction}
\label{sec:reduction}

We now prove \cref{thm:sketch-lower}. \amat{We retain the standing assumption that $R$ is nonempty and non-constant-satisfiable, and let $d=\cdeg(R)$.} 
We construct a hard distribution with frequencies bounded by $\amat{O_k}(1)$ and show that a linear sketch for it requires $\amat{\Omega_k}(n^d)$ bits per \amat{sketch}.
To prove the hardness, we reduce one-way \Index{} to satisfiability. In \Index, Alice receives a Boolean string $X=(X_t)_{t\in[n]^k}$, and Bob receives an index $j\in[n]^k$. Alice sends one message, after which Bob must recover $X_j$.

Roughly, we construct an instance on $kn+4k=\amat{\Theta_k}(n)$ variables that is unsatisfiable if and only if $X_j=1$. Alice sends $\amat{O_k}(n^{k-d})$ \amat{sketches}. Bob combines selected \amat{sketches} from her message with \amat{sketches} he computes himself to obtain the sketch of this instance. The $\Omega(n^k)$ communication lower bound for \Index{} then implies an $\amat{\Omega_k}(n^k/n^{k-d})=\amat{\Omega_k}(n^d)$ lower bound on the size of one \amat{sketch}.

Recall from \cref{def:capturing,def:selection} that a capturing polynomial is zero precisely on satisfying tuples, while a selection polynomial is nonzero at one chosen unsatisfying tuple and zero at all the others. By \cref{eq:max-degree,lem:capturing-selection}, some tuple $b\in\bar R$ has a selection polynomial $q$ of degree $k-d$ with $q(b)>0$. For the entire lower bound proof, we will fix this tuple $b$ and the selection polynomial $q$, and define $s=k-d$. They depend only on $R$ and are known to both parties before their inputs arrive.

\amat{Fix a linear sketch as defined in \cref{sec:sketches}. Since its seed is public, we omit $\omega$ and write $L=L_\omega$ for the sketching map. For a constraint $C$, the sketch of one copy is $L(e_C)$, which both parties can compute from its scope. Additivity gives $L(f)=\sum_C f_C L(e_C)$, where $f_C$ is the frequency of $C$.}

\subsubsection{Reduction}
\label{sec:construction}
We now define the reduction from \Index{} to $\CSP(R)$. We create $k$ disjoint groups of variables $V_a=\{x_{a,u}:u\in[n]\}$, one for each coordinate $a\in[k]$, and define $V=\bigcup_a V_a$. Each array entry $t=(t_1,\ldots,t_k)$ corresponds to the constraint
\(
C_t=R(x_{1,t_1},\ldots,x_{k,t_k}).
\)

\paragraph{Alice.}
Alice constructs and sends the following \amat{sketches}.
\begin{enumerate}[label=(\roman*)]
\item \emph{The full instance.} She sends
\begin{equation}
\label{eq:M0}
M_0=\sum_{t\in[n]^k}X_tL(e_{C_t}).
\end{equation}
This is the sketch of the instance containing one copy of $C_t$ whenever $X_t=1$.
\item \emph{Subinstances incident on at most $k-d$ specified variables.} For every nonempty $S\subseteq[k]$ with $|S|\leq s$, where $s=k-d$, and every $u\in[n]^S$, she sends
\begin{equation}
\label{eq:MS}
M_{S,u}=\sum_{t:\,t|_S=u}X_tL(e_{C_t}).
\end{equation}
Here $t|_S$ denotes the tuple $t$ restricted to the coordinates in $S$, and $[n]^S$ is the set of possible restricted tuples. This is the sketch of the constraints in Alice's instance that contain every specified variable $x_{a,u_a}$, for $a\in S$. There are $\amat{O_k}(n^s)$ such \amat{sketches}: the number of choices of $S$ depends only on $R$, and each has at most $n^s$ choices of $u$.
\end{enumerate}

Thus Alice sends $\amat{O_k}(n^s)=\amat{O_k}(n^{k-d})$ \amat{sketches} in total, as required.

\paragraph{Bob.}
Bob combines Alice's \amat{sketches} with \amat{sketches} of constraints he constructs himself. His goal is to obtain an instance that is satisfiable if and only if the queried bit is zero. 

The high-level idea is that when $X_j=1$, Bob needs the query constraint to rule out every satisfying assignment. \amat{By \cref{lem:boolean-core}, our standing assumption makes $R$ a core.} The pinning lemma, \cref{lem:pinning}, \amat{therefore} lets him force a chosen assignment $\alpha_j$ on $V$ up to a common automorphism, under which the query constraint is unsatisfied. For the other direction, when $X_j=0$, all remaining constraints must be satisfied by $\alpha_j$. Bob uses the selection polynomial to remove every other constraint unsatisfied by this assignment.

In particular, Bob constructs the following three \amat{sketches}.
\begin{enumerate}[label=(\roman*)]
\item \emph{Pinning instance.} Bob defines the assignment $\alpha_j:V\to\{0,1\}$ by
\begin{equation}
\label{eq:assignment}
\alpha_j(x_{a,u})=
\begin{cases}
b_a,&u=j_a,\\
1-b_a,&u\ne j_a.
\end{cases}
\end{equation}
Thus the queried constraint $C_j$ is assigned the unsatisfying tuple $b$. Bob constructs the pinning instance $P_{\alpha_j}$ from \cref{lem:pinning}. Its constraints force the assignment on $V$ to equal $\alpha_j$, up to a common automorphism of $R$. Such an automorphism preserves both satisfying and unsatisfying tuples, and as we will see later this freedom does not affect the reduction. The pinning instance uses $4k$ additional dummy variables, and Bob computes
\[
M_{\mathrm{pin}}=L(P_{\alpha_j}).
\]

Note that pinning already gives us one implication: if Bob simply added $M_{\mathrm{pin}}$ to $M_0$, then $X_j=1$ would force unsatisfiability, because $C_j$ would be present and violated under every assignment allowed by pinning. However, the converse can fail: even when $X_j=0$, another constraint from Alice's instance might be unsatisfied by $\alpha_j$. Bob's next step removes every such constraint while retaining the query constraint exactly when $X_j=1$.

\item \emph{Selecting the queried bit.} For each $t\in[n]^k$, Bob defines
\begin{equation}
\label{eq:beta}
\beta_{j,t}=q(C_t(\alpha_j)),
\end{equation}
where $C_t(\alpha_j)$ is the Boolean tuple assigned to the variables of $C_t$. To express these coefficients using Alice's message, define $z_a(t,j)=\one[t_a=j_a]$. This indicator is one exactly when $C_t$ uses the queried variable $x_{a,j_a}$ in coordinate $a$, and zero otherwise. By \cref{eq:assignment}, that coordinate is assigned $b_a$ in the first case and $1-b_a$ in the second. We can express both cases as
\[
(C_t(\alpha_j))_a=(1-b_a)+(2b_a-1)z_a(t,j).
\]
Substitute these expressions, one for each coordinate, into the fixed selection polynomial $q$. Each expression is affine in one indicator, so the resulting polynomial still has degree at most $s=k-d$. We therefore obtain an integer linear combination of products of at most $s$ indicators:
\begin{equation}
\label{eq:expansion}
\beta_{j,t}=a_{\varnothing}
+\sum_{\substack{\varnothing\ne S\subseteq[k]\\|S|\leq s}}
a_S\prod_{a\in S}\one[t_a=j_a]. 
\end{equation}

Note that the coefficients $a_S$ depend only on $R,b,q$. The product indexed by $S$ is one exactly for the array entries with $t|_S=j|_S$. Recall that Alice's \amat{sketch} $M_{S,j|_S}$ sums over exactly these entries. So, Bob uses these \amat{sketches} and forms the sketch
\begin{equation}
\label{eq:MBob}
M_{\mathrm{Bob}}=a_{\varnothing}M_0
+\sum_{\substack{\varnothing\ne S\subseteq[k]\\|S|\leq s}}
a_SM_{S,j|_S}
=\sum_t\beta_{j,t}X_tL(e_{C_t}).
\end{equation}
The last equality follows by substituting \cref{eq:M0,eq:MS} and collecting the coefficient of each $X_tL(e_{C_t})$. 

The crucial property is that $\beta_{j,t}=0$ whenever $t\ne j$ and $C_t(\alpha_j)$ is unsatisfying. Indeed, only $C_j$ is assigned $b$, and $q$ is zero at every other unsatisfying tuple. Thus the signed frequencies $\beta_{j,t}X_t$ in \cref{eq:MBob} eliminate every non-query constraint violated by $\alpha_j$. The query coefficient is $\beta_{j,j}=q(b)>0$, so its signed frequency is positive exactly when $X_j=1$.

\item \emph{Correcting negative frequencies.}  \amat{Some coefficients $\beta_{j,t}$ may be negative, so $M_{\mathrm{Bob}}$ need not yet be the sketch of a valid instance. If $\beta_{j,t}<0$, the current frequency $\beta_{j,t}X_t$ is either zero or $\beta_{j,t}$. Adding $-\beta_{j,t}$ copies of $C_t$ changes it to $-\beta_{j,t}$ or zero, respectively. Bob can do this without knowing $X_t$, since $\beta_{j,t}$ depends only on his query and the fixed polynomial $q$. He therefore computes}
\begin{equation}
\label{eq:Mpos}
M_{\mathrm{pos}}=\sum_{t\in[n]^k}(-\beta_{j,t})_+L(e_{C_t}),
\end{equation}
\amat{where $v_+=\max\{v,0\}$. Thus every negative coefficient now gives frequency $(-\beta_{j,t})(1-X_t)\geq0$, and nonnegative coefficients are unchanged.}
\amat{A coefficient can be negative only when $\alpha_j$ satisfies $C_t$, since $q$ is zero at every unsatisfying tuple other than $b$ and is positive at $b$. The added copies therefore introduce no constraint violated by $\alpha_j$. Constraints with coefficient zero remain absent, and the queried constraint still has frequency $q(b)X_j$.}

\end{enumerate}

Bob adds $M_{\mathrm{pin}}$, $M_{\mathrm{Bob}}$, and $M_{\mathrm{pos}}$, feeds the combined \amat{sketch} to the decoder, and outputs one if it reports unsatisfiable and zero otherwise. This completes the reduction.

\subsubsection{Correctness}
\label{sec:correctness}
Define the resulting frequency vector by
\begin{equation}
\label{eq:fB}
 f_B=f_B(X,j)=P_{\alpha_j}+\sum_{t\in[n]^k}c_{j,t}(X)e_{C_t},
 \qquad c_{j,t}(X)=\beta_{j,t}X_t+(-\beta_{j,t})_+.
\end{equation}
It consists of the pinning constraints together with the constraints $C_t$ at these frequencies. Additivity gives
\[
 L(f_B)=M_{\mathrm{pin}}+M_{\mathrm{Bob}}+M_{\mathrm{pos}}.
\]

To prove correctness of the reduction, we establish two claims: (1) $f_B$ has bounded nonnegative integer frequencies, and (2) $f_B$ is unsatisfiable if and only if $X_j=1$.

\begin{claim}
\label[claim]{clm:validity}
\amat{%
$\|f_B\|_1=O_k(n^k)$, where $\|f_B\|_1=\sum_C(f_B)_C$ is the total number of constraint copies.}
\end{claim}
\begin{proof}
From \cref{eq:fB},
\begin{equation}
\label{eq:frequencies}
c_{j,t}(X)=(-\beta_{j,t})_++\beta_{j,t}X_t
=\begin{cases}
\beta_{j,t}X_t,&\beta_{j,t}\geq0,\\
(-\beta_{j,t})(1-X_t),&\beta_{j,t}<0.
\end{cases}
\end{equation}
Since $X_t\in\{0,1\}$ and $\beta_{j,t}$ is an integer, both cases are nonnegative integers. Every pinning constraint contains an additional variable outside $V$, whereas every $C_t$ has all its variables in $V$. The pinning constraints therefore do not coincide with any $C_t$, and each contributes frequency one. Set $F_R=\max\{1,\max_{a\in\{0,1\}^k}|q(a)|\}$. Each $c_{j,t}(X)$ is at most $|\beta_{j,t}|\leq F_R$, so every frequency is bounded by $F_R=\amat{O_k}(1)$. There are $n^k$ constraints $C_t$ and $\amat{O_k}(n+1)$ pinning constraints. \amat{These bounds give $\|f_B\|_1=O_k(n^k)$.}
\end{proof}

\begin{claim}
\label[claim]{clm:guarantee}
For every $X,j$, the instance $f_B(X,j)$ is unsatisfiable if and only if $X_j=1$.
\end{claim}
\begin{proof}
Under the pinned assignment $\alpha_j$, only $C_j$ is assigned the tuple $b$ (for every other $t\ne j$, some coordinate has $t_a\ne j_a$, and \cref{eq:assignment} assigns the opposite Boolean value in that coordinate). If $t\ne j$ and $C_t(\alpha_j)$ is unsatisfying, the selection polynomial gives $\beta_{j,t}=0$, so $c_{j,t}(X)=0$. Thus every active constraint $C_t$ with $t\ne j$ is satisfied by $\alpha_j$. For the query index $j$, the frequency is 
$
c_{j,j}(X)=q(b)X_j,
$
so $C_j$ is active exactly when $X_j=1$.

If $X_j=0$, every active constraint $C_t$ is satisfied by $\alpha_j$. The pinning lemma ensures that $\alpha_j$ extends to an assignment satisfying $P_{\alpha_j}$. This extension satisfies all of $f_B$.

If $X_j=1$, every assignment satisfying the pinning constraints equals $g\circ\alpha_j$ on $V$, for some common automorphism $g\in\Aut(R)$. The active query constraint is then assigned $g(b)$, which is still unsatisfying. No assignment can satisfy both the query constraint and the pinning constraints. This proves the equivalence.
\end{proof}

\begin{restate}{thm:sketch-lower}
\sketchlowerboundstatement
\end{restate}
\begin{proof}
We combine \cref{clm:validity,clm:guarantee} with the communication lower bound for \Index. Alice sends $1+\sum_{r=1}^{s}\binom kr n^r=\amat{O_k}(n^{k-d})$ \amat{sketches}. By definition, $B$ is the size in bits of one \amat{sketch}, so each can be encoded in at most $B+1$ bits. When $X$ and $j$ are independent and uniform, $f_B(X,j)$ has a fixed distribution independent of the public seed. By \cref{clm:guarantee}, Bob recovers $X_j$ exactly when the sketch decoder correctly decides satisfiability of $f_B(X,j)$. Hence a sketch with average error at most $1/3$ on this distribution yields a public-coin \Index{} protocol with the same error and communication $\amat{O_k}(n^{k-d}(B+1))$. The $\Omega(n^k)$ one-way communication lower bound implies
\[
B+1=\amat{\Omega_k}(n^d).
\]
For sufficiently large $n$, this gives $B=\amat{\Omega_k}(n^d)$ on $kn+4k=\amat{\Theta_k}(n)$ variables. For an arbitrary sufficiently large total number $n'$ of variables, take $n=\lfloor(n'-4k)/k\rfloor$ and leave the remaining variables unused. The same construction and communication argument then give $B=\amat{\Omega_k}((n')^d)$. 
\end{proof}

\subsection{Lower bounds for Boolean constraint languages}
\label{sec:lb-languages}
The reduction also applies to fixed finite Boolean constraint languages. We choose a relation of maximum capturing degree for Alice's constraints. That relation need not itself be a core and Bob can use the entire language to pin the assignment.

\begin{corollary}[Lower bounds for constraint languages]
\label[corollary]{cor:lb-languages}
Let $\Gamma$ be a fixed finite, non-constant-satisfiable Boolean language of nonempty proper relations of arity $k$, and let $d=\cdeg(\Gamma)$. There is a constant $F_\Gamma$ such that, for every sufficiently large $n$, there is a distribution on instances on $n$ variables with frequencies at most $F_\Gamma$ for which every public-coin linear sketch deciding satisfiability with average error at most $1/3$ has \amat{size} $\amat{\Omega_k}(n^d)$.

There is also a constant $c_\Gamma$ such that every randomized one-pass algorithm deciding $\CSP(\Gamma)$ with error at most $1/3$ on every strict-turnstile stream of length at most $n^{c_\Gamma}$ uses $\amat{\Omega_k}(n^d)$ bits, for all sufficiently large $n$.
\end{corollary}

\begin{proof}
Choose $R\in\Gamma$ with $\cdeg(R)=d$. By \cref{lem:capturing-selection}, some unsatisfying tuple $b$ has an integer selection polynomial of degree $k-d$ that is positive at $b$. This lemma requires only that $R$ be nonempty and proper. Thus we can use exactly the same assignment, coefficients, and sketches as above, with $n$ variables in each of $k$ groups. Alice sends $O_k(n^{k-d})$ sketches to encode her $n^k$ input bits.

The only change is to the pinning instance. Since $\Gamma$ is not constant-satisfiable, it is a core. There is an analogous version of \cref{lem:pinning} for constraint languages (see \cite[Appendix A]{BPSV26}). Applying this lemma, Bob can enforce $\alpha_j$ up to one common automorphism of $\Gamma$. Such an automorphism preserves $R$, so every allowed assignment still violates the query constraint when $X_j=1$. When $X_j=0$, $\alpha_j$ satisfies the constraints obtained from Alice's message and extends to a solution of the pinning instance. Hence the final instance is unsatisfiable exactly when $X_j=1$.

\amat{Each data constraint has frequency at most $\max_{a\in\{0,1\}^k}|q(a)|$. Bob includes each distinct pinning constraint only once, so pinning increases any frequency by at most one. Thus all final frequencies are bounded by a constant $F_\Gamma$. For uniform independent $X$ and $j$, the $\Index$ lower bound and the number of sketches in Alice's message give $\Omega_k(n^d)$ bits per sketch. The construction uses $kn+O_k(1)$ variables, so the same bound holds in terms of the total number of variables, as in the proof of \cref{thm:sketch-lower}.}

Finally, the streaming-to-sketching result of Jiang, Liu, and Yu~\cite[Corollary 7.2.2]{JLY26} applies to the universe $\mathcal U_n^\Gamma$, whose size is polynomial in $n$. For a sufficiently large polynomial stream budget, an $S$-bit streaming algorithm therefore yields a sketch of size $O_k(S)$. The sketching lower bound gives $S=\Omega_k(n^d)$.
\end{proof}

\amat{When a language is padded to arity $k$, dropping the dummy coordinates preserves satisfiability and is linear on frequency vectors, so any linear sketch for the original language can be used in the same reduction. Although some frequencies may add, the projected vector $f$ still satisfies $\|f\|_2\leq\|f\|_1=O_k(n^k)$, so the same streaming lower bound follows from Jiang, Liu, and Yu~\cite[Corollary 7.2.2]{JLY26}.}

\adir{Can someone write a few sentences here why exact sparsification takes $\Omega_k (n^d/\log n)$ bits? I do not understand the lower bound enough to write something confidently here.\amati{done}}

\subsection{Lower bound for exact sparsification}
\label{sec:lb-exact-sparsification}

The lower bound in \cref{cor:lb-languages} extends to exact sparsification
even when $\Gamma$ is constant-satisfiable. 
Recall that an \emph{exact sparsifier} of a $\Gamma$-instance $f$ is a set
$\mathcal S\subseteq\supp(f)$ such that $\Sat_\Gamma(\mathcal S)=\Sat_\Gamma(f)$.
Thus it preserves the entire set of satisfying assignments, rather than
only whether the instance is satisfiable.

\amat{We use the same construction with $k$ variable groups. Dropping any dummy coordinates preserves whether each individual assignment satisfies the instance.} The
non-constant-satisfiability assumption was used only to pin Bob's assignment
$\alpha_j$. For exact sparsification, Bob omits the pinning instance and
keeps the frequencies $c_{j,t}(X)$ from \cref{eq:frequencies}. Every active
constraint other than $C_j$ is satisfied by $\alpha_j$, while $C_j$ is
violated by $\alpha_j$ and is active exactly when $X_j=1$. Hence
$\alpha_j$ satisfies this instance if and only if $X_j=0$, and Bob can
recover $X_j$ by testing whether $\alpha_j$ satisfies the returned
sparsifier. The same \Index{} reduction therefore gives an
$\Omega_k(n^d)$ sketching lower bound, where $d=\cdeg(\Gamma)$.
\amat{A constant number of independent runs gives a list in which, with probability at least $0.99$, a strict majority of the candidates are exact sparsifiers. Corollary 7.2.2 of Jiang, Liu, and Yu~\cite{JLY26} applies to arbitrary output tasks, including this guarantee on the list, and gives a sketch of $O_k(S)$ bits from an $S$-bit streaming algorithm. Bob tests $\alpha_j$ on each candidate and takes the majority answer, giving the following corollary.}

\begin{corollary}[Lower bound for exact sparsification]
\label[corollary]{cor:lb-exact-sparsification}
Let $\Gamma$ be a fixed finite nonempty Boolean language of nonempty proper
relations of arity at most $k$, and let $d=\cdeg(\Gamma)$. There is a constant
$c_\Gamma$ such that every randomized one-pass algorithm that, on every
strict-turnstile stream of length at most $n^{c_\Gamma}$ on $n$ variables,
outputs an exact sparsifier of the final instance with probability at least
$2/3$ uses $\Omega_k(n^d)$ bits, for all sufficiently large $n$.
\amatr{i dont think we need a formal proof here, in case needed it’s commented out.}
\end{corollary}

%% file: upperbound.tex
\section{Upper bound}
\label{sec:upper}
In this section, we prove that the integer capturing degree of \amat{$R$} also gives an upper bound on the space required to decide satisfiability \amat{in the strict-turnstile model}. \adi{Throughout this section we drop the requirement that $R$ be non-constant-satisfiable.} %

\begin{theorem}[Main upper bound]
\label[theorem]{thm:upper}
Let $R\subseteq\{0,1\}^k$ be a fixed nonempty proper
Boolean relation of arity $k$, and let \amat{$d=\cdeg(R)$}. Given a bound $\mu\geq1$ on every prefix multiplicity and $0<\delta<1/2$, there is a randomized one-pass strict-turnstile algorithm deciding $\CSP(R)$ on $n$ variables with error at most $\delta$ using
\[
 \amat{O_k}\!\left(n^d\polylog(2n\mu/\delta)\right)
\]
bits, including decoding workspace. \amat{The} algorithm also recovers a set of active constraints with exactly the original satisfying assignments.
\end{theorem}

We first express the constraints as polynomial equations and represent their coefficients as rows of a matrix. \amat{Let $d=\cdeg(R)$ and choose} an integer capturing polynomial of degree $d$. Substituting a constraint's variables into this polynomial gives a polynomial that vanishes precisely when the constraint is satisfied.

Every resulting polynomial uses products of at most $d$ distinct variables, since repeated powers can be removed on Boolean inputs. Thus all these polynomials can be written using one common list of $D=\amat{O_k}(n^d)$ monomials, including the constant monomial. We give the matrix one column for each monomial and one row for each constraint type, with the row containing that polynomial's coefficients. Each row has only constantly many nonzero integer entries because the capturing polynomial is fixed. Given an assignment, evaluating the monomials gives a Boolean vector; taking its inner product with a coefficient row evaluates the corresponding constraint polynomial.

The coefficient matrix is fixed. To represent the current instance, we form another matrix by multiplying each coefficient row by its constraint's frequency. An insertion or deletion changes the corresponding row by adding or subtracting its coefficient vector. On a satisfying assignment, every scaled polynomial evaluates to zero. On any other assignment, the polynomial of at least one active constraint evaluates to a nonzero integer, even after multiplication by its positive integer frequency. Consequently, the sum of the squared evaluations is either zero, when the assignment satisfies the instance, or at least one otherwise. We call this sum the assignment's \emph{energy}. The following lemma expresses this sum as a quadratic form in the monomial vector. The number of monomials will determine our space bound, and the gap between zero and one will let us preserve satisfaction using an approximate matrix.

\begin{lemma}[Polynomial representation]
\label[lemma]{lem:ub-representation}
Let $R$ be a fixed nonempty proper Boolean relation and $d=\cdeg(R)$. For every $n$, there exist an integer $D$ with $n+1\leq D=\amat{O_k}(n^d)$, a map $\phi:\{0,1\}^n\to\{0,1\}^D$ of monomial values, and an integer matrix $B\in\Z^{\mathcal U_n\times D}$. Let $b_C^\top$ be the row indexed by constraint $C$. For a nonnegative integer frequency vector $f$, set
\[
 A_f=\diag(f)B,\qquad
 G_f=A_f^\top A_f=B^\top\diag(f)^2B.
\]
Then every assignment $\alpha\in\{0,1\}^n$ satisfies
\[
 \phi(\alpha)^\top G_f\phi(\alpha)
 \begin{cases}
 =0,&\alpha\in\Sat_R(f),\\
 \geq1,&\alpha\notin\Sat_R(f)
 \end{cases}.
\]
Each row has $\amat{O_k}(1)$ nonzero entries and squared norm $\amat{O_k}(1)$, and can be generated from its scope in polynomial time and $\amat{O_k}(\log(n+2))$ workspace.
\end{lemma}

\begin{proof}
Choose an integer multilinear capturing polynomial $p$ of minimum degree $d$, so $p(a)=0$ precisely when $a\in R$. Let $\phi(\alpha)$ contain all multilinear monomials in the assignment variables of degree at most $d$, including the constant monomial:
\[
 \phi(\alpha)=\left(\prod_{j\in T}\alpha_j\right)_{T\subseteq[n],\ |T|\leq d},
 \qquad D=\sum_{j=0}^{\min(d,n)}\binom nj=\amat{O_k}(n^d).
\]
Every coordinate is zero or one, so $\|\phi(\alpha)\|_2^2\leq D$. Since $d\geq1$, we also have $D\geq n+1$.

For a constraint $C$, substitute its scope into $p$ and use $x_i^r=x_i$ for $r\geq1$ on Boolean inputs. Let $b_C$ be the coefficient vector of the resulting polynomial $p_C$ in the common monomial coordinates. Then $p_C(\alpha)=b_C^\top\phi(\alpha)$. The fixed polynomial $p$ has a bounded number of terms and bounded integer coefficients, so every row has $\amat{O_k}(1)$ nonzero entries and squared norm $\amat{O_k}(1)$. Repeated variables in a scope can only merge a fixed number of terms. To generate a row, enumerate the terms of $p$, identify the variable set in each substituted term, and combine matching sets. This takes polynomial time and $\amat{O_k}(\log(n+2))$ workspace.

For every nonnegative integer frequency vector $f$,
\[
 \phi(\alpha)^\top G_f\phi(\alpha)
 =\left\|A_f\phi(\alpha)\right\|_2^2
 =\sum_{C\in\mathcal U_n}f_C^2p_C(\alpha)^2.
\]
Every summand is a nonnegative integer. Their sum is zero exactly when every active constraint is satisfied. Otherwise, a violated active constraint contributes at least one, proving the lemma.
\end{proof}

\begin{example}[Polynomial representation of $4$-XOR]
\label[example]{ex:ub-four-xor}
Consider odd parity on four variables, which accepts tuples with one or three ones. For $r=z_1+z_2+z_3+z_4$, the polynomial $p(z)=(r-1)(r-3)$ captures this relation. On Boolean inputs it equals
\[
 p(z)=3-3\sum_{a=1}^4z_a+2\sum_{a<b}z_az_b.
\]
The common monomial coordinates are the constant, the $n$ individual variables, and their pairwise products, giving $D=1+n+\binom n2$. A constraint on four distinct variables has coefficient $3$ on the constant coordinate, $-3$ on its four variable coordinates, $2$ on its six pairwise-product coordinates, and zero elsewhere. At tuples with zero, two, or four ones, the polynomial has values $3,-1,3$, respectively. Thus every violated active constraint contributes at least one to the energy.
\end{example}

The matrix $B$ is fixed by the relation and the variable set; only the row frequencies change during the stream. We call $G_f=A_f^\top A_f$ the \emph{Gram matrix} of $A_f$. Our goal is to recover a small set $S$ of active rows, those with positive final frequency. Every original solution satisfies the corresponding constraints, since they belong to the instance. We must ensure that keeping only these constraints introduces no new solutions.

To see what matrix guarantee suffices, let $W=\sum_{C\in S}w_Cb_Cb_C^\top$ for positive weights $w_C$. An assignment $\alpha$ satisfying the retained constraints has zero energy in $W$. If $G_f\preceq W+I/(4D)$, its energy in the original instance is therefore at most $\|\phi(\alpha)\|_2^2/(4D)\leq1/4$, because the monomial vector has $D$ Boolean coordinates. The identity term thus contributes less than the energy of any violating assignment, which is at least one. This inequality guarantees that the retained constraints have exactly the original solutions.

The following corollary obtains this inequality using space proportional to $D$ up to polylogarithmic factors. We derive it from \cref{thm:ub-recovery} in \cref{sec:ub-recovery}. That stronger theorem adapts the linear-sketching methods of Chen, Khanna, and Li (CKL)~\cite[Lemma 6.5 and Section 6.4]{CKL22} and Kapralov, Lee, Musco, Musco, and Sidford (KLMMS)~\cite[Section 7]{KLMMS17}; its proof is in \cref{sec:ub-algorithm}.

\begin{corollary}[Active-row sparsification from \cref{thm:ub-recovery}]
\label[corollary]{cor:ub-active}
Let $B\in\Z^{m\times D}$ be a fixed matrix, where $m,D\geq1$. Each row $b_i^\top$ has at most $s_0$ nonzero entries and squared norm at most $b_0$, for fixed constants $s_0,b_0$, and can be generated from $i$ in polynomial time and polylogarithmic workspace. Given $\mu\geq1$ and $0<\delta<1/2$, a one-pass algorithm on a strict-turnstile stream maintaining integer frequencies $0\leq f_i\leq\mu$ recovers a set $S\subseteq\supp(f)$ of $O_{s_0,b_0}(D\polylog(2mD\mu/\delta))$ indices and positive weights $w_i$ such that, with probability at least $1-\delta$,
\[
 B^\top\diag(f)^2B
 \preceq\sum_{i\in S}w_ib_ib_i^\top+\frac1{4D}I.
\]
It uses $O_{s_0,b_0}(D\polylog(2mD\mu/\delta))$ bits, including recovery workspace and stored output. Recovery takes polynomial time in $m,D,\mu,\delta^{-1}$ and the row-generation time.
\end{corollary}

The polynomial representation and sparsification guarantee together imply the upper bound.
\begin{proof}[Proof of \cref{thm:upper}]

Let $f$ be the final frequency vector. Apply \cref{lem:ub-representation} to obtain $B$ and $\phi$, and let $G=B^\top\diag(f)^2B$. By \cref{cor:ub-active}, with probability at least $1-\delta$ we recover active constraint types $S$ and positive weights satisfying
\[
 G\preceq W+\frac1{4D}I,
 \qquad W=\sum_{C\in S}w_Cb_Cb_C^\top.
\]
Regard $S$ as an unweighted instance. Since its constraints are active (i.e., final frequency is non-zero), every solution of $f$ satisfies $S$. Conversely, if $\alpha$ satisfies $S$, each retained equation has $b_C^\top\phi(\alpha)=0$, and therefore $\phi(\alpha)^\top W\phi(\alpha)=0$. The vector $\phi(\alpha)$ has Boolean coordinates, so $\|\phi(\alpha)\|_2^2\leq D$. The matrix inequality gives
\[
 0\leq\phi(\alpha)^\top G\phi(\alpha)
 \leq\phi(\alpha)^\top W\phi(\alpha)
       +\frac{\|\phi(\alpha)\|_2^2}{4D}
 \leq\frac14.
\]
The integer gap in \cref{lem:ub-representation} forces this energy to be zero: an assignment violating any active constraint would have energy at least one. Hence $\Sat_R(S)=\Sat_R(f)$.

Here $m=|\mathcal U_n|=n^k$ and $D=\amat{O_k}(n^d)$, so recovery uses $\amat{O_k}(n^d\polylog(2n\mu/\delta))$ bits, including its workspace and output. After recovery, discard the weights and enumerate Boolean assignments to decide whether $S$ is satisfiable. The additional $O(n)$ work bits fit within $D\geq n+1$. The equality of solution sets proves the decision guarantee. Recovery takes polynomial time while the final satisfiability computation may take exponential time.
\end{proof}

It remains to prove the active-row sparsification guarantee in \cref{cor:ub-active}. We establish a stronger approximation guarantee for Gram matrices with an added positive identity term.

\subsection{Streaming row sparsification}
\label{sec:ub-recovery}
Our streaming algorithm maintains linear sketches of the multiplicity-scaled row matrix and recovers a small set of active rows from them after the stream. Throughout this subsection, a \emph{sparse row dictionary} is a fixed integer matrix satisfying the sparsity, norm, and row-generation assumptions of \cref{cor:ub-active}. The theorem adds a positive identity term $\lambda I$, called a \emph{regularizer}, to the Gram matrix so that inverse queries are well-defined. This term remains unchanged in the output.

\begin{theorem}[{Regularized dictionary recovery, adapting \cite{CKL22,KLMMS17}}]
\label[theorem]{thm:ub-recovery}
Let $B\in\Z^{m\times D}$ be a sparse row dictionary, and let a strict-turnstile stream maintain integer frequencies $0\leq f_i\leq\mu$. Set $G=B^\top\diag(f)^2B$. Given $0<\varepsilon\leq1/8$, $0<\delta<1/2$, and $\lambda=2^{-L}$ for an integer $L\geq0$, a randomized one-pass algorithm recovers active indices $S$ and positive weights forming $W=\sum_{i\in S}w_ib_ib_i^\top$ such that, with probability at least $1-\delta$,
\begin{equation}
\label{eq:ub-target}
 (1-\varepsilon)(G+\lambda I)
 \preceq W+\lambda I
 \preceq(1+\varepsilon)(G+\lambda I).
\end{equation}
For $Q=2+mD\mu/(\varepsilon\delta\lambda)$, the output has $O_{s_0,b_0}(D\varepsilon^{-2}\polylog Q)$ indices, and the algorithm uses $O_{s_0,b_0}(D\varepsilon^{-4}\polylog Q)$ bits, including stored randomness, output, and working space, for every choice of random seeds. The time per update and the time after the stream are polynomial in $m,D,\mu,\varepsilon^{-1},\delta^{-1},\lambda^{-1}$ and the time needed to obtain a row.
\end{theorem}

We prove this theorem in \cref{sec:ub-algorithm}. First, we derive \cref{cor:ub-active} from its statement.

\begin{proof}[Proof of \cref{cor:ub-active} from \cref{thm:ub-recovery}]
Choose a power of two $\lambda_D$ with $1/(16D)<\lambda_D\leq1/(8D)$ and apply the theorem with $\varepsilon=1/8$. If its row sum is $W_0$, then $G\preceq2(W_0+\lambda_DI)\preceq2W_0+I/(4D)$. Doubling the returned weights gives the required inequality. Since $\lambda_D^{-1}=O(D)$, the space and time bounds become those in the corollary.
\end{proof}

\subsubsection{Technical overview of \Cref{thm:ub-recovery}}
\label{sec:ub-overview}
Our proof essentially adapts the sampling and sketching arguments of CKL~\cite[Section 6.4]{CKL22} and KLMMS~\cite[Sections 4 and 7]{KLMMS17} to sparse integer rows. We first explain the classical offline sampling algorithm with access to the final instance, and then describe how linear sketches let us implement it after a stream. The complete algorithm and analysis appear in \cref{sec:ub-algorithm}.

\paragraph*{The offline algorithm.}
Let us first suppose that we know all final frequencies and can compute with the full matrix $A=\diag(f)B$ and its Gram matrix $G=A^\top A$. This assumption lets us understand which rows to keep before addressing how to find them in a stream. We follow the iterative sampling approach of Koutis~\cite{Koutis14}, as used by Chen, Khanna, and Li~\cite[Sections 6.3--6.4]{CKL22}. If $G=0$, the empty output is exact. Otherwise, the range of $G$ is the span of the active row vectors. Its pseudoinverse $G^\dagger$ inverts $G$ on this subspace and is zero on its orthogonal complement.

We would like to keep rows whose removal could substantially change the matrix. The \emph{leverage score} $\tau_i=f_i^2b_i^\top G^\dagger b_i$ measures precisely this possibility: it is the largest fraction $f_i^2(b_i^\top x)^2/(x^\top Gx)$ of the total energy that row $i$ contributes in any direction $x$ with $x^\top Gx>0$. Every score is at most one because the row's contribution is part of the total energy. Moreover, $\sum_i\tau_i=\tr(G^\dagger G)=\rank G\leq D$.\footnote{For graph incidence rows, this identity is Foster's theorem: the sum of edge weight times effective resistance equals the number of vertices minus the number of connected components.} Thus only a few rows can have large scores.

For graph incidence rows with edge weights $w_i=f_i^2$, the leverage score is $w_i$ times the corresponding effective resistance~\cite{SS11}. A simpler offline algorithm would sample directly according to these scores, as Spielman and Srivastava~\cite{SS11} do for graph sparsification. We follow the iterative approach because it only needs to test whether scores exceed a threshold. As we explain below, coordinate and norm sketches let us perform these tests without knowing the row frequencies.

The algorithm keeps rows with large scores and randomly samples the others. Each repetition is a \emph{sampling level}. Let $a_i^{(t)}$ be the scalar multiplying row $b_i$ at the start of level $t$, initially $a_i^{(0)}=f_i$, and let $J_t=\sum_i(a_i^{(t)})^2b_ib_i^\top$ be the current matrix. Its row scores are $s_i^{(t)}=(a_i^{(t)})^2b_i^\top G^\dagger b_i$. We continue to measure them relative to the original $G$, even as the multipliers change.

We take $T_{\mathrm{off}}=O(\log(2m/\delta))$ levels and a threshold with $\theta_{\mathrm{off}}^{-1}=O(\varepsilon^{-2}T_{\mathrm{off}}^2\log(2DT_{\mathrm{off}}/\delta))$, choosing sufficiently large constant factors in the number of levels and the reciprocal threshold. Starting with all active rows and an empty retained set, we repeat the following for $t=0,\ldots,T_{\mathrm{off}}-1$:
\begin{enumerate}[label=(\roman*)]
\item Retain each not-yet-retained row whose score is at least $\theta_{\mathrm{off}}$, and keep its current multiplier fixed thereafter.
\item Independently keep each remaining row with probability $1/4$. Double its multiplier if it survives and remove it otherwise. Leave retained rows unchanged.
\end{enumerate}
After the final level, we return the retained rows, weighted by their squared multipliers. We declare failure if an unretained row remains.

The sampling rule preserves the matrix in expectation: a sampled row contributes four times as much if it survives, which happens with probability one quarter. To prove correctness, we need three guarantees: (i) the changes over all levels leave the matrix within relative error $\varepsilon$ of $G$, (ii) no unretained row remains at the end, and (iii) at most $O(D/\theta_{\mathrm{off}})$ rows are retained.

For the first guarantee, we sample only rows with small scores. Each sampled row therefore has a small contribution in every direction, allowing matrix Bernstein to bound the change at one level; our threshold makes the errors small enough to add over all levels. For the second, an unretained row can remain only if it survives every sampling step, an event whose probability decreases geometrically. Finally, every retained row has score at least $\theta_{\mathrm{off}}$. Once the final matrix is close to $G$, its total score is $O(D)$, which gives the third guarantee. The following claim combines these arguments.

\begin{claim}[{Classical sampling, adapting \cite[Section 6.4.1]{CKL22}}]
\label[claim]{clm:ub-offline}
With probability at least $1-\delta$, after $T_{\mathrm{off}}$ levels only retained rows remain. There are $O(D/\theta_{\mathrm{off}})$ such rows, and their sum, weighted by the squared retained multipliers, lies between $(1-\varepsilon)G$ and $(1+\varepsilon)G$.
\end{claim}
\begin{proof}
We first bound the change at one level. Fix all decisions preceding that level's sampling, and suppose $J_t\preceq2G$. All matrices in the next calculation act on the range of $G$. For each row to be sampled, define $Q_i=(a_i^{(t)})^2(G^\dagger)^{1/2}b_ib_i^\top(G^\dagger)^{1/2}$. Its operator norm is the row score, so $0\preceq Q_i\preceq\theta_{\mathrm{off}}I$, and $\sum_iQ_i\preceq2I$.

Let $\xi_i$ indicate whether row $i$ survives this level. The normalized change is $\sum_i(4\xi_i-1)Q_i$. Its summands are independent and have mean zero and norm at most $3\theta_{\mathrm{off}}$. Their variance matrix is $3\sum_iQ_i^2\preceq6\theta_{\mathrm{off}}I$, since $Q_i^2\preceq\theta_{\mathrm{off}}Q_i$. Matrix Bernstein~\cite[Theorem 6.1]{Tropp12}, applied to both tails, therefore gives the following conditional bound.
\[
 \Pr\!\left[
   \left\|\sum_i(4\xi_i-1)Q_i\right\|_{\mathrm{op}}
   >\frac{\varepsilon}{T_{\mathrm{off}}}
 \right]
 \leq 2D\exp\!\left(
   -\frac{\varepsilon^2}
          {14T_{\mathrm{off}}^2\theta_{\mathrm{off}}}
 \right)
 \leq\frac{\delta}{2T_{\mathrm{off}}}.
\]
The last inequality follows by choosing the constant in $\theta_{\mathrm{off}}^{-1}$ sufficiently large. Returning to the original coordinates, the change lies between $-(\varepsilon/T_{\mathrm{off}})G$ and $(\varepsilon/T_{\mathrm{off}})G$ outside this failure event. We now prove by induction that $(1-t\varepsilon/T_{\mathrm{off}})G\preceq J_t\preceq(1+t\varepsilon/T_{\mathrm{off}})G$. The base case is $J_0=G$. Assuming the bound at level $t$, we have $J_t\preceq2G$, so the preceding estimate applies and establishes the bound at level $t+1$. A union bound over all levels gives the desired approximation to $G$ with failure probability at most $\delta/2$.

Next, an unretained row still present at the end must have survived all $T_{\mathrm{off}}$ independent sampling decisions. The probability that any such row remains is at most $m4^{-T_{\mathrm{off}}}\leq\delta/2$. Otherwise, the retained rows form the entire final matrix.

Finally, each retained row keeps a score of at least $\theta_{\mathrm{off}}$, since both its multiplier and the reference $G$ remain fixed. On the approximation event, their total score is at most $\tr(G^\dagger J_{T_{\mathrm{off}}})\leq2D$. Hence there are at most $2D/\theta_{\mathrm{off}}$ retained rows. The two failure probabilities sum to at most $\delta$, proving all three guarantees.
\end{proof}
In particular, the output contains $O(D\varepsilon^{-2}\polylog(2mD/\delta))$ rows.

\paragraph*{Linear sketches.}
We now return to the streaming setting. The offline algorithm assumes access to both the frequencies and $G$, but storing all $m$ frequencies or all $D^2$ entries of $G$ can exceed our space bound. These difficulties also arise in KLMMS~\cite[Sections 3 and 7]{KLMMS17} and CKL~\cite[Section 6.4]{CKL22}. We first explain how to find important rows without knowing their frequencies, assuming a suitable approximation to $G$ is available. We then explain how to obtain that approximation.

Our stored information consists of \emph{linear sketches} $Y=\Phi A$, where $\Phi$ is a random matrix with few rows, acting on the $m$ row indices of $A$. A unit update $(i,\sigma)$ changes $Y$ by $\sigma(\Phi e_i)b_i^\top$, where $e_i$ is the $i$th unit vector. We can make this change using the row index and update alone, without knowing $f_i$. After the stream, for any vector $v$ we can compute $Yv=\Phi(Av)$. Thus a small matrix maintained during the stream gives us a sketch of $Av$ even when we choose $v$ afterward. The question is which vectors to query, and which information about them suffices to find important rows.

\paragraph*{Detecting rows from a coarse approximation.}
For the moment, we add a known positive term $\ell I$ to $G$. This modification is called \emph{regularization}: it adds $\ell\|x\|_2^2$ to the energy in every direction $x$, so the resulting matrix $G+\ell I$ is invertible even when $G$ is singular. We call $\ell$ the regularization parameter. We will later vary $\ell$ to obtain the approximation needed for the row tests. First, let us see how the tests work if the exact matrix $H=G+\ell I$ is given.

For a nonzero row $b_i$, define $v_i=H^{-1}b_i$ and $q_i^*=b_i^\top H^{-1}b_i>0$. The vector $y^{(i)}=Av_i$ has $i$th coordinate $f_iq_i^*$, so this coordinate contains the unknown frequency multiplied by a known scalar. Moreover, its total energy, including the added identity term, is $\|y^{(i)}\|_2^2+\ell\|v_i\|_2^2=v_i^\top Hv_i=q_i^*$. Consequently,
\begin{equation}
\label{eq:ub-heavy-identity}
 \frac{|y_i^{(i)}|^2}{\|y^{(i)}\|_2^2+\ell\|v_i\|_2^2}
 =f_i^2b_i^\top(G+\ell I)^{-1}b_i.
\end{equation}
The right-hand side is the leverage score with $(G+\ell I)^{-1}$ in place of $G^\dagger$, called the \emph{regularized leverage score}. The identity tells us how to test whether this score is large: we compare one squared coordinate with the total energy. If the coordinate is large enough to estimate accurately, dividing it by $q_i^*$ also estimates the row's frequency.

This is the reason for choosing an iterative sampling algorithm. At each level it only needs to detect rows above a score threshold and estimate their multipliers; small rows can continue to be sampled. The required estimates come from two standard streaming techniques: CountSketch~\cite{CCF04} and the sketch of Alon, Matias, and Szegedy (AMS)~\cite{AMS99}. A coordinate is \emph{$\ell_2$-heavy} at threshold $\theta$ if its squared value is at least $\theta$ times the squared norm of the vector. CountSketch estimates a specified coordinate with additive error proportional to that norm. For a heavy coordinate, sufficiently accurate estimates therefore have small relative error. AMS estimates the squared norm. Together, the two sketches let us estimate the numerator and denominator of \cref{eq:ub-heavy-identity}; the additional term $\ell\|v_i\|_2^2$ is already known. We examine row indices one at a time, testing their scores and estimating the multipliers of the rows we retain.

We do not need the exact matrix $G+\ell I$ for these tests. Suppose instead that we are given a \emph{coarse approximation} $H$, by which we mean $\frac12H\preceq G+\ell I\preceq2H$. With $v_i=H^{-1}b_i$, the coordinate fraction in \cref{eq:ub-heavy-identity} agrees with $f_i^2b_i^\top H^{-1}b_i$ within constant factors. This is enough to distinguish scores above and below constant multiples of a threshold. The same reasoning applies as sampling changes the multipliers, provided their current Gram matrix plus $\ell I$ stays within constant factors of $H$. Bounding the error at each sampling level, as in the offline argument, establishes this comparison.

Let us now specify the sketches that supply these estimates at successive sampling levels. Before the stream, we choose nested random subsets $E_0\supseteq E_1\supseteq\cdots$ of the row indices, starting with $E_0=[m]$ and keeping each member independently with probability $1/4$ at the next level. Thus a row belongs to $E_t$ with probability $4^{-t}$. For each level, we maintain $Y_t=\Phi_t\diag(f\one_{E_t})B$, where $f\one_{E_t}$ agrees with $f$ on $E_t$ and is zero elsewhere. The matrix $\Phi_t$ consists of a CountSketch and an AMS sketch, each applied to every column of the sampled row matrix. An update to row $i$ affects this pair only if $i\in E_t$, and its contribution is linear in the update. In particular, membership in the sample concerns a row index, not its unknown number of copies.

After the stream, $Y_tv$ supplies both a coordinate estimate and a norm estimate for $\diag(f\one_{E_t})Bv$. A row that survives $t$ sampling steps without being retained has multiplier $2^tf_i$; multiplying the sampled sketches by $2^t$ accounts for this rescaling. The decoder also maintains the small list of rows already retained and their estimated multipliers. We keep independent copies of the sketch pair for successive estimates, so each queried vector is determined independently of the sketch used to estimate it. All these copies are maintained during the same pass. This describes the sampled measurements underlying the algorithm; \cref{sec:ub-algorithm} gives the full decoder and proves that, given a coarse $H$, it returns a weighted row sum $W$ such that $W+\ell I$ approximates $G+\ell I$ to relative error $\varepsilon$.

\paragraph*{Successive approximations through regularization.}
We have reduced the problem to obtaining a coarse approximation $H$. There is an apparent circularity: the row tests need an approximation in order to produce one. The useful distinction is that they need only a constant-factor approximation, but return one with the much smaller relative error $\varepsilon$. Regularization lets us use this improvement repeatedly. We start with a large identity term, where a coarse approximation is immediate, and gradually reduce that term while maintaining an accurate approximation to the new target.

For the starting point, choose a power of two $\lambda_0=O_{b_0}(m\mu^2)$ large enough that $G\preceq\lambda_0I$ and $\lambda_0\geq1$. Such a choice follows from $\|G\|_{\mathrm{op}}\leq\tr G\leq b_0m\mu^2$. Then $\lambda_0I\preceq G+\lambda_0I\preceq2\lambda_0I$, so the known matrix $H_0=\lambda_0I$ is already a coarse approximation. Applying the decoder with this reference produces $W_0+\lambda_0I$ with relative error $\varepsilon$. We have therefore learned enough about $G$ to move to a smaller identity term.

Suppose we have reached some parameter $\ell$ and obtained $W+\ell I$ with relative error $\varepsilon$. We would next like to approximate $G+(\ell/2)I$. A natural reference is $H'=W+(\ell/2)I$: we keep the recovered rows and halve only the identity term. The difference between the reference and its new target is still $W-G$. The target itself has decreased by at most a factor of two, since $G+\ell I\preceq2(G+(\ell/2)I)$. Thus the relative error has increased to at most $2\varepsilon$, which is still a coarse approximation. We can now apply the decoder again, at parameter $\ell/2$, to restore the relative error to $\varepsilon$.

This is the full cycle: an accurate approximation at one parameter gives a coarse approximation at half that parameter, and another decoding step makes it accurate again. We repeat until the parameter reaches the prescribed $\lambda$, taking $O_{b_0}(\log(2m\mu^2/\lambda))$ such steps. The error does not double repeatedly: each new call returns to relative error $\varepsilon$ by querying fresh sketches of the original rows. The previous approximation only determines the vectors used in those queries. We maintain the independent sketch collections for all parameters during the single streaming pass, and decode them in this order afterward.

The final matrix $W+\lambda I$ approximates $G+\lambda I$ as required. In the CSP application, we stop at $\lambda$ of order $1/D$, where the identity term is small enough for the integer-gap argument to preserve all satisfying assignments. This construction follows the regularization argument of KLMMS~\cite[Section 4]{KLMMS17}, combined with CKL's recovery for squared row multipliers~\cite[Lemma 6.5 and Section 6.4]{CKL22}. In \cref{sec:ub-algorithm}, we verify these adaptations for sparse integer rows and account for the space needed by the sketches and computations.

\paragraph*{Removing the regularizer.}
A known condition-number bound also lets us approximate $G$ itself, which is not needed in our main upper bound argument but is an additional result we can obtain. Here the condition number is the ratio of the largest eigenvalue of $G$ to its smallest positive eigenvalue, so it is measured on the range of $G$. The bound lets us choose a regularizer whose contribution on that range can be absorbed into the relative error. Since the output contains only active rows, it also vanishes on the kernel of $G$. We obtain the following consequence, proved in \cref{sec:ub-unregularized}.

\begin{corollary}
\label[corollary]{cor:ub-unregularized}
\unregularizedsparsificationstatement
\end{corollary}

\subsection{Upper bounds for Boolean constraint languages}
\label{sec:ub-languages}
For a constraint language, we choose one capturing polynomial for each relation and express all substituted polynomials in the same monomial coordinates. The largest degree determines the number of coordinates, so the matrix sparsification argument applies without any change.

\begin{corollary}[Upper bound for constraint languages]
\label[corollary]{cor:ub-languages}
Let $\Gamma$ be a fixed finite nonempty Boolean language of nonempty proper relations of arity $k$, and put $d=\cdeg(\Gamma)$. Given a bound $\mu\geq1$ on every prefix multiplicity and $0<\delta<1/2$, a randomized one-pass strict-turnstile algorithm decides $\CSP(\Gamma)$ on $n$ variables with error at most $\delta$ using $\amat{O_k}(n^d\polylog(2n\mu/\delta))$ bits, including decoding workspace. With probability at least $1-\delta$, it also recovers a set $S$ of active constraints with $\Sat_\Gamma(S)=\Sat_\Gamma(f)$, where $f$ is the final instance.
\end{corollary}
\begin{proof}
For each $R\in\Gamma$, fix an integer multilinear capturing polynomial $p_R\in\Z[z_1,\ldots,z_{k}]$ of degree at most $d$. Let $\phi(\alpha)$ contain all multilinear monomials of degree at most $d$ on the $n$ variables. Its dimension is $D=\amat{O_k}(n^d)$, and $\|\phi(\alpha)\|_2^2\leq D$. For a constraint $C$ using relation $R$, substitute its scope into $p_R$ and let $b_C$ be the resulting coefficient vector. As in \cref{lem:ub-representation}, these rows form a matrix $B$ with constant row sparsity and norm, and each row can be generated in polynomial time and logarithmic workspace. The constants can be chosen uniformly over the fixed finite language.

Put $G=B^\top\diag(f)^2B$. The energy $\phi(\alpha)^\top G\phi(\alpha)=\sum_C f_C^2(b_C^\top\phi(\alpha))^2$ is zero exactly on satisfying assignments and at least one otherwise. By \cref{cor:ub-active}, with probability at least $1-\delta$ we recover active rows with positive weights and weighted sum $W$ satisfying $G\preceq W+I/(4D)$. Any assignment satisfying the retained constraints has zero energy in $W$, and therefore energy at most $1/4$ in $G$. The integer gap forces it to satisfy the original instance. The converse holds because every retained constraint is active.

The number of rows is $|\mathcal U_n^\Gamma|=\sum_{R\in\Gamma}n^{k}\leq|\Gamma|n^k=n^{\amat{O_k}(1)}$, so the sparsification uses the stated space, including the recovered constraints and working memory. Enumerating assignments to decide satisfiability adds only $O(n)$ bits, which fit within this bound since $d\geq1$.
\end{proof}

Together with \cref{cor:lb-languages}, this gives matching bounds up to polylogarithmic factors for non-constant-satisfiable languages when algorithms must handle every stream up to a sufficiently large polynomial length. For an arbitrary fixed finite Boolean language, we first handle empty and full relations as in \cref{sec:instances}. The remaining language either admits a common constant satisfying assignment, making its decision problem trivial, or is covered by these two corollaries.

%% file: larger_domains.tex
\section{Extension to Larger Domains}\label{sec:larger-domains}

\amat{We extend the upper bound to relations over arbitrary fixed finite domains by first restricting to a core. We then show that the Boolean degree identity used in the lower bound can fail on a three-element domain, so the same argument does not give a matching lower bound.}

\subsection{Upper bounds over finite domains}
\label{sec:ub-finite-domain}
\amat{Let $R\subseteq D^k$ be a nonempty relation over the fixed finite domain $D=\{0,\ldots,q-1\}$.} The first step is to replace $R$ by its core. Recall from \cref{sec:cores} that a relation is a core if every map of its domain preserving the relation is an automorphism. Brakensiek, Putterman, Sharma, and Velusamy~\cite[Lemma 4.2]{BPSV26} show that deciding satisfiability over a relation and over its core have the same streaming space complexity. Their reduction replaces the relation on each constraint's scope, and therefore also preserves multiplicities and strict-turnstile updates. It thus suffices to give an algorithm for the core.

To specify this reduction on $\amat{D}$, choose a map $h:\amat{D}\to\amat{D}$ with $h(R)\subseteq R$ and smallest image $C$, where $h$ acts coordinatewise on tuples. Its restriction to $C$ is a permutation: otherwise $h^2$ would have a smaller image. A suitable positive power $e$ of $h$ therefore fixes $C$ pointwise, and $e(R)=R\cap C^k$. This restriction is a core. Indeed, a nonbijective map of $C$ preserving it would compose with $e$ to give a map preserving $R$ with smaller image. Fix this choice and write $R^\star=\Core(R)=R\cap C^k$, considered as a relation on $C$.

\amat{The core domain $C$ retains its labels from $D$, and all degrees below refer to these inherited labels.} When $|C|\geq2$, let \amat{$d=\cdeg(R^\star)$} be the minimum total degree of an integer polynomial whose zeros on $C^k$ are exactly $R^\star$. \amat{Such a polynomial exists by interpolating the function that is zero on $R^\star$ and one elsewhere on $C^k$, then clearing its rational denominators. Interpolation also gives $d\leq k(|C|-1)\leq k(q-1)$. By \cref{obs:coefficient-fields}, allowing rational or real coefficients gives the same minimum degree.} If $|C|=1$, set $\cdeg(R^\star)=0$, as in the Boolean case. The degree below refers to the chosen core with these integer labels. Restricting a capturing polynomial for $R$ to $C^k$ captures $R^\star$, so passing to the core cannot increase the degree.

After restricting to $C$, we apply the polynomial representation from \cref{lem:ub-representation}. The degree again bounds the number of monomials, and integer coefficients give the same gap between satisfaction and violation. The monomial values are no longer Boolean, but they remain bounded by a constant because the domain and polynomial are fixed. Choosing a smaller regularizer accounts for this change and gives the following bound.

\begin{corollary}[Finite-domain upper bound]
\label[corollary]{cor:ub-finite-domain}
Let $R\subseteq \amat{D}^k$ be a fixed nonempty relation over a fixed finite domain \amat{of size $q$}, with the core and integer labels chosen above, and put $d=\cdeg(\Core(R))$. Given $\mu\geq1$ and $0<\delta<1/2$, a randomized one-pass strict-turnstile algorithm decides $\CSP(R)$ on $n$ variables with error at most $\delta$ using
\[
 \amat{O_{k,q}}\!\left(n^d\polylog(2n\mu/\delta)\right)
\]
bits, including its working space, when every prefix multiplicity is at most $\mu$. It can also recover active constraints preserving exactly the satisfying assignments with values in the core domain $C$. \amat{The constants depend only on $k$ and the original domain size $q$.} If $d=0$, the decision problem uses constant space.
\end{corollary}

\begin{proof}
We first justify restricting the assignments to $C$. If an assignment satisfies an $R$-instance, composing it with $e$ gives a satisfying assignment over $R^\star$. Conversely, every $R^\star$-solution is an $R$-solution. Replacing $R$ by $R^\star$ on each scope therefore preserves satisfiability, variables, multiplicities, and the entire update sequence. This is the same argument as in \cref{lem:core-reduction}, and it applies to the finite domain $\amat{D}$ without change. If $C=\{a\}$, nonemptiness gives $R^\star=\{a^k\}$, so the constant assignment satisfies every instance. We may always report satisfiable and return the empty set of constraints.

Assume henceforth that $|C|\geq2$. The nonempty core $R^\star$ is proper, since a full relation would admit a constant map, so $d\geq1$. Choose an integer polynomial $p$ of degree $d$ capturing $R^\star$. For an assignment $\alpha\in C^n$, let $\phi(\alpha)$ be the vector of all monomials of total degree at most $d$, including the constant monomial. We include repeated powers, which cannot in general be removed on $C$. There are $\amat{N}=\binom{n+d}{d}=\amat{O_{k,q}}(n^d)$ such monomials. Put $M=\max(\{1\}\cup\{|a|:a\in C\})$ and $K=M^{2d}$. \amat{Since $M\leq\max\{1,q-1\}$ and $d\leq k(q-1)$, the constant $K$ is bounded in terms of $k,q$ alone.} Each squared monomial value is at most $K$, so $\|\phi(\alpha)\|_2^2\leq K\amat{N}$.

Substituting a constraint's scope into $p$ gives an integer coefficient row $b_i$ on these monomials. The polynomial is fixed, so every row has constantly many nonzero entries and bounded squared norm. Repeated variables in a scope may add exponents or combine terms, but do not increase total degree or the number of terms. A row can be generated from its scope in polynomial time and $\amat{O_{k,q}}(\log(n+2))$ space. Thus the resulting matrix $B\in\Z^{m\times \amat{N}}$, where $m=n^k$, meets the assumptions of \cref{thm:ub-recovery}.

For the final frequency vector $f$, set $G=B^\top\diag(f)^2B$. As in the Boolean proof, $\phi(\alpha)^\top G\phi(\alpha)=\sum_i f_i^2p_i(\alpha)^2$, where $p_i$ is the substituted polynomial for constraint $i$. All evaluations are integers because the domain labels and polynomial coefficients are integers. The energy is therefore zero on a satisfying assignment and at least one on any other assignment in $C^n$.

We now choose the regularizer small enough for the bound $\|\phi(\alpha)\|_2^2\leq K\amat{N}$. Apply \cref{thm:ub-recovery} with $\varepsilon=1/8$ and a power of two $\lambda$ satisfying $1/(16K\amat{N})<\lambda\leq1/(8K\amat{N})$. On its success event, let $W_0$ be the returned weighted row sum. The lower approximation bound gives $G\preceq2(W_0+\lambda I)$. Doubling the returned weights gives $W=2W_0$ and hence $G\preceq W+I/(4K\amat{N})$. If $\alpha\in C^n$ satisfies all retained constraints, its energy in $W$ is zero, and
\[
 0\leq\phi(\alpha)^\top G\phi(\alpha)
 \leq\frac{\|\phi(\alpha)\|_2^2}{4K\amat{N}}
 \leq\frac14.
\]
The integer gap forces this energy to be zero. Conversely, every original solution in $C^n$ satisfies the retained constraints because they are active. Thus the retained constraints preserve exactly the solutions on the core domain, and deciding whether they have such a solution decides the original instance over $\amat{D}$.

Finally, $K$ is constant and $\lambda^{-1}=\amat{O_{k,q}}(\amat{N})$. Substituting $m=n^k$ and $\amat{N}=\amat{O_{k,q}}(n^d)$ into \cref{thm:ub-recovery} gives the stated space bound, including stored output and working space. After discarding the weights, we can enumerate assignments in $C^n$ and test the retained constraints. This needs $O(n\log|C|)=\amat{O_{k,q}}(n)$ additional bits, which fit within the bound since $d\geq1$. The failure probability is at most $\delta$.
\end{proof}

For the empty relation over any domain, satisfiability is equivalent to having no active constraints. Maintaining the total number of constraint copies gives an $\amat{O_{k,q}}(\log(n\mu+2))$-bit algorithm, as in the Boolean case.

\subsection{Limitations of the lower-bound argument}
\label{sec:ternary-counterexample}
\amat{The core reduction above also applies to the lower-bound question, so we may assume that $R$ is already a core.}
\amat{The preceding corollary gives an $O_{k,q}(n^d\polylog(2n\mu/\delta))$-bit upper bound, where $d=\cdeg(R)$, $\mu$ bounds prefix multiplicities, and $\delta$ is the failure probability. The degree is taken with respect to the fixed integer labels for the domain. To obtain a matching lower bound by our Boolean argument, we would need a selection polynomial of degree $k-d$ that is nonzero at one unsatisfying tuple and zero at every other unsatisfying tuple. The following example shows that such a polynomial need not exist over a larger domain.}

Consider
\[
 R=\{(x,y,z)\in\{0,1,2\}^3:x+y+z\equiv1\pmod3\}.
\]

\begin{observation}
\label[observation]{obs:ternary-counterexample}
The relation $R$ is a core with capturing degree two. Every selection polynomial has degree at least four, and degree four is attained for the forbidden tuple $000$. In particular, $d_R(000)+s_R(000)=6$, rather than the arity three.
\end{observation}
\begin{proof}
Let's first verify that $R$ is already a core. If $h:\{0,1,2\}\to\{0,1,2\}$ preserves $R$, the tuples $001,022\in R$ imply $2h(0)+h(1)\equiv1$ and $h(0)+2h(2)\equiv1\pmod3$. Thus $h(a)\equiv h(0)+a\pmod3$, so $h$ is a translation and hence an automorphism. Therefore $R$ is a core.

Next, let's compute the capturing degree of $R$. Let $S=x+y+z$. The polynomial $p=(S-1)(S-4)$ captures $R$. Conversely, an affine linear polynomial vanishing at $100,010,001$ must be a multiple of $S-1$; vanishing also at $022$ forces it to be zero. Hence $\cdeg(R)=2$, and indeed $d_R(b)=2$ for every unsatisfying tuple $b$.

Let $q$ select an unsatisfying tuple $b$. Then $pq$ is nonzero only at $b$ on the grid $\{0,1,2\}^3$. The unique polynomial with individual degrees at most two representing this function is a nonzero multiple of
\[
 \prod_{i=1}^{3}\ \prod_{\substack{a\in\{0,1,2\}\\a\ne b_i}}
 \frac{x_i-a}{b_i-a},
 \qquad (x_1,x_2,x_3)=(x,y,z),
\]
which has total degree six. Reducing $pq$ using $x_i^3=3x_i^2-2x_i$ preserves its values on the grid and cannot increase total degree. Consequently, $6\leq2+\deg q$, giving $\deg q\geq4$.

Finally, $q_0=(S-2)(S-3)(S-5)(S-6)$ selects $000$: it is nonzero there and vanishes at every other unsatisfying tuple, whose sum is two, three, five, or six. Thus $s_R(000)=4$.
\end{proof}

For this relation, the finite-domain upper bound gives $\amat{O}(n^2\polylog(2n\mu/\delta))$ bits. However, extending our Boolean lower-bound argument directly would require a selection polynomial of degree $k-d=1$, whereas every selection polynomial has degree at least four. Since $R$ is already a core, reducing to the core does not remove this obstruction. The example identifies a limitation of our current proof techniques; it does not rule out a matching lower bound by a different argument.

%% file: upperbound-proof.tex
\section{Streaming algorithm and analysis}
\label{sec:ub-algorithm}

The algorithm and proof in this section essentially follow CKL~\cite[Section 6.4]{CKL22} and KLMMS~\cite[Sections 4 and 7]{KLMMS17}, with slight adaptations to the present integer rows of constant sparsity and bounded squared norm. We include the details to verify correctness in this setting and to account for the space needed to store randomness and perform the computations. 

We give the complete algorithm in \cref{sec:ub-construction} and prove its guarantees in \cref{sec:ub-correctness}, completing the proof of \cref{thm:ub-recovery}. Throughout, $B\in\Z^{m\times D}$ is the given matrix, with at most $s_0$ nonzero entries and squared norm at most $b_0$ in each row. The constants $s_0,b_0$ are fixed. We have access to each row in polynomial time and polylogarithmic workspace. The stream maintains integer frequencies $0\leq f_i\leq\mu$. Let $G=B^\top\diag(f)^2B$ be the matrix at its end. Without loss of generality, we ignore zero rows but keep the original row indices (if all rows are zero, we return the empty list).

\subsection{Algorithm}
\label{sec:ub-construction}
\Cref{alg:ub-main} gives the complete algorithm. The next two subsections describe its sketching and decoding subroutines, respectively. Its inputs are $B$, the stream, $\mu\geq1$, $0<\varepsilon\leq1/8$, $0<\delta<1/2$, and $\lambda=2^{-L}$ for an integer $L\geq0$. We specify all remaining parameters from these inputs in \cref{par:ub-parameters}. In particular, $K$ is the number of regularization levels and $\lambda_j=2^{-j}\lambda_0$, with $\lambda_{K-1}=\lambda$. 

We represent each weighted row sum by a list of indices and positive weights. A multiplier $a_i$ scales row $b_i^\top$, so its contribution to the Gram matrix is $a_i^2b_ib_i^\top$.

\begin{algorithm}[H]
\caption{Streaming approximation of $G+\lambda I$}
\label{alg:ub-main}
\begin{algorithmic}[1]
\Require $B$, the stream, $\mu,\varepsilon,\delta,\lambda$.
\Ensure A weighted row list representing $W$, or failure.
\State Compute the parameters in \cref{par:ub-parameters}.
\State Maintain the sketches in \cref{alg:ub-encode} during the stream.\alglabel{line:ub-main-sketch}
\State After the stream, set $W\gets0$.
\For{$j=0,\ldots,K-1$}
 \State Set $H\gets W+\lambda_jI$.
 \State Apply \cref{alg:ub-decode} to $H$, $\ell=\lambda_j$, and the sketches indexed by $j$.\alglabel{line:ub-main-decode}
 \State Return failure if the call fails; otherwise set $W$ to its output.
\EndFor
\State Return the list representing $W$.
\end{algorithmic}
\end{algorithm}

During the stream, \cref{alg:ub-main} maintains a separate collection of sketches for each regularization level. Once the stream ends, it starts with an empty row list and calls the decoder for $\lambda_0,\lambda_0/2,\ldots,\lambda$. At each call, it forms $H$ by adding the current identity term to the preceding weighted row sum, then replaces that sum with the decoder's output. All calls operate on the stored sketches; none requires another pass over the stream.

Intuitively, the decreasing regularizers make it possible to supply the decoder with the approximation it needs. Initially, $H=\lambda_0I$ is within a factor of two of $G+\lambda_0I$, since $G\preceq\lambda_0I$. The decoder improves this to relative error $\varepsilon$. Halving the identity term increases that error by at most a factor of two, so the resulting matrix is accurate enough for the next call. Repeating this argument gives the final approximation to $G+\lambda I$; we prove the induction in \cref{sec:ub-correctness}. There are $K=1+\log_2(\lambda_0/\lambda)=O_{b_0}(\log(2m\mu^2/\lambda))$ calls, and only two consecutive row lists need to coexist. This construction follows Miller and Peng~\cite[Section 4]{MP12}, as developed by KLMMS~\cite[Theorem 2 and Section 4]{KLMMS17}. The next two subsections describe the sketches and the decoder.

\subsubsection{Sketching}
\label{sec:ub-sketch-algorithm}
During the stream, we maintain the sampled linear sketches in \cref{alg:ub-encode}. Let $T$ be the number of sampling levels, $h$ control the independence of the sampling choices, $\eta$ be a threshold for testing rows, $\varepsilon_{\mathrm a}$ be the desired relative accuracy of multiplier estimates, and $\beta_s$ be the failure probability allowed for one scan.

To indicate their sizes, we define $Q=2+mD\mu/(\varepsilon\delta\lambda)$. Roughly, we take $T,h=O_{b_0}(\log Q)$, with $\varepsilon_{\mathrm a}^{-1}=O_{b_0}(\varepsilon^{-1}\log Q)$ and $\eta^{-1}=O_{b_0}(\varepsilon^{-2}\log^3Q)$. The failure budget satisfies $\beta_s^{-1}=O_{b_0}(\delta^{-1}\log^3Q)$. The exact choices are in \cref{par:ub-parameters}.

We index the sketches by the regularization level $j$, sampling level $t$, and copy $u$ used for successive decoding scans. For each triple, the map $\Phi_{j,t,u}$ combines a CountSketch~\cite{CCF04} and an AMS sketch~\cite{AMS99}, applied to every column of the sampled row matrix. CountSketch estimates a specified coordinate of a vector $y$ to additive error at most $(\varepsilon_{\mathrm a}\sqrt\eta/100)\|y\|_2$, and AMS estimates its squared norm to relative error at most $1/10$. The probability that any of the estimates in one scan of at most $m$ queries fails is at most $\beta_s$.

For each $j,t$, a \emph{sampling mask} is a random function $\xi_{j,t}:[m]\to\{0,1\}$: the value one keeps an index in the next sample, and zero removes it. Each $\xi_{j,t}(i)$ is Bernoulli with mean $1/4$. Within a mask, these variables are $h$-wise independent, meaning that every set of at most $h$ entries has the joint distribution of independent variables. Different masks are independent. All sketch pairs are mutually independent and independent of the masks. We store seeds specifying these functions and evaluate them when needed. Here $\mathbf e_i$ denotes the $i$th unit vector, and $f\one_E$ agrees with $f$ on $E$ and is zero elsewhere.

\begin{algorithm}[H]
\caption{Maintaining the linear sketches}
\label{alg:ub-encode}
\begin{algorithmic}[1]
\Require $B$ and the stream.
\Statex \textbf{Parameters:} $K,T,h,\eta,\varepsilon_{\mathrm a},\beta_s$.
\Ensure Stored sketches and their seeds.
\For{$j=0,\ldots,K-1$}
 \State Set $E_{j,0}\gets[m]$.
 \For{$t=0,\ldots,T-1$}
  \State Choose $\xi_{j,t}$ and define $E_{j,t+1}=\{i\in E_{j,t}:\xi_{j,t}(i)=1\}$.
  \State Choose $\Phi_{j,t,0},\ldots,\Phi_{j,t,T}$ and initialize every $Y_{j,t,u}\gets0$.
 \EndFor
\EndFor
\For{each update $(i,\sigma)$ with $b_i\ne0$}
 \For{each $j,t,u$ with $i\in E_{j,t}$}
  \State $Y_{j,t,u}\gets Y_{j,t,u}+\sigma(\Phi_{j,t,u}\mathbf e_i)b_i^\top$.
 \EndFor
\EndFor
\State Return the stored sketches and seeds.
\end{algorithmic}
\end{algorithm}

\Cref{alg:ub-encode} first chooses the nested samples and the independent sketch pairs. Starting with all row indices in $E_{j,0}$, each sampling mask keeps an index with probability $1/4$, so an index belongs to $E_{j,t}$ with probability $4^{-t}$. The algorithm initializes one pair $Y_{j,t,u}$ for each $j,t,u$. For an update $(i,\sigma)$, it adds $\sigma(\Phi_{j,t,u}\mathbf e_i)b_i^\top$ to every pair whose sample contains $i$. This keeps $Y_{j,t,u}=\Phi_{j,t,u}\diag(f\one_{E_{j,t}})B$ at every stream prefix.

These updates require the row and the change in its frequency, but not the frequency itself. Sampling keeps an entire row type, including all its current copies. After the stream, multiplying a stored pair by $v$ gives coordinate and norm estimates for $\diag(f\one_{E_{j,t}})Bv$. We keep several independent copies because the decoder will change row multipliers using answers to earlier queries: a later copy then provides fresh estimates for the modified matrix. The decoder below specifies the role of each copy. This construction follows CKL~\cite[Figure 6]{CKL22}.

\subsubsection{Decoding at one regularizer}
\label{sec:ub-decode-algorithm}
At a fixed regularizer $\ell$, we are given a matrix $H$ satisfying $\frac12H\preceq G+\ell I\preceq2H$. From the stored sketches, we seek a small weighted set of active rows whose sum $W$ satisfies $(1-\varepsilon)(G+\ell I)\preceq W+\ell I\preceq(1+\varepsilon)(G+\ell I)$. \Cref{alg:ub-decode} obtains this approximation by saving rows as the sampled set shrinks; its probability of success is proved in \cref{lem:ub-refinement}.

Let $\theta$ be the threshold for saving new rows and $M_{\max}$ the maximum number of saved rows. We take $\theta=1024\eta$, so $\theta^{-1}=O_{b_0}(\varepsilon^{-2}\log^3Q)$, and $M_{\max}=O_{b_0}(D\varepsilon^{-2}\log^3Q)$. The remaining parameters control rounding, vector computation, and limits on stored numbers; we introduce them with the corresponding subroutines. All exact values are fixed in \cref{par:ub-parameters}.

We keep the matrix $H$ fixed during one decoder call and omit the index $j$ from its sketches and sampled sets. For each saved row, the list $F$ contains its index $i$, the level $c_i$ at which it was saved, and an estimated multiplier $\widehat a_i$.

\begin{algorithm}[H]
\caption{Decoding at regularizer $\ell$}
\label{alg:ub-decode}
\begin{algorithmic}[1]
\Require The sketches for one regularizer, $H$, $\ell$, and $B$.
\Statex \textbf{Parameters:} $T,\eta,\theta,g_a,U,g_v,r_{\mathrm{inv}},M_{\max},W_{\max},C_{\max}$.
\Ensure A weighted row list, or failure.
\State Set $F\gets\varnothing$.
\For{$t=0,\ldots,T-1$}
 \State Obtain $Z_0,\ldots,Z_T$ and the integers $d_i$ from \cref{alg:ub-prepare}.\alglabel{line:ub-decode-prepare}
 \State Apply \cref{alg:ub-correct} at level $t$ to these sketches and integers.\alglabel{line:ub-decode-correct}
 \For{each nonzero row $i$ outside $F$}\alglabel{line:ub-decode-select-start}
  \State Test $i$ with $Z_T$ and threshold $\theta$ by \cref{alg:ub-test}.\alglabel{line:ub-decode-test}
  \If{the test returns an estimate $\widehat a_i$}
   \State Return failure if $|F|=M_{\max}$.
   \State Add $(i,c_i=t,\widehat a_i)$ to $F$.\alglabel{line:ub-decode-add}
  \EndIf
 \EndFor\alglabel{line:ub-decode-select-end}
 \State After the queries, consult $\xi_t$ to determine the next sampled set.\alglabel{line:ub-decode-sample}
\EndFor
\State Return failure if $E_T\ne\varnothing$.
\State Return $F$ with weights $\widehat a_i^2$, representing $W=\sum_{i\in F}\widehat a_i^2b_ib_i^\top$.
\end{algorithmic}
\end{algorithm}

At each sampling level, \cref{alg:ub-decode} first prepares working copies $Z_0,\ldots,Z_T$ of that level's sketches (\cref{line:ub-decode-prepare}). It then adjusts the contributions of rows already in the saved list $F$ (\cref{line:ub-decode-correct}). These two subroutines are described next. Once they finish, the decoder tests every row outside $F$ using $Z_T$ and saves those for which the test returns a multiplier estimate (\crefrange{line:ub-decode-select-start}{line:ub-decode-select-end}). Only after these tests does it use the next sampling mask (\cref{line:ub-decode-sample}). At the end, it checks that the sample is empty and returns the saved rows, each weighted by the square of its estimated multiplier.

To understand the working sketches, consider an unsaved row that is still in the sample at level $t$. Its multiplier is $2^tf_i$. If it survives the next sampling step, the multiplier doubles; otherwise, its contribution disappears. Since survival has probability $1/4$, the squared multiplier is preserved in expectation: $\frac14(2a)^2=a^2$. A row saved at level $c_i$, however, should keep the multiplier $2^{c_i}f_i$ from then onward. The preparation and adjustment subroutines make the sketches represent both kinds of rows together.

There are $T+1$ sketch pairs for each regularization and sampling level. The adjustment subroutine queries $Z_0,\ldots,Z_{T-1}$ in succession, modifying only copies that have not yet been queried. The final pair $Z_T$ is reserved for testing new rows. These independent copies are what allow earlier estimates to determine changes to a matrix while later queries still have the usual sketch guarantees.

Saving rows in \cref{line:ub-decode-add} before revealing the next mask makes the choice of rows independent of that sampling step. The proof will show that every row left unsaved has a small contribution, so sampling those rows introduces only a small matrix error. We choose $T$ large enough that the final sample is empty with high probability; when it is empty, the saved list contains all remaining contributions. Saving and adjusting rows follows CKL~\cite[Figure 6 and Lemma 6.5]{CKL22}, with graph rows replaced by the rows of $B$. The identity term $\ell I$ remains unchanged throughout.

Every subroutine returns failure if one of its calls fails. We also check the specified bounds on stored numbers after each modification. These limits ensure that the space bound holds even when a random estimate is inaccurate. We prove in \cref{sec:ub-implementation} that the checks do not interrupt the computation while the estimates and sampling errors satisfy the bounds used in the proof.

\paragraph{Preparing the working sketches}
At the beginning of level $t$, we scale the sampled rows to their current multipliers and restore saved rows that have left the sample. \Cref{alg:ub-prepare} makes these changes in working copies of the stored sketches. For a saved row still in the sample, let $d_i$ be the number of halvings needed to restore the multiplier it had when saved. The parameter $C_{\max}=Q^{O_{b_0}(1)}$ bounds the magnitude of each sketch entry; its exact value is fixed in \cref{par:ub-parameters}.

\begin{algorithm}[H]
\caption{Preparing the sketches at level $t$}
\label{alg:ub-prepare}
\begin{algorithmic}[1]
\Require $t$, the level-$t$ sketches and seeds, $F$, and $B$.
\Statex \textbf{Parameters:} $T,C_{\max}$.
\Ensure $Z_0,\ldots,Z_T$ and $d_i$ for $i\in F$, or failure.
\State Set $Z_u\gets2^tY_{t,u}$ for $u=0,\ldots,T$.\alglabel{line:ub-prepare-scale}
\For{$i\in F$}
 \If{$i\notin E_t$}
  \State Add $\widehat a_i(\Phi_{t,u}\mathbf e_i)b_i^\top$ to every $Z_u$ and set $d_i\gets0$.\alglabel{line:ub-prepare-restore}
 \Else
  \State Set $d_i\gets t-c_i$.
 \EndIf
\EndFor
\State Return failure if any modified sketch entry exceeds $C_{\max}$ in magnitude.
\State Return the working sketches and the integers $d_i$.
\end{algorithmic}
\end{algorithm}

\Cref{alg:ub-prepare} multiplies every level-$t$ sketch by $2^t$, giving each sampled row its current multiplier $2^tf_i$ (\cref{line:ub-prepare-scale}). It then goes through the saved list. If a saved row has left the sample, the algorithm adds its saved estimate $\widehat a_i$ to every working sketch and sets $d_i=0$ (\cref{line:ub-prepare-restore}). If the row is still sampled, it already occurs with multiplier $2^tf_i$, whereas its intended multiplier is $2^{c_i}f_i$. The algorithm leaves this contribution in place and sets $d_i=t-c_i$ for the next subroutine.

After preparation, every saved row is represented, including those absent from the sample. The counts $d_i$ specify the halvings still needed for saved rows that remain sampled. Each sampling level starts from copies of its original stored sketches, so changes made at earlier levels are not carried forward. The proof uses this to control the accumulation of estimation errors. The additions of saved rows follow CKL~\cite[Figure 6]{CKL22}.

\paragraph{Correcting selected rows}
We next restore the multipliers of saved rows that remain in the sample. \Cref{alg:ub-correct} performs the halvings specified by the counts $d_i$. Each scan queries one sketch copy and modifies only later copies.

\begin{algorithm}[H]
\caption{Correcting the multipliers of selected rows}
\label{alg:ub-correct}
\begin{algorithmic}[1]
\Require $t$, $Z_0,\ldots,Z_T$, their seeds, $F$, the integers $d_i$, $H$, $\ell$, and $B$.
\Statex \textbf{Parameters:} $T,\eta,g_a,U,g_v,r_{\mathrm{inv}},W_{\max},C_{\max}$.
\Ensure Modified working sketches, or failure.
\For{$u=0,\ldots,T-1$}
 \For{each $i\in F$ with $d_i>0$}
  \State Test $i$ with $Z_u$ and threshold $\eta$ by \cref{alg:ub-test}.\alglabel{line:ub-correct-test}
  \If{the test returns $\widehat z_i$}
   \State Subtract $(\widehat z_i/2)(\Phi_{t,v}\mathbf e_i)b_i^\top$ from every $Z_v$ with $v>u$.\alglabel{line:ub-correct-subtract}
   \State Set $d_i\gets d_i-1$.\alglabel{line:ub-correct-decrement}
  \EndIf
 \EndFor
 \State Return failure if any modified sketch entry exceeds $C_{\max}$ in magnitude.
\EndFor
\State Return failure if some $d_i>0$; otherwise return the modified sketches.
\end{algorithmic}
\end{algorithm}

\Cref{alg:ub-correct} makes $T$ scans through the saved rows. In scan $u$, it queries $Z_u$ for every row with $d_i>0$ (\cref{line:ub-correct-test}). If the row test returns an estimate $\widehat z_i$ of its current multiplier, the algorithm subtracts half that estimate from every later copy $Z_v$, $v>u$, and decreases $d_i$ by one (\crefrange{line:ub-correct-subtract}{line:ub-correct-decrement}). Thus a multiplier $z_i$ becomes $z_i-\widehat z_i/2$, which is close to $z_i/2$. The queried copy is unchanged during the scan. The subroutine succeeds only if every count reaches zero.

Estimating the current multiplier at each halving keeps the error small relative to the contribution being adjusted. Subtracting the entire excess using only the old saved estimate could instead magnify that estimate's error by the large factor $2^{t-c_i}$. The smaller threshold $\eta$, compared with the threshold $\theta$ for saving new rows, lets us estimate rows whose contributions still need adjustment.

A test need not detect every row with $d_i>0$ in every scan. The proof shows that it detects all rows with the largest remaining count. That count therefore decreases after each scan, and $T$ scans suffice to complete the halvings. Using a new copy for each scan is also essential: its represented matrix is determined by answers from earlier copies, independently of its own seed. This lets us apply the coordinate and norm guarantees even though the preceding changes depend on earlier estimates. The repeated estimation and halving follow CKL~\cite[Figure 6 and Claims 6.20--6.21]{CKL22}.

\paragraph{Testing a row}
Correction and selection both require a test that detects a sufficiently large row contribution and estimates its multiplier. \Cref{alg:ub-test} performs this test, with a different threshold for the two uses. Let $\vartheta\in\{\eta,\theta\}$ be the threshold passed to this call, $g_a$ the rounding spacing for multiplier estimates, and $W_{\max}$ their maximum allowed value. Their sizes satisfy $g_a^{-1}=O_{b_0}(\varepsilon^{-1}\log Q)$ and $W_{\max}=Q^{O_{b_0}(1)}$. The parameters $U,g_v,r_{\mathrm{inv}}$ belong to the vector computation in \cref{alg:ub-solve}. All are fixed in \cref{par:ub-parameters}. For a positive power of two $g$, define $\operatorname{round}_g(x)$ as a nearest multiple of $g$, with fixed tie breaking; for vectors, round each coordinate separately.

\begin{algorithm}[H]
\caption{Testing a row and estimating its multiplier}
\label{alg:ub-test}
\begin{algorithmic}[1]
\Require A nonzero row index $i$, a working sketch pair $Z$ with its seed, $H$, and $\ell$.
\Statex \textbf{Parameters:} $\vartheta,g_a,U,g_v,r_{\mathrm{inv}},W_{\max}$.
\Ensure Rejection, an estimated multiplier, or failure.
\State Compute $v_i$ by \cref{alg:ub-solve} and set $q_i\gets b_i^\top v_i$.\alglabel{line:ub-test-solve}
\State Return failure if $q_i\leq0$.
\State From $Zv_i$, obtain a CountSketch estimate $\widehat y_i$ of coordinate $i$ and an AMS estimate $\widehat\nu_i$ of the squared norm.
\State Set $\widehat t_i^2\gets\widehat\nu_i+\ell\|v_i\|_2^2$.
\State Reject unless $\widehat y_i>0$ and $\widehat y_i^2\geq2\vartheta\widehat t_i^2$.\alglabel{line:ub-test-threshold}
\State Set $\widehat z_i\gets\operatorname{round}_{g_a}(\widehat y_i/q_i)$.\alglabel{line:ub-test-multiplier}
\State Return failure unless $0<\widehat z_i\leq W_{\max}$; otherwise return $\widehat z_i$.\alglabel{line:ub-test-return}
\end{algorithmic}
\end{algorithm}

\Cref{alg:ub-test} first computes a vector $v_i$ close to $H^{-1}b_i$ and the scalar $q_i=b_i^\top v_i$ (\cref{line:ub-test-solve}). After checking that $q_i>0$, it multiplies the sketch pair by $v_i$ to estimate one coordinate and the squared norm of the resulting vector. The test then compares the squared coordinate estimate with $\widehat\nu_i+\ell\|v_i\|_2^2$, rejecting if the coordinate is nonpositive or too small (\cref{line:ub-test-threshold}). Otherwise, it divides the coordinate estimate by $q_i$, rounds the quotient, and returns it as the multiplier estimate (\crefrange{line:ub-test-multiplier}{line:ub-test-return}).

The reason for this quotient is an exact identity. Suppose the working sketch represents $\diag(z)B$, where $z_i$ is row $i$'s current multiplier. The queried vector is $y=\diag(z)Bv_i$, whose $i$th coordinate is $y_i=z_iq_i$. We need not know or store the multipliers $z_i$ to query this vector. Dividing an accurate coordinate estimate by the known scalar $q_i$ therefore estimates $z_i$, even though $v_i$ only approximates $H^{-1}b_i$.

The comparison with the norm identifies the coordinates for which that estimate is accurate enough. CountSketch gives an additive error proportional to $\|y\|_2$; once a coordinate is sufficiently large, this becomes a small relative error. Moreover, when $H$ approximates the represented Gram matrix plus $\ell I$, the fraction $y_i^2/(\|y\|_2^2+\ell\|v_i\|_2^2)$ is comparable to the row's leverage score $z_i^2b_i^\top H^{-1}b_i$. Thus the same test detects important rows and estimates their multipliers. The regularizer's contribution is computed directly from $v_i$ and $\ell$.

Under the sketch guarantees, a row with zero multiplier cannot pass the test. Rounding the returned estimate keeps its bit length bounded. The threshold is $\eta$ when adjusting saved rows and $\theta$ when testing new rows. This adapts CKL's coordinate-to-energy test~\cite[Lemma 6.6 and Figure 7]{CKL22}, with the regularizer included explicitly.

\paragraph{Computing the vector $v_i$}
The row test needs a vector close to $H^{-1}b_i$. \Cref{alg:ub-solve} computes it by iteration from the positive identity term and weighted row list representing $H$. Let $U$ bound its largest eigenvalue, $g_v$ be the rounding spacing for vector coordinates, and $r_{\mathrm{inv}}$ be the number of iterations. All three, including $g_v^{-1}$ in place of $g_v$, are bounded by $Q^{O_{b_0}(1)}$; their exact values are fixed in \cref{par:ub-parameters}.

\begin{algorithm}[H]
\caption{Computing $v_i$}
\label{alg:ub-solve}
\begin{algorithmic}[1]
\Require $H=\ell I+\sum_a w_ab_{i_a}b_{i_a}^\top$ as a weighted row list, and a nonzero $b_i$.
\Statex \textbf{Parameters:} $U,g_v,r_{\mathrm{inv}}$.
\Ensure A vector $v_i$.
\State Set $v\gets0$.
\For{$r=1,\ldots,r_{\mathrm{inv}}$}
 \State Compute $Hv=\ell v+\sum_a w_ab_{i_a}(b_{i_a}^\top v)$ by scanning the list.\alglabel{line:ub-solve-product}
 \State Set $v\gets\operatorname{round}_{g_v}\bigl(v+U^{-1}(b_i-Hv)\bigr)$.\alglabel{line:ub-solve-update}
\EndFor
\State Return $v_i=v$.
\end{algorithmic}
\end{algorithm}

\Cref{alg:ub-solve} starts from $v=0$. At each iteration, it computes $Hv$ by scanning the weighted row list, adds $U^{-1}(b_i-Hv)$ to $v$, and rounds each coordinate to the nearest multiple of $g_v$ (\cref{line:ub-solve-update}). After the prescribed number of iterations, it returns the current vector.

The difference $b_i-Hv$ measures the error in the equation we want to solve. Adding a scaled copy of this difference reduces that error: $U$ bounds the largest eigenvalue, while the positive term $\ell I$ bounds every eigenvalue away from zero. Multiplication by $H$ needs only the list, since each row contributes $w_ab_{i_a}(b_{i_a}^\top v)$ and the identity term contributes $\ell v$. We therefore avoid storing a dense matrix or its inverse.

Rounding bounds the number of bits per coordinate. We choose the spacing small enough that accumulated rounding error remains below the accuracy needed by the row test, and enough iterations to reduce the initial error to the same scale. The computation needs only a constant number of length-$D$ work vectors. We prove the accuracy and space bounds in \cref{clm:ub-inverse-accuracy,clm:ub-scalar-bits}.

\subsubsection{Parameter choices}
\label{par:ub-parameters}
We now fix all quantities used by the algorithms. We may first round $\varepsilon$ and $\delta$ down to powers of two, keeping the same notation. This strengthens the requested guarantees and changes the bounds by at most constant factors. Let $\varepsilon_{\mathrm a}$ be the relative error allowed in multiplier estimates, $\varepsilon_{\mathrm s}$ the error allowed in one sampling step, and $\varepsilon_{\mathrm v}$ the relative error allowed when computing $v_i$.

\begin{enumerate}[label=(\roman*)]
\item We begin with a regularizer large enough that $G\preceq\lambda_0I$, and halve it until reaching $\lambda$. We choose $\lambda_0$ as the least power of two at least $\max\{1,b_0m\mu^2\}$, set $K=1+\log_2(\lambda_0/\lambda)$, and set $\lambda_j=2^{-j}\lambda_0$ for $0\leq j<K$.
\item We choose enough sampling levels that the final sample is empty with the required probability. We set $T=\lceil\log_4(8Km/\delta)\rceil$. Each of the $K$ decoding calls has at most $T$ sampling steps and $T(T+1)$ scans. We allow failure probability $\beta_c=\delta/(8KT)$ for one sampling step and $\beta_s=\delta/(8KT(T+1))$ for the estimates in one scan.
\item We choose the independence of each mask to control its sampling error, setting $h=2\lceil\log_2(4D^2/\beta_c)\rceil$. The errors from successive sampling steps and multiplier corrections must remain small in total. We therefore set $\varepsilon_{\mathrm s}=\varepsilon/(64T)$ and $\varepsilon_{\mathrm a}=\varepsilon/(1000(T+1))$. A fixed relative accuracy suffices for computing each $v_i$, so we set $\varepsilon_{\mathrm v}=1/100$.
\item The selection threshold must be small enough that discarding unselected rows introduces little error. The correction threshold is smaller, so previously selected rows can still be detected while their multipliers are restored. We set $\eta=\varepsilon_{\mathrm s}^2/(2^{14}\cdot1000h)$ and $\theta=1024\eta$, so $16\theta=\varepsilon_{\mathrm s}^2/(1000h)$. For rounding multiplier estimates, we choose $g_a$ as the largest power of two at most $\varepsilon_{\mathrm a}/100$.
\item We bound the saved list length, the multiplier estimates, and the magnitudes of sketch entries, respectively, by $M_{\max}=\lceil12D/\theta\rceil$, $W_{\max}=2^{T+3}\mu$, and $C_{\max}=4m(T+2)W_{\max}(b_0+1)$. These limits bound storage even when a random estimate is inaccurate. We prove below that they do not cause failure when the estimates and sampling steps satisfy their required bounds.
\item The list and multiplier limits give an upper bound on the largest eigenvalue of $H$. We choose $U$ as the least power of two at least $\lambda_0+b_0M_{\max}W_{\max}^2$. To control the rounding error in computing $v_i$, we choose $g_v$ as the largest power of two at most $\varepsilon_{\mathrm v}\lambda/(8U^2D)$. To reduce the error from the initial vector, we set $r_{\mathrm{inv}}=(U/\lambda)\lceil\log_2(8U(b_0+1)/(\varepsilon_{\mathrm v}\lambda))\rceil$.
\end{enumerate}
Each sketch pair uses $O((\varepsilon_{\mathrm a}^2\eta)^{-1}\log(2m/\beta_s))$ entries per column. We implement the coordinate and norm estimates with coefficients in $\{-1,0,1\}$. Recall that $Q=2+mD\mu/(\varepsilon\delta\lambda)$. All the quantities above have $O_{b_0}(\log Q)$-bit rational descriptions. The implementation of the remaining arithmetic, including the tests before exceeding a storage limit, is given in \cref{sec:ub-implementation}.

\subsection{Correctness}
\label{sec:ub-correctness}
We prove \cref{thm:ub-recovery} by analyzing \cref{alg:ub-main}. We restate the guarantee here before giving the three supporting lemmas.

\begin{restate}{thm:ub-recovery}
Let $B\in\Z^{m\times D}$ have at most $s_0$ nonzero entries and squared norm at most $b_0$ in each row, where $s_0,b_0$ are fixed constants. Suppose that a row can be obtained from its index in polynomial time and polylogarithmic space. Let $\mu\geq1$, $0<\varepsilon\leq1/8$, $0<\delta<1/2$, and $\lambda=2^{-L}$ for an integer $L\geq0$, and define $Q=2+mD\mu/(\varepsilon\delta\lambda)$.

For every fixed strict-turnstile stream maintaining integer frequencies $0\leq f_i\leq\mu$, \cref{alg:ub-main} returns, with probability at least $1-\delta$, a set $S\subseteq\{i:f_i>0\}$ and positive weights forming $W=\sum_{i\in S}w_ib_ib_i^\top$ such that
\[
 (1-\varepsilon)(G+\lambda I)\preceq W+\lambda I\preceq(1+\varepsilon)(G+\lambda I),
 \qquad G=B^\top\diag(f)^2B.
\]
The set has $O_{s_0,b_0}(D\varepsilon^{-2}\polylog Q)$ indices. For every choice of seeds, the algorithm uses $O_{s_0,b_0}(D\varepsilon^{-4}\polylog Q)$ bits, including all stored sketches, seeds, output, and working space. The time per update and the time after the stream are polynomial in $m,D,\mu,\varepsilon^{-1},\delta^{-1},\lambda^{-1}$ and the time needed to obtain a row.
\end{restate}

We use three lemmas. The sketching lemma guarantees accurate coordinate and norm estimates for queries fixed independently of the sketch seed. The decoding lemma shows how these estimates produce a small weighted set of active rows approximating $G+\ell I$ at one regularizer $\ell$. The space and time lemma bounds the computation for every choice of seeds, including choices for which decoding fails. Their parameters always have the values fixed in \cref{par:ub-parameters}. We state the lemmas first and then prove \cref{thm:ub-recovery}, including the short induction over the regularizers. Their proofs follow in \cref{sec:ub-sketch-proof,sec:ub-stage-proof,sec:ub-implementation}, respectively.

\begin{lemma}[Sketching]
\label[lemma]{lem:ub-sketching}

Let $\beta_s=\delta/(8KT(T+1))$ be the failure probability allowed for all estimates in one scan of at most $m$ row queries. Let $\Phi$ be a sketch pair constructed as in \cref{alg:ub-encode}, and let $A\in\R^{m\times D}$ and $v_1,\ldots,v_m\in\R^D$ be fixed independently of its seed. Given $\Phi A$, the estimates for $y^{(i)}=Av_i$ satisfy, simultaneously for every queried $i$,
\begin{equation}
\label{eq:ub-estimator-event}
 |\widehat y_i-y_i^{(i)}|\leq\frac{\varepsilon_{\mathrm a}\sqrt\eta}{100}\|y^{(i)}\|_2,
 \qquad 0.9\|y^{(i)}\|_2^2\leq\widehat\nu_i\leq1.1\|y^{(i)}\|_2^2,
\end{equation}
with probability at least $1-\beta_s$.
\end{lemma}

The independence assumption in \cref{lem:ub-sketching} needs care during decoding, because earlier estimates determine changes to later working sketches. We will verify that each queried matrix is determined independently of the seed used to estimate it in \cref{clm:ub-scan-independence}. The next lemma gives the resulting guarantee for a complete decoding call.

\begin{lemma}[Decoding]
\label[lemma]{lem:ub-refinement}
Fix the main inputs and their parameter values, a frequency vector $f\in\{0,\ldots,\mu\}^m$, and $j\in\{0,\ldots,K-1\}$. Put $\ell=\lambda_j$. Suppose $H=\ell I+\sum_{r=1}^{p}a_r^2b_{i_r}b_{i_r}^\top$, where $p\leq M_{\max}$ and each $a_r\in(0,W_{\max}]$ is a multiple of $g_a$. Suppose also that $\frac12H\preceq G+\ell I\preceq2H$ and that $H$ is independent of the masks and sketch seeds used in this call.

With probability at least $1-\delta/K$, \cref{alg:ub-decode} returns a list $F$ of at most $M_{\max}$ active rows with multipliers $\widehat a_i\in(0,W_{\max}]$, each a multiple of $g_a$. The weighted sum $W=\sum_{i\in F}\widehat a_i^2b_ib_i^\top$ satisfies $(1-\varepsilon)(G+\ell I)\preceq W+\ell I\preceq(1+\varepsilon)(G+\ell I)$.
\end{lemma}

\begin{lemma}[Space and time]
\label[lemma]{lem:ub-space}
Let $Q=2+mD\mu/(\varepsilon\delta\lambda)$. For every choice of random seeds, \cref{alg:ub-main} uses $O_{s_0,b_0}(D\varepsilon^{-4}\polylog Q)$ bits, including the stored sketches, seeds, output, and working space. Its output contains at most $O_{s_0,b_0}(D\varepsilon^{-2}\polylog Q)$ indices. The time per update and the time after the stream are polynomial in $m,D,\mu,\varepsilon^{-1},\delta^{-1},\lambda^{-1}$ and the time needed to obtain a row.
\end{lemma}

\begin{proof}[Proof of \cref{thm:ub-recovery}]
Fix a legal stream and let $f$ be its final frequency vector. During the stream, an update $(i,\sigma)$ changes each sampled row matrix by $\sigma\one[i\in E_{j,t}]\mathbf e_i b_i^\top$. The corresponding change to its sketch is exactly the update in \cref{alg:ub-encode}. Since the sketches start at zero, linearity gives $Y_{j,t,u}=\Phi_{j,t,u}\diag(f\one_{E_{j,t}})B$ at the end of the single pass. The same linearity permits additions and subtractions during decoding: changing a row by $ab_i^\top$ changes its sketch by $a(\Phi\mathbf e_i)b_i^\top$.

We prove by induction over the decoding calls that each returned matrix $W_j$ satisfies $(1-\varepsilon)M_j\preceq W_j+\lambda_jI\preceq(1+\varepsilon)M_j$, where $M_j=G+\lambda_jI$. This induction follows the sequence of regularizers of Miller and Peng~\cite[Section 4]{MP12}, as presented by KLMMS~\cite[Theorem 2 and Section 4]{KLMMS17}. For the first call, the row-norm and frequency bounds give $G\preceq\tr(G)I\preceq b_0m\mu^2I\preceq\lambda_0I$, and hence $H_0=\lambda_0I\preceq M_0\preceq2H_0$.

For a later call, suppose by induction that the previous calls returned the required approximations. We set $H_j=W_{j-1}+\lambda_jI$. Since $\lambda_{j-1}=2\lambda_j$, we have $M_j\preceq M_{j-1}\preceq2M_j$. The new matrix has exactly the previous approximation error, since $H_j-M_j=(W_{j-1}+\lambda_{j-1}I)-M_{j-1}$. By the induction hypothesis, this error lies between $-\varepsilon M_{j-1}$ and $\varepsilon M_{j-1}$, and therefore between $-2\varepsilon M_j$ and $2\varepsilon M_j$. Thus $(1-2\varepsilon)M_j\preceq H_j\preceq(1+2\varepsilon)M_j$. As $\varepsilon\leq1/8$, these inequalities imply $\frac12H_j\preceq M_j\preceq2H_j$.

The decoder is therefore given a constant-factor approximation at every call. With probability at least $1-\delta/K$, it returns at most $M_{\max}$ active rows with positive weights whose sum $W_j$ satisfies $(1-\varepsilon)M_j\preceq W_j+\lambda_jI\preceq(1+\varepsilon)M_j$. This is proved in \cref{lem:ub-refinement}, which also guarantees that the returned multipliers lie in $(0,W_{\max}]$ and are multiples of $g_a$. Thus the initial empty list and every successful output meet the representation requirements for the next call. Moreover, $H_j$ depends only on earlier calls, independently of this call's masks and sketch seeds. Thus the same success probability holds after fixing any earlier successful outputs. Summing the probabilities that each call is the first to fail gives total failure probability at most $K(\delta/K)=\delta$.

The last regularizer is $\lambda_{K-1}=\lambda$, so the last output gives the claimed matrix inequality and contains only active rows. The parameter choices give $M_{\max}=O_{b_0}(D\varepsilon^{-2}\polylog Q)$. Finally, all the sketches, seeds, row lists, and working vectors fit within $O_{s_0,b_0}(D\varepsilon^{-4}\polylog Q)$ bits on every random outcome, and each update and the computation after the stream take the stated polynomial time, as proved in \cref{lem:ub-space}.
\end{proof}

\subsubsection{Sketching}
\label{sec:ub-sketch-proof}
We prove \cref{lem:ub-sketching} by combining the standard coordinate and norm estimates from CountSketch and AMS~\cite{CCF04,AMS99}. We state these guarantees without proof, then allocate their failure probabilities so that all estimates in one scan are accurate together.

\begin{claim}[{\cite{CCF04}}]
\label[claim]{clm:ub-countsketch}
For any fixed $y\in\R^m$, coordinate $i$, $0<\alpha<1$, and $0<\beta<1/2$, a CountSketch with $O(\alpha^{-2}\log(1/\beta))$ entries returns an estimate satisfying $\Pr[|\widehat y_i-y_i|>\alpha\|y\|_2]\leq\beta$.
\end{claim}

\begin{claim}[{\cite{AMS99}}]
\label[claim]{clm:ub-ams}
For any fixed $y\in\R^m$ and $0<\beta<1/2$, an AMS sketch with $O(\log(1/\beta))$ entries returns $\widehat\nu$ satisfying $\Pr[\widehat\nu\notin[0.9\|y\|_2^2,1.1\|y\|_2^2]]\leq\beta$.
\end{claim}

\begin{proof}[Proof of \cref{lem:ub-sketching}]
The budget $\beta_s=\delta/(8KT(T+1))$ covers a whole scan. We allocate failure probability $\beta_s/(2m)$ to each of its at most $m$ coordinate estimates and $m$ norm estimates. For a fixed query vector $v_i$, multiplying the stored matrix by $v_i$ gives $(\Phi A)v_i=\Phi(Av_i)$. CountSketch estimates its $i$th coordinate with error at most $(\varepsilon_{\mathrm a}\sqrt\eta/100)\|Av_i\|_2$, with probability at least $1-\beta_s/(2m)$, by \cref{clm:ub-countsketch}. The AMS estimate lies between $0.9\|Av_i\|_2^2$ and $1.1\|Av_i\|_2^2$ with the same success probability, by \cref{clm:ub-ams}. The independence assumed in the lemma permits both estimates for each of the fixed vectors. A union bound over at most $m$ coordinate estimates and $m$ norm estimates gives simultaneous success with probability at least $1-\beta_s$, proving \cref{eq:ub-estimator-event}.
\end{proof}

\subsubsection{Decoding}
\label{sec:ub-stage-proof}
We prove \cref{lem:ub-refinement}. We fix one call, with regularizer $\ell$ and matrix $H$, and put $M_\ell=G+\ell I$. We adapt CKL's proof of squared-weight recovery~\cite[Lemmas 6.5--6.6 and 6.19]{CKL22}. The argument below uses the general rows of $B$, the fixed identity term, and the computed vectors $v_i$.

The target $M_\ell$ and its coarse approximation $H$ remain fixed throughout this call. To analyze the changing row multipliers, we consider two further matrices. The matrix $J_t$ follows the decoder's actual selection and sampling decisions with exact multipliers. The matrix $M_z$ describes the multipliers actually represented by a working sketch, including unfinished corrections and estimation errors. Comparing $J_t$ with $M_\ell$ will bound the sampling error; comparing $M_z$ with $H$ will justify the row tests. Both matrices include the fixed regularizer $\ell I$.

We first define $J_t$ and state the induction used to control it.

Let $F_t$ be the saved set before level $t$. For a row selected at level $c_i$, let $a_i=2^{c_i}f_i$. We set $a_i^{(t)}=a_i$ for $i\in F_t$, $a_i^{(t)}=2^tf_i$ for $i\in E_t\setminus F_t$, and $a_i^{(t)}=0$ otherwise, and define $J_t=\ell I+\sum_i(a_i^{(t)})^2b_ib_i^\top$. Thus saved rows keep their exact selection multipliers, and unsaved sampled rows have their current multipliers. Saving a row fixes its multiplier at its current value, so only sampling changes this exact matrix. Let $q_i^*=b_i^\top H^{-1}b_i$ be the leverage of row $b_i$ relative to $H$.

We prove by induction on $t$ that the following three properties hold whenever the estimates used so far satisfy \cref{eq:ub-estimator-event} and the previous sampling steps satisfy \cref{clm:ub-thinning}:
\begin{enumerate}[label=(\roman*)]
\item $-t\varepsilon_{\mathrm s}H\preceq J_t-M_\ell\preceq t\varepsilon_{\mathrm s}H$;
\item $a_i^2q_i^*\geq\theta/4$ for every $i\in F_t$;
\item $|\widehat a_i-a_i|\leq\varepsilon_{\mathrm a}a_i$ for every $i\in F_t$.
\end{enumerate}
At level zero, $J_0=M_\ell$ and $F_0=\varnothing$. For the induction step, we assume the three properties at the beginning of level $t$. We first prove that the corrections finish and selection preserves the last two properties. We then bound the change from the next sample, which proves the first property at level $t+1$.

We next define the quantities needed to analyze a row test. For a working sketch representing $\diag(z)B$, let $z_i$ be its multiplier for row $i$ and define $M_z=\ell I+\sum_i z_i^2b_ib_i^\top$. Its query vector is $y^{(i)}=\diag(z)Bv_i$, whose $i$th coordinate is $z_iq_i$, where $q_i=b_i^\top v_i$. Define $R_i=(z_iq_i)^2/(v_i^\top M_zv_i)$, the fraction of the query energy, including the identity term, contributed by that coordinate. The quantities $J_t$, $z$, $M_z$, and $q_i^*$ are used only for the proof and are not stored.

We use the following four claims. We state them together, finish the proof of the lemma, and then prove each claim.

\begin{claim}
\label[claim]{clm:ub-detector}
For every nonzero row, $q_i>0$ and $0.96q_i^*\leq q_i^2/(v_i^\top Hv_i)\leq q_i^*$. Fix $z\geq0$ and suppose that the estimates in a scan satisfy \cref{eq:ub-estimator-event}. At threshold $\vartheta\in\{\eta,\theta\}$, every row with $R_i\geq4\vartheta$ passes the acceptance tests, and every row passing them has $R_i\geq\vartheta$ and $z_i>0$. If $z_i\geq0.99$, its rounded estimate satisfies $|\widehat z_i-z_i|\leq\varepsilon_{\mathrm a}z_i$ and is positive.
\end{claim}

\begin{claim}
\label[claim]{clm:ub-one-level}
Fix $0\leq t<T$. Suppose the three inductive properties hold at the beginning of level $t$ and all its estimates satisfy \cref{eq:ub-estimator-event}. Then all corrections finish within $T$ scans. Every row saved after selection is active and satisfies $a_i^2q_i^*\geq\theta/4$ and $|\widehat a_i-a_i|\leq\varepsilon_{\mathrm a}a_i$. There are at most $12D/\theta$ saved rows, and every remaining row satisfies $(a_i^{(t)})^2q_i^*<16\theta$.
\end{claim}

\begin{claim}
\label[claim]{clm:ub-thinning}
Fix the random choices used through the correction and selection at level $t$, and suppose the hypotheses and conclusions of \cref{clm:ub-one-level} hold. With probability at least $1-\beta_c$ over the next sampling mask,
\begin{equation}
\label{eq:ub-thinning-bound}
 -\varepsilon_{\mathrm s}H\preceq J_{t+1}-J_t\preceq\varepsilon_{\mathrm s}H.
\end{equation}
\end{claim}

\begin{claim}
\label[claim]{clm:ub-stage-failure}
With probability at least $1-\delta/K$, every scan satisfies \cref{eq:ub-estimator-event}, every sampling step satisfies \cref{eq:ub-thinning-bound}, and $E_T=\varnothing$.
\end{claim}

\begin{proof}[Proof of \cref{lem:ub-refinement}]
With probability at least $1-\delta/K$, every coordinate and norm estimate has the stated accuracy, every sampling step changes $J_t$ by at most $\varepsilon_{\mathrm s}H$ in either matrix direction, and $E_T$ is empty. This joint probability is proved in \cref{clm:ub-stage-failure}. We fix choices for which these conclusions hold and verify the decoder's output.

We prove the three properties above by induction on the sampling level. Initially, $F_0=\varnothing$ and $J_0=M_\ell$, so all three hold. Suppose they hold at level $t$. The corrections finish within $T$ scans. Each newly saved row has $f_i>0$, $a_i^2q_i^*\geq\theta/4$, and $|\widehat a_i-a_i|\leq\varepsilon_{\mathrm a}a_i$, and the saved list has at most $12D/\theta$ rows; these are the conclusions of \cref{clm:ub-one-level}. The previously saved multipliers and estimates remain unchanged. Saving a row also leaves $J_t$ unchanged, because its multiplier is fixed at its current value. The next sampling step contributes at most $\varepsilon_{\mathrm s}H$ to the error in either direction. Adding this to the previous bound $t\varepsilon_{\mathrm s}H$ proves the first property at level $t+1$, and the saved-row guarantees prove the other two.

After $T$ levels, we have $-T\varepsilon_{\mathrm s}H\preceq J_T-M_\ell\preceq T\varepsilon_{\mathrm s}H$. Since $H\preceq2M_\ell$ and $T\varepsilon_{\mathrm s}=\varepsilon/64$, this gives $(1-\varepsilon/32)M_\ell\preceq J_T\preceq(1+\varepsilon/32)M_\ell$. The final sample is empty, so $J_T$ consists exactly of the saved rows with their exact squared multipliers, together with $\ell I$. Replacing each multiplier by its saved estimate gives
\[
 (1-\varepsilon_{\mathrm a})^2J_T\preceq W+\ell I
 \preceq(1+\varepsilon_{\mathrm a})^2J_T.
\]
Combining these inequalities and $\varepsilon_{\mathrm a}\leq\varepsilon/1000$ yields $(1-\varepsilon)M_\ell\preceq W+\ell I\preceq(1+\varepsilon)M_\ell$.

Every saved row is active, and the list has at most $M_{\max}$ rows. Each saved multiplier is rounded to a multiple of $g_a$ and checked to lie in $(0,W_{\max}]$ by \cref{alg:ub-test}; its square is the output weight. Under the same estimate and sampling bounds, the numerical and list limits cannot interrupt the computation; \cref{clm:ub-guards}, proved in \cref{sec:ub-implementation}, verifies these limits from the bounds within each level. Thus the decoder returns both the required approximation and the stated multiplier representation with probability at least $1-\delta/K$.
\end{proof}

\paragraph{Computing and testing a row}
We first justify the guarantees for a single row test in \cref{clm:ub-detector}. These guarantees will let us analyze both correction and selection. We bound the error of the rounded iteration, then compare the coordinate and norm estimates as in CKL~\cite[Lemma 6.6]{CKL22}. For a vector $v$, define $\|v\|_H=(v^\top Hv)^{1/2}$.

Although $H$ is available as a weighted row list, forming its inverse could require too much space. \Cref{alg:ub-solve} instead approximates $H^{-1}b_i$ using repeated products with $H$. The first claim shows that this approximation is accurate enough for the row test. Without rounding, each iteration shrinks the error. Rounding introduces a small new error, but subsequent iterations shrink that error as well. We bound these two contributions and then use the fact that $b_i$ is a nonzero integer row to turn the resulting absolute error into a relative bound.

\begin{claim}
\label[claim]{clm:ub-inverse-accuracy}
For every $H$ satisfying the representation assumptions of \cref{lem:ub-refinement}, $\lambda I\preceq H\preceq UI$. For every nonzero row $b_i$, \cref{alg:ub-solve} returns $v_i$ satisfying $\|v_i-H^{-1}b_i\|_H\leq\varepsilon_{\mathrm v}\|H^{-1}b_i\|_H$, where $\varepsilon_{\mathrm v}=1/100$.
\end{claim}
\begin{proof}
First, the stored representation gives the required eigenvalue bounds. Its identity term satisfies $\lambda\leq\ell\leq\lambda_0$, and it contains at most $M_{\max}$ rows, each with squared norm at most $b_0$ and multiplier at most $W_{\max}$. Consequently, $\lambda I\preceq H\preceq(\lambda_0+b_0M_{\max}W_{\max}^2)I\preceq UI$.

Let $v_*=H^{-1}b_i$ be the exact solution. An unrounded update sends the error $v-v_*$ to $(I-H/U)(v-v_*)$, since $Hv_*=b_i$. Every eigenvalue of $I-H/U$ lies in $[0,1-\lambda/U]$, so the Euclidean error shrinks by a factor of at most $1-\lambda/U$. Rounding each coordinate to a multiple of $g_v$ then adds a vector of Euclidean norm at most $g_v\sqrt D$. After the prescribed $r_{\mathrm{inv}}$ iterations, we therefore have
\[
 \|v_i-v_*\|_2
 \leq
 \left(1-\frac{\lambda}{U}\right)^{r_{\mathrm{inv}}}\|v_*\|_2
 +g_v\sqrt D\sum_{r=0}^{r_{\mathrm{inv}}-1}
       \left(1-\frac{\lambda}{U}\right)^r.
\]
The first term is the error remaining from the initial vector $v=0$. We have $\|v_*\|_2\leq\sqrt{b_0}/\lambda$, and the choice of $r_{\mathrm{inv}}$ makes this term at most $\varepsilon_{\mathrm v}/(8U)$, using $(1-\lambda/U)^r\leq\exp(-r\lambda/U)$. If $U=\lambda$, it vanishes after one iteration. The second term sums the rounding errors after their subsequent contractions. The geometric sum is at most $U/\lambda$, and our choice $g_v\leq\varepsilon_{\mathrm v}\lambda/(8U^2D)$ makes this term at most $\varepsilon_{\mathrm v}/(8U)$. Thus $\|v_i-v_*\|_2\leq\varepsilon_{\mathrm v}/(4U)$.

It remains to express this error relative to the size of the exact solution in the $H$-norm. Since $H\preceq UI$, the absolute $H$-norm error is at most $\sqrt U\|v_i-v_*\|_2\leq\varepsilon_{\mathrm v}/(4\sqrt U)$. On the other hand, $\|v_*\|_H^2=b_i^\top H^{-1}b_i\geq\|b_i\|_2^2/U\geq1/U$. The last inequality is where we use that $b_i$ is a nonzero integer row. Comparing these bounds gives $\|v_i-v_*\|_H\leq\varepsilon_{\mathrm v}\|v_*\|_H$, as required.
\end{proof}

The row test uses the scalar $q_i=b_i^\top v_i$ and the energy $v_i^\top Hv_i$. The next claim translates the vector approximation into the ratio bound needed for that test.

\begin{claim}
\label[claim]{clm:ub-query-comparison}
If $\|v_i-H^{-1}b_i\|_H\leq\varepsilon_{\mathrm v}\|H^{-1}b_i\|_H$ with $\varepsilon_{\mathrm v}=1/100$, then $q_i>0$ and $0.96q_i^*\leq q_i^2/(v_i^\top Hv_i)\leq q_i^*$.
\end{claim}
\begin{proof}
By Cauchy--Schwarz in the $H$ inner product, $|q_i-q_i^*|\leq\varepsilon_{\mathrm v}q_i^*$. Also $v_i^\top Hv_i\leq(1+\varepsilon_{\mathrm v})^2q_i^*$. These inequalities give positivity and the lower bound, since $(1-\varepsilon_{\mathrm v})^2/(1+\varepsilon_{\mathrm v})^2>0.96$. Cauchy--Schwarz applied to $b_i^\top v_i$ gives the upper bound.
\end{proof}

\begin{claim}
\label[claim]{clm:ub-test}
Suppose $z\geq0$, $q_i>0$, and \cref{eq:ub-estimator-event} holds. At either threshold $\vartheta\in\{\eta,\theta\}$, the acceptance tests pass if $R_i\geq4\vartheta$ and fail if $R_i<\vartheta$. For a row passing them with $z_i\geq0.99$, the rounded estimate has relative error at most $\varepsilon_{\mathrm a}$.
\end{claim}
\begin{proof}
This is the threshold calculation from CKL~\cite[Lemma 6.6]{CKL22}, with the regularizer added exactly. Let $t_i^2=v_i^\top M_zv_i$. The norm estimate gives $0.9t_i^2\leq\widehat t_i^2\leq1.1t_i^2$, and the coordinate error is at most $(\varepsilon_{\mathrm a}\sqrt\vartheta/100)t_i$, since $\vartheta\geq\eta$.

If $R_i<\vartheta$, then $|\widehat y_i|<(1+\varepsilon_{\mathrm a}/100)\sqrt\vartheta\,t_i<\sqrt{1.8\vartheta}\,t_i$, so the test fails. If $R_i\geq4\vartheta$, the nonnegative multiplier and positive $q_i$ give $\widehat y_i\geq(2-\varepsilon_{\mathrm a}/100)\sqrt\vartheta\,t_i>\sqrt{2.2\vartheta}\,t_i>0$, so the test passes.

For a row passing the tests, $R_i\geq\vartheta$. The coordinate error is then at most $(\varepsilon_{\mathrm a}/100)|z_iq_i|$. Dividing by the computed $q_i$ preserves this relative error. Rounding adds at most $g_a/2\leq\varepsilon_{\mathrm a}/200$, whose relative contribution for $z_i\geq0.99$ is at most $\varepsilon_{\mathrm a}/198$. The sum is smaller than $\varepsilon_{\mathrm a}$.
\end{proof}

\begin{proof}[Proof of \cref{clm:ub-detector}]
The computed vector has relative $H$-norm error at most $\varepsilon_{\mathrm v}$ by \cref{clm:ub-inverse-accuracy}. Applying \cref{clm:ub-query-comparison} therefore gives $q_i>0$ and $0.96q_i^*\leq q_i^2/(v_i^\top Hv_i)\leq q_i^*$.

With this positive $q_i$ and the assumed sketch estimates, \cref{clm:ub-test} shows that every row with $R_i\geq4\vartheta$ passes and every passing row has $R_i\geq\vartheta$. A passing row therefore has $z_i>0$, since $z\geq0$ and its energy fraction is positive. For $z_i\geq0.99$, the same claim gives $|\widehat z_i-z_i|\leq\varepsilon_{\mathrm a}z_i$, which also makes the estimate positive because $\varepsilon_{\mathrm a}<1$.
\end{proof}

\paragraph{Correction and selection at one level}
The row test detects a large fraction $R_i$ of the energy in a working sketch. For selection, however, we need to detect rows whose exact contribution $(a_i^{(t)})^2q_i^*$ relative to $H$ is large. Unfinished corrections can distort the working matrix and hence the fraction seen by the test. We will show that, once the corrections finish, $M_z$ is within a constant factor of $H$. For an unsaved row, whose multiplier is still exact, the row-test fraction then approximates $(a_i^{(t)})^2q_i^*$ within a constant factor. Thus selection saves every row whose contribution is too large for the next sampling step.

To prove \cref{clm:ub-one-level}, we therefore first show that the corrections finish, then establish the selection guarantee and bound the saved list. Throughout this argument we assume the three inductive properties at level $t$ and the estimates in \cref{eq:ub-estimator-event} for the scans of that level. The first property, together with $t\varepsilon_{\mathrm s}\leq1/64$, gives $(31/64)H\preceq J_t\preceq(129/64)H$.

We divide the argument into three claims. The correction argument adapts CKL~\cite[Claims 6.20--6.21]{CKL22}, and the selection argument adapts their Lemma 6.19(2)--(3).

\begin{claim}
\label[claim]{clm:ub-correction}
Every correction count becomes zero within $T$ scans. Every positive multiplier used in those scans is at least $0.99$. After the scans, $H/4\preceq M_z\preceq3H$.
\end{claim}

\begin{claim}
\label[claim]{clm:ub-retention}
Every newly selected row $i$ has $f_i>0$, $a_i^2q_i^*\geq\theta/4$, and $|\widehat a_i-a_i|\leq\varepsilon_{\mathrm a}a_i$. Every row left unselected has $(a_i^{(t)})^2q_i^*<16\theta$.
\end{claim}

\begin{claim}
\label[claim]{clm:ub-list-size}
After selection, $|F_{t+1}|\leq12D/\theta$.
\end{claim}

\begin{proof}[Proof of \cref{clm:ub-one-level}]
All correction counts become zero within $T$ scans, every positive working multiplier remains at least $0.99$, and the resulting matrix satisfies $H/4\preceq M_z\preceq3H$, by \cref{clm:ub-correction}. The selection scan therefore begins with the required comparison between its working matrix and $H$.

Each newly saved row is active, satisfies $a_i^2q_i^*\geq\theta/4$, and has a multiplier estimate of relative error at most $\varepsilon_{\mathrm a}$, while every remaining row satisfies $(a_i^{(t)})^2q_i^*<16\theta$, by \cref{clm:ub-retention}. Previously saved rows keep their multipliers and estimates, and $H$ is fixed, so their guarantees persist. The entire saved list has at most $12D/\theta$ rows by \cref{clm:ub-list-size}.
\end{proof}

Recall that $d_i$ counts the remaining halvings for a saved row; for this calculation, set $d_i=0$ for unselected rows. To prove \cref{clm:ub-correction}, we show that the largest positive count decreases in every scan. The first claim bounds the error accumulated during the halvings. The second shows that every row attaining the largest count is detected and has its count decreased.

\begin{claim}
\label[claim]{clm:ub-correction-error}
At the beginning of each correction scan, its represented multipliers satisfy $0.99\,2^{d_i}a_i^{(t)}\leq z_i\leq1.01\,2^{d_i}a_i^{(t)}$. The same bounds hold for later copies after each correction applied to them, with the updated counts $d_i$.
\end{claim}
\begin{proof}
We adapt the calculation in CKL~\cite[Claim 6.20]{CKL22}. Initially, a selected row in $E_t$ has exact multiplier $2^{d_i}a_i$, a selected row outside $E_t$ has multiplier $\widehat a_i$, and an unselected row has its ideal multiplier $a_i^{(t)}$.

Suppose the next estimate is $\widehat z_i=z_i(1+u_i)$ with $|u_i|\leq\varepsilon_{\mathrm a}$. Subtracting half that estimate changes the multiplier to $z_i(1-u_i)/2$ and decreases $d_i$ by one. Relative to an exact halving, this contributes a factor in $[1-\varepsilon_{\mathrm a},1+\varepsilon_{\mathrm a}]$. At most $T$ halvings and one saved-estimate error give a factor between $(1-\varepsilon_{\mathrm a})^{T+1}$ and $(1+\varepsilon_{\mathrm a})^{T+1}$, and hence in $[0.99,1.01]$.

This calculation also justifies the estimate used at the next correction. Every positive ideal multiplier is an integer at least one, so the inequalities keep a positive working multiplier at least $0.99$. We may therefore apply \cref{clm:ub-detector} at that correction. Induction over the corrections proves the claim, including intermediate modifications of later copies. At each new level we start from the original stored sketches, so these correction errors do not accumulate across levels.
\end{proof}

We next show why a row with the largest remaining correction count is still detectable. Up to the small errors just bounded, such a row has been amplified at least as much as every other row relative to its exact multiplier in $J_t$. Its squared contribution and the bound on the total query energy therefore contain the same amplification factor. These factors cancel when we compare the coordinate with the total energy, leaving a score large enough for detection because the row was selected earlier.

\begin{claim}
\label[claim]{clm:ub-max-correction}
If $\Delta=\max_i d_i>0$ at the beginning of a correction scan, every selected row with $d_i=\Delta$ has $R_i\geq4\eta$ in that scan.
\end{claim}
\begin{proof}
We use the cancellation in CKL~\cite[Claim 6.21]{CKL22}. By \cref{clm:ub-correction-error}, $M_z\preceq1.01^2 4^\Delta J_t\preceq3\cdot4^\Delta H$, including the unchanged identity term. A row with $d_i=\Delta$ has $z_i^2\geq0.99^2 4^\Delta a_i^2$. Consequently,
\[
 R_i\geq\frac{0.99^2}{3}a_i^2\frac{q_i^2}{v_i^\top Hv_i}
 \geq\frac14a_i^2q_i^*\geq\frac\theta{16}=64\eta>4\eta.
\]
We used \cref{clm:ub-query-comparison} and the induction hypothesis for saved rows. The queried copy is unchanged during the scan, so this estimate applies to every row with the stated initial count.
\end{proof}

\begin{proof}[Proof of \cref{clm:ub-correction}]
Suppose the largest count at the start of a scan is $\Delta>0$. Every row with count $\Delta$ has $R_i\geq4\eta$ by \cref{clm:ub-max-correction}, so the row test detects it by \cref{clm:ub-detector}. The algorithm subtracts half its estimated multiplier from the later copies and decreases its count by one. All other counts are already at most $\Delta-1$, and no count increases. Thus the largest count after the scan is at most $\Delta-1$. Initially all counts are at most $t<T$, so all become zero within $T$ scans.

The multiplier bounds in \cref{clm:ub-correction-error} keep every positive multiplier at least $0.99$ throughout these scans. When all counts reach zero, the same bounds give $0.99^2J_t\preceq M_z\preceq1.01^2J_t$. Combining this with $(31/64)H\preceq J_t\preceq(129/64)H$ gives $H/4\preceq M_z\preceq3H$, as required.
\end{proof}

\begin{proof}[Proof of \cref{clm:ub-retention}]
Correction leaves the multipliers of unselected rows unchanged, so $z_i=a_i^{(t)}$. The bound $H/4\preceq M_z\preceq3H$ from \cref{clm:ub-correction}, together with \cref{clm:ub-query-comparison}, gives $\frac14(a_i^{(t)})^2q_i^*\leq R_i\leq4(a_i^{(t)})^2q_i^*$. For the lower bound we used $0.96/3>1/4$. This comparison relates the row test to the exact multipliers in the induction.

A row passing the test has $R_i\geq\theta$ by \cref{clm:ub-detector}, so the upper comparison gives $a_i^2q_i^*\geq\theta/4$. Its multiplier is $2^tf_i>0$, so it is active. This multiplier is an integer at least one, and \cref{clm:ub-detector} bounds the estimate's relative error by $\varepsilon_{\mathrm a}$.

Conversely, an unselected row with $(a_i^{(t)})^2q_i^*\geq16\theta$ would have $R_i\geq4\theta$ by the lower comparison and would therefore be selected. This proves the bound for every remaining row.
\end{proof}

\begin{proof}[Proof of \cref{clm:ub-list-size}]
Every selected row has $a_i^2q_i^*\geq\theta/4$, by \cref{clm:ub-retention} or the induction hypothesis. All these contributions occur in $J_t$. Thus $|F_{t+1}|\theta/4\leq\sum_{i\in F_{t+1}}a_i^2q_i^*\leq\tr(H^{-1}J_t)\leq3D$. The last inequality follows by conjugating $J_t\preceq3H$ by $H^{-1/2}$ and taking traces. Rearranging proves the claim. The same bound holds for the list after any prefix of the selection scan.
\end{proof}

\paragraph{The change caused by sampling}
We have shown that correction finishes, the saved list stays small, and every remaining row has a small contribution. We now bound the change caused by sampling the remaining rows. A remaining row with multiplier $a$ survives with probability $1/4$ and has its multiplier doubled if it survives. Its squared contribution is therefore preserved in expectation, since $\frac14(2a)^2=a^2$. Saved rows and the identity term stay unchanged. Consequently, sampling preserves the entire matrix in expectation. Since selection has removed every individually large contribution from this sample, a concentration bound will show that the actual change is small as well.

To measure that change relative to $H$, set $A_i=(a_i^{(t)})^2H^{-1/2}b_ib_i^\top H^{-1/2}$ for each row left unselected, where $H^{-1/2}$ is the inverse of the positive definite square root of $H$. Multiplication on both sides by $H^{-1/2}$ turns a bound relative to $H$ into a bound relative to $I$. In particular, the operator norm of $A_i$ is exactly $(a_i^{(t)})^2q_i^*$, which selection has made small.

The sampling calculation adapts CKL~\cite[Lemma 6.19(4)]{CKL22}. Our masks have $h$-wise independent entries so that we can store them with short seeds. To justify this amount of independence, we bound an even $h$-th trace moment of the sampling error. Its expansion contains products involving at most $h$ distinct indicators, so its expectation is the same as for fully independent sampling. We obtain the latter bound from Tropp's matrix Bernstein inequality~\cite[Theorem 6.1(ii)]{Tropp12}. We state the resulting concentration bound next, apply it to prove \cref{clm:ub-thinning}, and then give its proof.

\begin{claim}
\label[claim]{clm:ub-thinning-concentration}
Let $A_i\succeq0$, $\sum_iA_i\preceq3I$, and $\|A_i\|_{\mathrm{op}}\leq\tau$. Suppose the indicators $\xi_i$ are $h$-wise independent with mean $1/4$, where $h$ is even and $h\geq\ln(4D^2/\beta)$. If $0<\varepsilon_{\mathrm s}\leq1$ and $\tau\leq\varepsilon_{\mathrm s}^2/(1000h)$, then $\Pr[\|\sum_i(4\xi_i-1)A_i\|_{\mathrm{op}}>\varepsilon_{\mathrm s}]\leq\beta$.
\end{claim}

\begin{proof}[Proof of \cref{clm:ub-thinning}]
Fix the choices made through correction and selection. For every row left unselected, $A_i\succeq0$ and $\|A_i\|_{\mathrm{op}}=(a_i^{(t)})^2q_i^*<16\theta$ by \cref{clm:ub-retention}. Their sum is bounded by $H^{-1/2}J_tH^{-1/2}\preceq3I$, so these matrices meet the size bounds in \cref{clm:ub-thinning-concentration}.

The queries so far use only the masks defining $E_t$ and sketches at that level or earlier. They do not use the next mask $\xi_t$, whose seed is independent. A remaining row disappears if $\xi_t(i)=0$ and has its multiplier doubled otherwise, so its squared contribution is multiplied by $4\xi_t(i)$. Saved rows and $\ell I$ stay unchanged. Hence $H^{-1/2}(J_{t+1}-J_t)H^{-1/2}=\sum_i(4\xi_t(i)-1)A_i$.

We can now apply \cref{clm:ub-thinning-concentration} with $\tau=16\theta$ and $\beta=\beta_c$, since the parameter choices give $16\theta=\varepsilon_{\mathrm s}^2/(1000h)$ and the required independence. With probability at least $1-\beta_c$, the sum has operator norm at most $\varepsilon_{\mathrm s}$ and hence lies between $-\varepsilon_{\mathrm s}I$ and $\varepsilon_{\mathrm s}I$. Multiplying on both sides by $H^{1/2}$ gives the required bound $-\varepsilon_{\mathrm s}H\preceq J_{t+1}-J_t\preceq\varepsilon_{\mathrm s}H$.
\end{proof}

To prove \cref{clm:ub-thinning-concentration}, let $Y=\sum_i(4\xi_i-1)A_i$. We first bound an even trace moment for independent indicators and then retain that bound under $h$-wise independence.

\begin{claim}
\label[claim]{clm:ub-thinning-moment}
For fully independent indicators, $\mathbb E\tr(Y^h)\leq4D^2\max\{\sqrt{18h\tau},4h\tau\}^{\,h}$.
\end{claim}
\begin{proof}
The centered summands $X_i=(4\xi_i-1)A_i$ have norm at most $R_0=3\tau$. Also $\mathbb EX_i^2=3A_i^2$, so their variance satisfies $\sigma^2=\|\sum_i\mathbb EX_i^2\|_{\mathrm{op}}\leq9\tau$. Applying matrix Bernstein~\cite[Theorem 6.1(ii)]{Tropp12} to $Y$ and $-Y$ gives
\[
 \Pr(\|Y\|_{\mathrm{op}}\geq u)
 \leq2D\left(\exp\!\left(-\frac{u^2}{4\sigma^2}\right)+\exp\!\left(-\frac{3u}{4R_0}\right)\right).
\]
We integrate against $hu^{h-1}$. The first integral is $(2\sigma)^h\Gamma(h/2+1)\leq(\sqrt{2h}\,\sigma)^h$, and the second is $(4R_0/3)^h h!\leq(4R_0h/3)^h$. Using $\tr(Y^h)\leq D\|Y\|_{\mathrm{op}}^h$ and the bounds on $\sigma^2,R_0$ proves the claim. If the variance is zero, the conclusion is immediate.
\end{proof}

\begin{proof}[Proof of \cref{clm:ub-thinning-concentration}]
For fully independent indicators, \cref{clm:ub-thinning-moment} gives $\mathbb E\tr(Y^h)\leq4D^2\max\{\sqrt{18h\tau},4h\tau\}^{\,h}$. This bound also holds under $h$-wise independence: expanding $\tr(Y^h)$ as a polynomial in the indicators and replacing $\xi_i^r$ by $\xi_i$ leaves monomials with at most $h$ distinct indicators. Each has the same expectation as in the fully independent case. The matrices $A_i$ need not commute for this argument.

Since $h$ is even, $\|Y\|_{\mathrm{op}}^h\leq\tr(Y^h)$. Markov's inequality therefore gives $\|Y\|_{\mathrm{op}}\leq\exp(1)\max\{\sqrt{18h\tau},4h\tau\}$ with probability at least $1-4D^2\exp(-h)\geq1-\beta$. The assumption $\tau\leq\varepsilon_{\mathrm s}^2/(1000h)$ makes this threshold at most $\varepsilon_{\mathrm s}$, as required.
\end{proof}

\paragraph{Probability of failure}
These arguments prove the required progress at each level when the estimates and sampling errors satisfy their bounds. It remains to show that all these bounds hold together and that the final sample is empty, as stated in \cref{clm:ub-stage-failure}.

We first verify the independence needed to apply \cref{lem:ub-sketching}. In \cref{alg:ub-correct}, answers from one sketch copy modify only later copies; the copy being queried stays unchanged throughout its scan. Thus earlier answers determine the matrix represented by the next copy before we query it, independently of that copy's seed. The final copy used for selection is also unchanged during its scan. This is the reason for keeping independent copies, following CKL~\cite[Figure 6]{CKL22}. The next claim makes the argument precise.

\begin{claim}
\label[claim]{clm:ub-scan-independence}
Before a scan, fix the input stream, the previous calls' outputs, the masks defining $E_t$, and the seeds of sketch pairs already queried. These choices determine the represented matrix and query vectors independently of the current pair's seed. Conditional on these choices, the probability that the scan violates \cref{eq:ub-estimator-event} is at most $\beta_s$.
\end{claim}
\begin{proof}
The matrix $H$ depends only on earlier calls. At level $t$, the initial represented multipliers depend on $E_t$ and the previously saved estimates. Every subsequent modification to the current copy is determined by answers from earlier copies. Once those earlier seeds are fixed, all these modifications are fixed independently of the current seed. The represented matrix remains unchanged throughout the scan, and each $v_i$ depends only on $H,b_i$ and the deterministic rounding choices.

We may therefore apply \cref{lem:ub-sketching} to this matrix and all its possible row queries simultaneously. The conditional probability that any estimate in the scan is inaccurate is at most $\beta_s$, regardless of which indices the algorithm inspects. By \cref{clm:ub-guards} below, storage checks cannot stop the computation before one of the stated estimate or sampling bounds fails; we do not condition on the entries of an unqueried sketch.
\end{proof}

\begin{proof}[Proof of \cref{clm:ub-stage-failure}]
We consider the first scan or sampling step to violate its required inequality. Until then, \cref{clm:ub-one-level,clm:ub-thinning} preserve the three inductive properties, and \cref{clm:ub-guards} ensures that no storage check interrupts the computation. Before a scan we fix the choices in \cref{clm:ub-scan-independence}; the conditional probability of an inaccurate estimate is at most $\beta_s$. After selection we fix those choices and answers before revealing the independent next mask. With these choices fixed, \cref{clm:ub-thinning} bounds its probability of violating \cref{eq:ub-thinning-bound} by $\beta_c$.

It remains to bound the probability that the final sample is nonempty. A fixed index belongs to $E_T$ precisely when all $T$ independently seeded masks retain it, each with probability $1/4$. Thus $\Pr[i\in E_T]=4^{-T}$, and a union bound over the $m$ indices gives $\Pr[E_T\ne\varnothing]\leq m4^{-T}$.

There are at most $T(T+1)$ scans and $T$ sampling steps. Adding their failure probabilities and the probability of a nonempty final sample gives at most $T(T+1)\beta_s+T\beta_c+m4^{-T}\leq3\delta/(8K)<\delta/K$. This proves the claim and completes the proof of \cref{lem:ub-refinement}.
\end{proof}

\subsubsection{Space and time analyses}
\label{sec:ub-implementation}

We prove \cref{lem:ub-space}, with $Q=2+mD\mu/(\varepsilon\delta\lambda)$. The main storage consists of the sketches, whose dimensions are fixed before the stream. We also need space for their seeds, the saved row lists, and the vectors used in decoding. Before counting these objects, we justify that each number needs only $O_{b_0}(\log Q)$ bits and that the storage limits do not interrupt the correctness argument.

\begin{claim}
\label[claim]{clm:ub-scalar-bits}
Every stored scalar and arithmetic operand needs $O_{b_0}(\log Q)$ bits, with failure declared before exceeding this bound. Computing $v_i$ needs $O_{b_0}(D\log Q)$ additional bits, besides the space needed to obtain rows.
\end{claim}
\begin{proof}
All magnitude bounds and reciprocals of rounding spacings in \cref{par:ub-parameters} are polynomial in $Q$. During the stream, each sketch entry is an integer signed sum of magnitude at most $m\mu\sqrt{b_0}$. This depends on the current frequencies, so cancelling updates do not increase the required precision. During decoding, working entries lie on the fixed grid of spacing $g_a/2$ and are bounded by $C_{\max}$. Saved multipliers lie on the grid of spacing $g_a$ and are bounded by $W_{\max}$. Thus these numbers have $O_{b_0}(\log Q)$-bit representations.

For the vector computation, $\lambda I\preceq H\preceq UI$ follows from the list and multiplier limits. The contraction argument in \cref{clm:ub-inverse-accuracy} bounds every iterate by a polynomial in $Q$, and each coordinate is rounded to a multiple of $g_v$. Multiplication by $H$ scans its weighted row list, so a constant number of length-$D$ vectors suffices.

The remaining operations are sums, products, medians, and comparisons of these numbers. We choose power-of-two group sizes for the AMS averages, and round $\widehat y_i/q_i$ immediately to the $g_a$ grid after checking $q_i>0$. Hence denominators do not accumulate across operations. The intermediate sums and products also have $O_{b_0}(\log Q)$ bits. Choosing a sufficiently large constant in this bit bound and checking it before storing each result gives the guarantee for every choice of seeds.
\end{proof}

\begin{claim}
\label[claim]{clm:ub-guards}
Consider a call satisfying the hypotheses of \cref{lem:ub-refinement}. Before the first violation of \cref{eq:ub-estimator-event,eq:ub-thinning-bound}, no limit on the saved list, multiplier estimates, sketch entries, or arithmetic is exceeded.
\end{claim}
\begin{proof}
Assume that the estimates and sampling steps so far satisfy their inequalities. The induction in \cref{clm:ub-one-level,clm:ub-thinning} then applies to every completed level, and \cref{clm:ub-list-size} bounds each prefix of the saved list by $M_{\max}$.

At level $t$, every working multiplier is at most $1.01\,2^t\mu$, including those in partially modified sketch copies, by \cref{clm:ub-correction-error}. Its reported estimate is therefore below $W_{\max}=2^{T+3}\mu$. Preparing a sketch adds at most one saved contribution per row; subsequent corrections subtract at most $T$ half-estimates per row. Each sketch coefficient has magnitude at most one, and each row coefficient at most $\sqrt{b_0}$. The triangle inequality therefore bounds every intermediate sketch entry by $C_{\max}=4m(T+2)W_{\max}(b_0+1)$. This also bounds copies not yet queried. Finally, the preceding claim bounds the vector computation and remaining arithmetic. Thus the storage checks introduce no failure while the inequalities used in the correctness proof hold.
\end{proof}

\begin{proof}[Proof of \cref{lem:ub-space}]
\textbf{Sketches and seeds.}
There are $KT(T+1)=O_{b_0}(\log^3Q)$ sketch pairs. Each pair is applied to $D$ columns and has $O((\varepsilon_{\mathrm a}^2\eta)^{-1}\log(2m/\beta_s))$ entries per column. Since $\varepsilon_{\mathrm a}^{-1}=O(\varepsilon^{-1}\log Q)$ and $\eta^{-1}=O(\varepsilon^{-2}\log^3Q)$, all stored sketches together have $O_{b_0}(D\varepsilon^{-4}\polylog Q)$ entries. The $T+1$ working copies needed at one sampling level fit within the same bound. Each entry needs $O_{b_0}(\log Q)$ bits by \cref{clm:ub-scalar-bits}, giving the claimed space bound for the sketches.

The seeds take only $\polylog Q$ additional bits. An $h$-wise independent mask has an $O(h\log(2m))$-bit description by the standard polynomial construction~\cite[Construction 3.32 and Proposition 3.33]{Vadhan12}. Each CountSketch and AMS pair also has a $\polylog Q$-bit description~\cite{CCF04,AMS99}, and there are only polylogarithmically many masks and pairs. These descriptions can be constructed and evaluated in polynomial time and polylogarithmic workspace; the sampled sets are never stored.

\textbf{Row lists and working space.}
Each list has at most $M_{\max}=\lceil12D/\theta\rceil=O_{b_0}(D\varepsilon^{-2}\polylog Q)$ rows. We store an index, a selection level, and a multiplier for each row, taking $O_{b_0}(\log Q)$ bits per entry. Two consecutive lists suffice, since a new output replaces the preceding list. Computing $v_i$ adds only $O_{b_0}(D\log Q)$ work bits, and rows of $B$ are obtained one at a time. Thus the lists, output, and working space also fit within $O_{s_0,b_0}(D\varepsilon^{-4}\polylog Q)$ bits.

The algorithm checks all storage limits before exceeding them, so this bound holds even when an estimate is inaccurate. The output has the claimed size because its list never exceeds $M_{\max}$. The preceding claim ensures that these checks do not affect the correctness guarantee.

\textbf{Time.}
Every loop bound is polynomial in $m,D,\mu,\varepsilon^{-1},\delta^{-1},\lambda^{-1}$. An update evaluates the masks and sketch maps and modifies the entries corresponding to the nonzero coordinates of its row. After the stream, each scan examines at most $m$ rows. Each vector iteration computes $Hv$ by scanning a list of at most $M_{\max}$ rows, and the prescribed number $r_{\mathrm{inv}}$ of iterations is polynomial in the same parameters. All arithmetic involves $O_{b_0}(\log Q)$-bit numbers. Including seed construction and row access, both the time per update and the time after the stream are therefore polynomial as claimed.
\end{proof}

\subsection{Removing the regularizer}
\label{sec:ub-unregularized}
We now prove the corollary stated at the end of \cref{sec:upper}, which uses a condition-number bound to remove the identity term.

\begin{restate}{cor:ub-unregularized}
\unregularizedsparsificationstatement
\end{restate}

\begin{proof}
Apply \cref{thm:ub-recovery} with $\varepsilon=\varepsilon_0/8$ and $\lambda$ the largest power of two at most $1/\kappa$. On its success event, the returned rows are active and the regularized matrices satisfy the stated approximation. If $G=0$, no nonzero row is active, so the output is empty and the conclusion holds. Assume henceforth that $G\ne0$.

Some active row $f_ib_i$ is a nonzero integer vector. Since $G\succeq f_i^2b_ib_i^\top$, we have $\lambda_{\max}(G)\geq\|f_ib_i\|_2^2\geq1$, and the assumed eigenvalue ratio gives $\lambda_{\min}^+(G)\geq1/\kappa$. Thus $\lambda I\preceq G$ on the range of $G$. Subtracting $G+\lambda I$ from the regularized approximation shows that $-\varepsilon(G+\lambda I)\preceq W-G\preceq\varepsilon(G+\lambda I)$. On this range, it follows that $-2\varepsilon G\preceq W-G\preceq2\varepsilon G$.

If $x$ is in the kernel of $G$, then $0=x^\top Gx=\sum_i f_i^2(b_i^\top x)^2$, so $b_i^\top x=0$ for every active row. The output contains only active rows, hence $Wx=0$ as well. Both symmetric matrices vanish on the orthogonal complement of the range, and the previous inequalities therefore hold on the whole space. Since $2\varepsilon\leq\varepsilon_0$, they imply the desired approximation.

Finally, $1/(2\kappa)<\lambda\leq1/\kappa$, so $\lambda^{-1}=O(\kappa)$ and the parameter $Q$ of \cref{thm:ub-recovery} is $O(Q_\kappa)$. Substitution gives the output-size, space, and time bounds. The success probability remains at least $1-\delta$.
\end{proof}